\documentclass[useAMS,usenatbib]{mnras}

\pdfoutput=0

\usepackage{amssymb}
\usepackage{amsmath}
\usepackage{times}
\usepackage{graphicx}
\usepackage{wasysym}

\usepackage[utf8]{inputenc}

\DeclareFontFamily{U}{gbsn5d}{}
\DeclareFontFamily{U}{gbsn66}{}
\DeclareFontShape{U}{gbsn5d}{m}{n}{<-> gbsnu5d}{}
\DeclareFontShape{U}{gbsn66}{m}{n}{<-> gbsnu66}{}

\DeclareUnicodeCharacter{5DDD}{{\usefont{U}{gbsn5d}{m}{n}\char"DD}}
\DeclareUnicodeCharacter{666E}{{\usefont{U}{gbsn66}{m}{n}\char"6E}}

\def\aap{A\&A}
\def\apj{ApJ}

\def\apjl{ApJ}
\def\pasa{PASA}
\def\mnras{MNRAS}
\def\araa{ARA\&A}

\def\nat{Nat}

\def\apjs{ApJS}

\newcommand{\mincir}{\raise
-2.truept\hbox{\rlap{\hbox{$\sim$}}\raise5.truept 
\hbox{$<$}\ }}
\newcommand{\magcir}{\raise
-2.truept\hbox{\rlap{\hbox{$\sim$}}\raise5.truept
\hbox{$>$}\ }}
\newcommand{\minmag}{\raise-2.truept\hbox{\rlap{\hbox{$<$}}\raise
6.truept\hbox
{$>$}\ }}

\newcommand{\be}{\begin{equation}}
\newcommand{\ee}{\end{equation}}
\newcommand{\ba}{\begin{eqnarray}}
\newcommand{\ea}{\end{eqnarray}}
\newcommand{\brr}{\begin{array}}
 
\newcommand{\err}{\end{array}}
\newcommand{\bc}{\begin{center}}
\newcommand{\ec}{\end{center}}

\DeclareMathAlphabet{\mathsc}{OT1}{cmr}{m}{sc}
\def\testbx{bx}%
\DeclareRobustCommand{\ion}[2]{%
\relax\ifmmode
\ifx\testbx\f@series
{\mathbf{#1\,\mathsc{#2}}}\else
{\mathrm{#1\,\mathsc{#2}}}\fi
\else\textup{#1\,{\mdseries\textsc{#2}}}%
\fi}

\title[{\rm The gamma growth of Dark matter haloes}]{Modelling the mass accretion histories of dark matter haloes using a Gamma formalism}

\author[A. Katsianis et al.]{Antonios Katsianis$^{1, 2, 3 }$ \thanks{E-mail:
    katsianis@mail.sysu.edu.cn}, Xiaohu Yang$^{2,3,4}$, Matthew Fong$^2$ and Jie Wang$^{5,6}$ \\
  $^1$ School of Physics and Astronomy, Sun Yat-sen University, Zhuhai Campus, 2 Daxue Road, Xiangzhou District, Zhuhai, P. R. China \\
  $^2$  Department of Astronomy, Shanghai Key Laboratory for Particle Physics and Cosmology, Shanghai Jiao Tong University, Shanghai 200240, China \\
  $^3$ Key Laboratory for Particle Physics, Astrophysics and Cosmology, Ministry of Education, Shanghai Jiao Tong University, Shanghai 200240, China \\
   $^4$ Tsung-Dao Lee Institute, Shanghai Jiao Tong University, Shanghai 200240, China \\
   $^5$Key Laboratory for Computational Astrophysics,National Astronomical Observatories, Chinese Academy of Sciences, 20A Datun Road, Beijing 100101, China \\
   $^6$School of Astronomy and Space Science, University of Chinese Academy of Sciences, Beijing 100039, China \\
} 

\begin{document}

\maketitle

\begin{abstract}
We present a physical model of the Mass Accretion Histories (MAH) of haloes in concordance with the {\it observed} cosmic star formation rate density (CSFRD). We model the MAHs of dark matter haloes using a Gamma ($\Gamma$) functional form: $M_h(T) =  \frac{M_0}{f_{0}} \, \times \frac{\gamma(\alpha_h, ~\beta_h  \times (T-Th))}{\Gamma(\alpha_h)}$, where $M_0$ is the halo mass at present time, $T$ is time, $\alpha_h$ and $\beta_h$ are parameters we explore, $f_{0}$ is the percentage of the mass of the halo at z = 0 with respect to the final mass of the halo achieved  at $T = \infty$. We use the MAHs of haloes obtained from cosmological simulations and analytical models to constrain our model. $f_{0}$ can be described by a power-law ($f_{0} = 1- c \times  M_{0}^{d}$). Haloes with small masses have already on average attained most of their final masses. The average $<f_{0}>$ of haloes in the Universe is $ > 0.95$ pointing to the direction that the cosmic MAH/CSFRD is saturated at our era. The average $<\beta_{h}>$ parameter (the depletion rate of the available dark matter for halo growth) is related to the dynamical timescales of haloes. The $\alpha$ parameter is a power-law index of $M_{0}$ and represents the early growth a halo experiences before the expansion of the Universe starts to slow it down. Finally, $T_{h}$ (the time that marks the co-evolution/growth of galaxies and haloes after the Big Bang) is found to be 150-300 million years. 
\end{abstract}


\section{Introduction}
\label{sec_intro}

In the current paradigm of structure and galaxy formation, a key concept of the buildup of structure in the Universe, is the hierarchical formation of dark matter (DM) haloes. Galaxies are assumed to form by the cooling and condensation of baryons within these haloes \citep[see][for an overview]{Mo2010}. Apart from the sophisticated hydro-dynamical simulation of galaxy formation \citep[e.g.][]{Schaye2015,Katsianis2016,Pillepich2018,dave2019,Zhaoka2020}, our understanding of the galaxy formation processes within dark matter haloes can be probed using semi-analytical \citep[e.g.][]{Mo1998, Somerville2008,Lagos2018} or empirical galaxy formation models \citep[e.g.][]{Lu2014, Li2016,Behroozi2019,Chen2021,Katsianis2021b}. In these theories, the backbone to model the galaxy birth and growth is the mass accretion histories of dark matter haloes, which can either be extracted from N-body simulations or be generated from analytical methods, e.g. the extended Press-Schechter formalism \citep{PS1974,Correa2015} theory or empirical fitting formulas.


Along this line, the mass accretion history (MAH), $M(z)$, of a dark matter halo along its main branch have been extensively studied during the past twenty years  \citep{Wechsler2002,vandenBosch2002,Zhao2003a,Zhao2003,Tasitsimi2004, Zhao2009, McBride2009,Fakhouri2009,Ludlow2013,Wu2013,Van2014,Correa2015b,Deboni2016,Puebla2016,Diemer2017,Mostoghiu2019,MonteroDorta2021,Hearin2021,Davies2021}. 

There is a census between numerous studies that typically dark matter haloes follow initially a high mass growth at early times followed by a turnover at later times. In order to quantify their results (which usually rely on the halo merger trees extracted from numerical simulations) different authors adopt a variety of MAH models. These typically are presented as a function of redshift $z$ and investigate the question: given the halo mass $M_{0}$ at $z \sim 0$ what is on average the mass accretion history path that was followed ?

Since galaxy formation and evolution is supposed to occur in growing dark matter haloes, MAHs are typically used as a backbone for modelling baryonic gas accretion (inflows, outflows), star formation, etc. \citep{Dutton2007,Dave2012,Yang2012, Yang2013,Dekel2014,PengMaolino2014,LuYi2020,Lopez2020,Sharma2020,Tejos2021,Yang2021,Fernandez-Figueroa2022,Wang2022,Li2022}. For example, \citet{vandenBosch2002} computed the average MAH of dark matter haloes and then used the former to follow the baryonic gas accretion to galaxies and the CSFRD. \citet{Yang2012, Yang2013} used the MAH generated from the analytical modeling of \citet{Yang2011} to predict the stellar mass growth and star formation histories of galaxies in haloes of different masses. \citet{PengMaolino2014} used the MAHs from cosmological simulations \citep{Faucher2011} and then followed the evolution of gas and stars within galaxies using simple analytic solutions. \citet{Sharma2020} formulated the Ikea galaxy formation model by using as a backbone the MAH given by \citet[][]{Correa2015b}. A questions arises. Can we follow the vice versa logic and construct a MAH model relying on {\it observations}? 

Traditionally, MAHs of dark matter haloes and SFHs of galaxies are modelled as a function of redshift $z$.  
However, as pointed out in a recent study by \citet{Katsianis2021b}, if modelled as a function of cosmic time, the {\it observed} CSFRD can be nicely  described by a form that resembles a $\Gamma$ distribution with only two free parameters. Compared to the traditional CSFRD models as a function of redshift $z$, this new $\Gamma$ formalism as a function of cosmic time, (besides that requires less parameters) is physically motivated as is closely related to the star formation physics itself, e.g., the gas accretion and depletion time scales. In addition, a ``$\Gamma$ formalism'' provides an opportunity to relate mathematically SFHs to other branches of science and engineering, as the $\Gamma$ distribution and $\Gamma$ function have a widespread use, from modeling the spread of infectious diseases to the deterioration of buildings \citep{Maghsoodloo2014,Vazquez2020,McCombs2020,ZiffandZiff}.  

Given the intrinsic connections between the growth of dark matter haloes and the growth of galaxies, in our current work we set out to probe a new MAH model, $M_h(T)$, in concordance with the CSFRD model constrained from observations. The model parameters are then constrained using the EAGLE/Bolshoi simulations and the sophisticated analytical models of \citet{Zhao2009}/\citet{Correa2015b}. By doing so, we are aiming to provide an interpretation or link between the MAHs of dark matter haloes and the CSFRD while gaining a further insight on the origin of the parameterization of the SFH given in \citet{Katsianis2021}. 


This paper is organized as follows: In section \ref{sec2} we briefly describe some models/parameterizations used in the literature to describe halo growth. In section \ref{sec_model} we provide the average MAH model parameterization in concordance with the {\it observed} CSFRD \citep{Katsianis2021b}. In section \ref{sec_data} we fit our model to MAHs of haloes with different masses $M_0$ using the EAGLE/Bolshoi simulations and the \citet{Zhao2009}/\citet{Correa2015b} models. In section \ref{sec_param} we present the relation of the parameters of our model to $M_{0}$, something that allows us to follow the MAH of a halo and gain a better understanding of the relation between CSFRD and average MAH. Last, in section \ref{sec_concl} we draw our conclusions. In the appendix \ref{GAMMAMotivation} we discuss further the motivation of our choice to work with $\Gamma$ forms and the connection to other branches of science.

\section{MAH models in the Literature}
\label{sec2}

In this section we briefly describe some models/parameterizations used in the literature to describe halo growth :

\begin{figure*}
\vspace{1.0cm}
\centering
\includegraphics[scale=0.45]{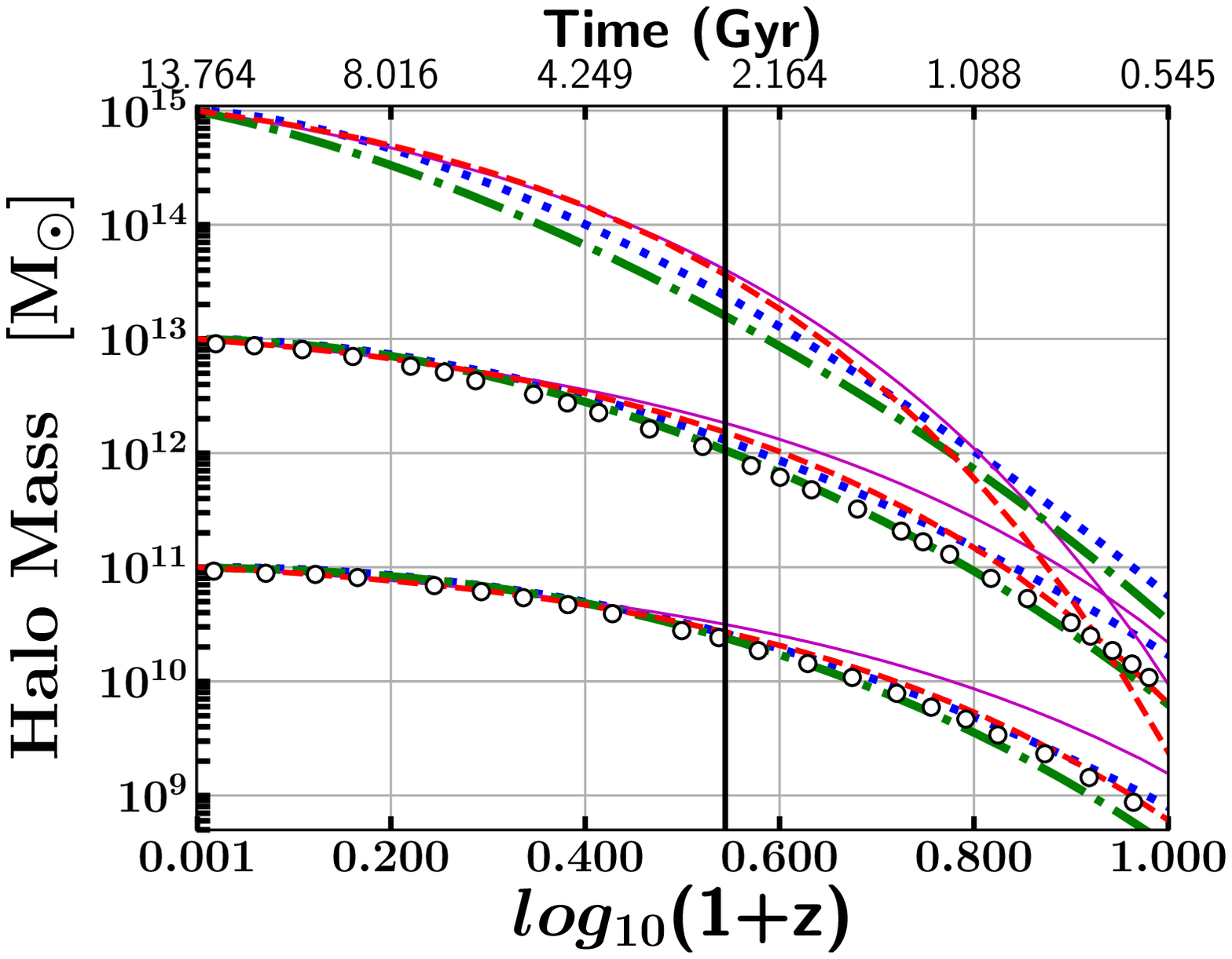}
\includegraphics[scale=0.45]{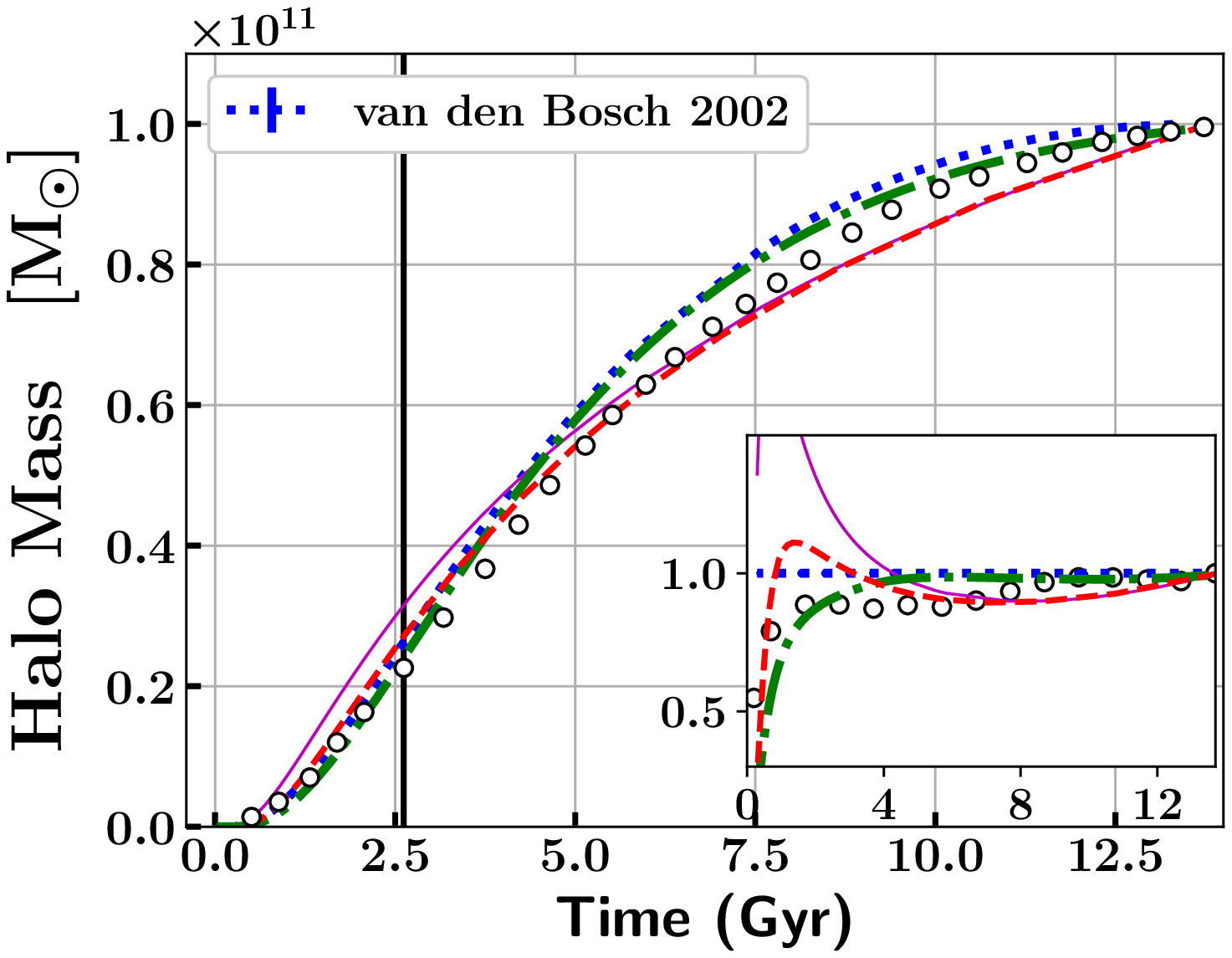}
\includegraphics[scale=0.45]{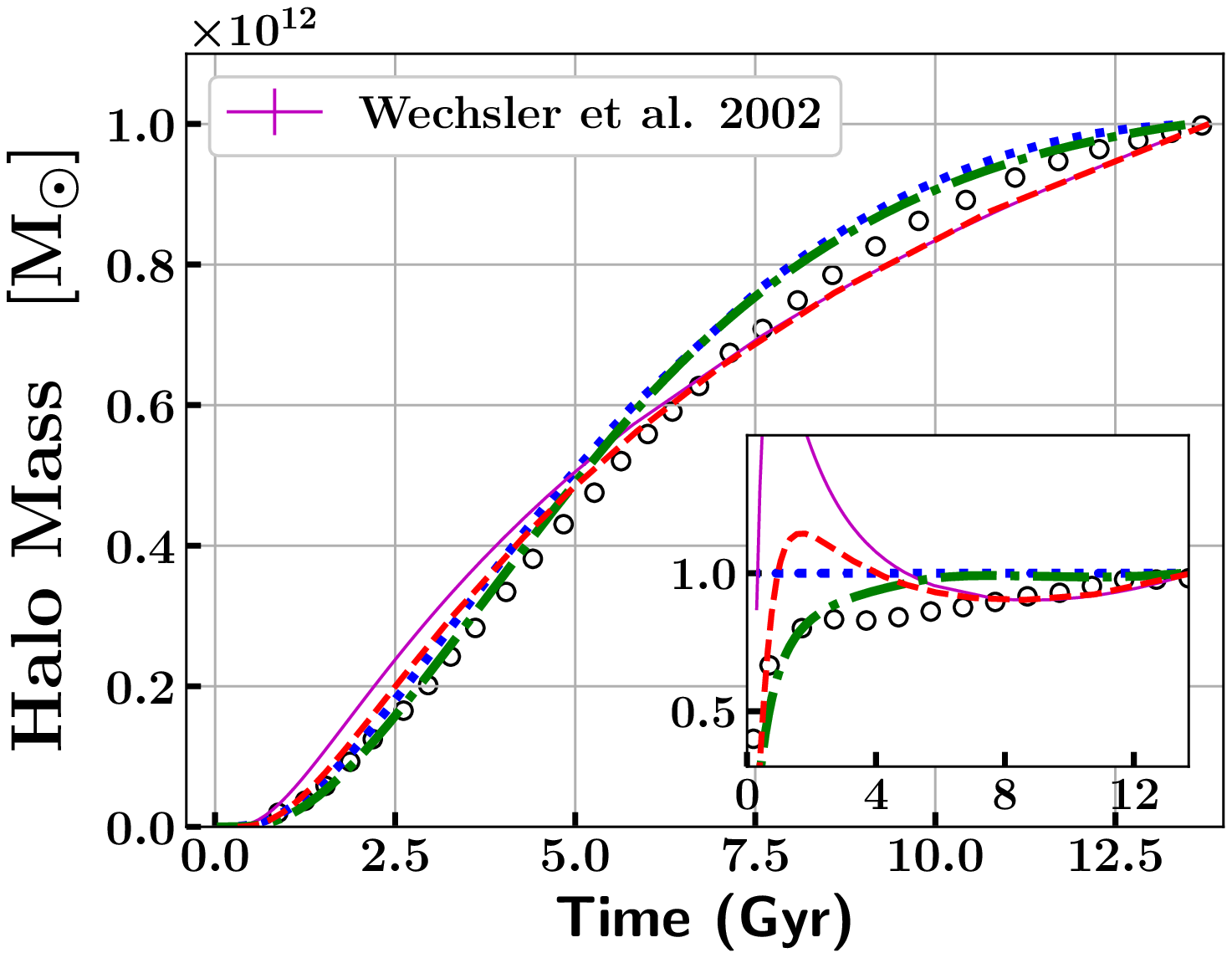}
\includegraphics[scale=0.45]{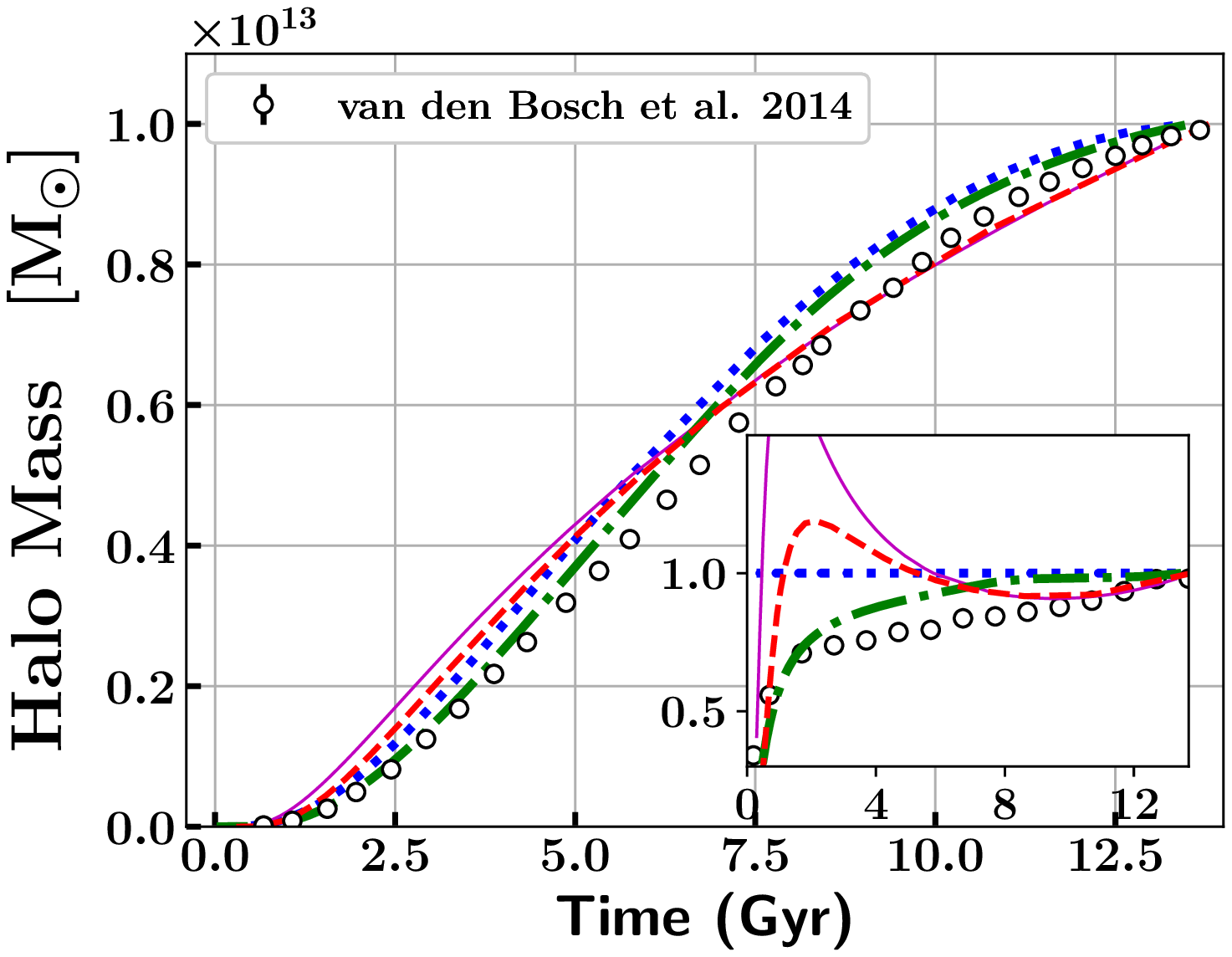}
\includegraphics[scale=0.45]{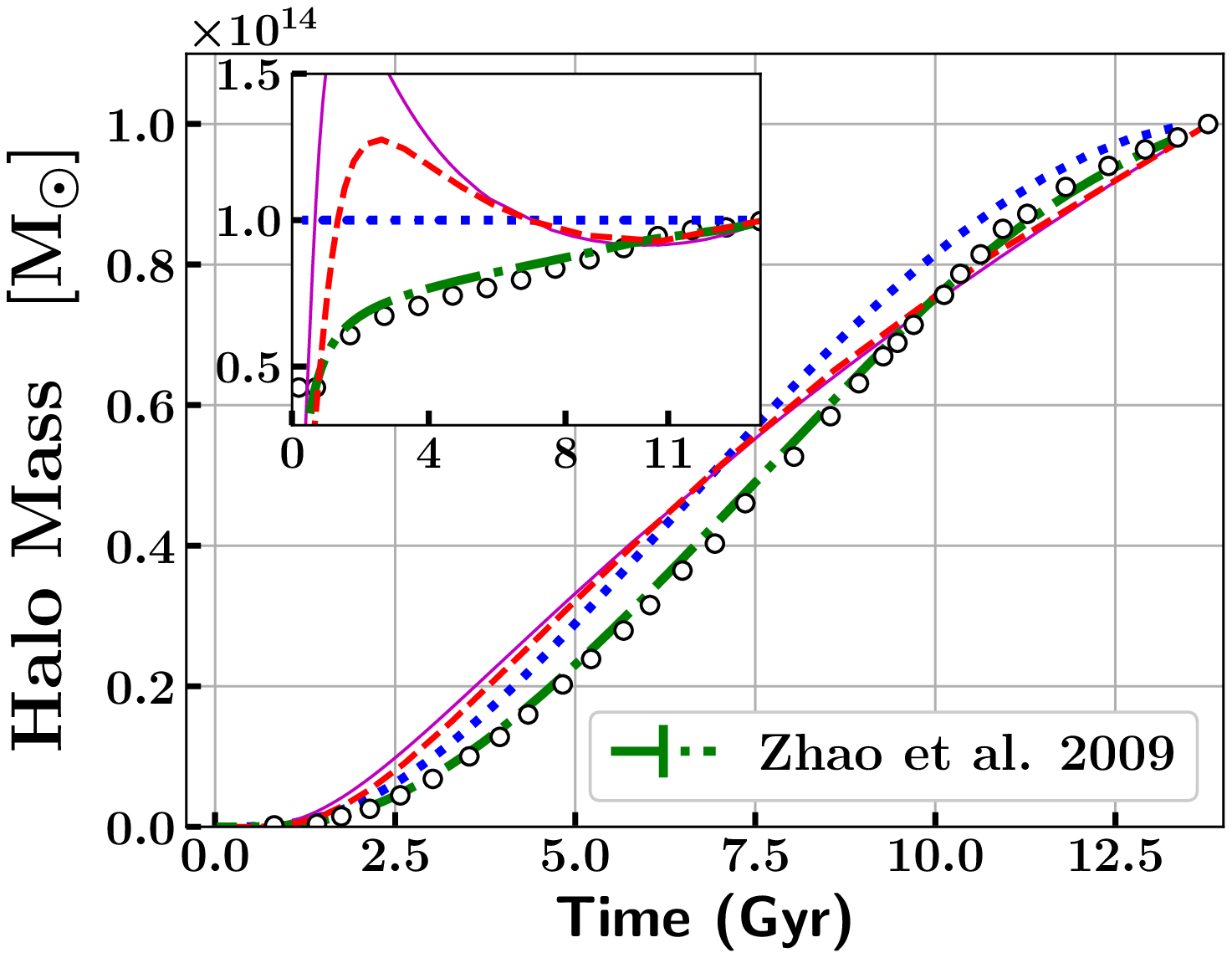}
\includegraphics[scale=0.45]{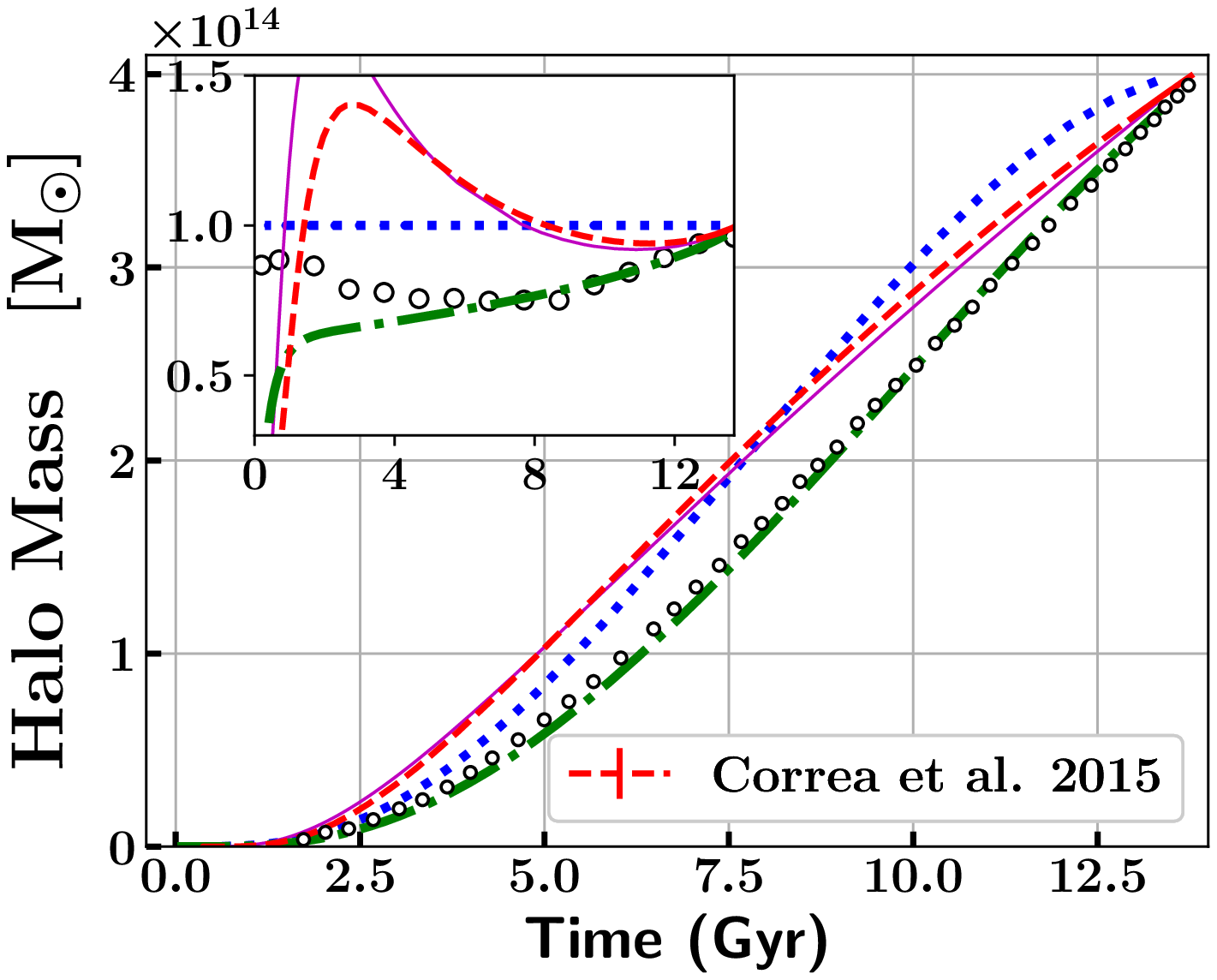}
\vspace{-1.0cm}
\caption{The growth of dark matter haloes at $z \sim 0-9$ given by the work of \citet[][magenta solid line]{Wechsler2002}, \citet[][blue dotted line]{vandenBosch2002}, \citet[][green dashed line]{Zhao2009}, \citet[][black open circles]{Van2014} and \citet[][red dashed line]{Correa2015b}. The MAHs given by different authors besides their successes demonstrate some differences. This effect is expected since different authors relied on different simulations or modelling. }
\label{figEpicG0}
\end{figure*}

\begin{itemize}
\item \citet{Wechsler2002} followed the MAHs of simulated haloes more massive than $M_0 = 10^{12} \, M_{\odot}$, where $M_0$ is the halo mass at present time. The result was a simple exponential parameterization of the form:
\begin{eqnarray}
\label{Wechsler20022}
M_h(z) = M_0 \times {e}^{-\alpha' z},
\end{eqnarray}
where $z$ is redshift and $\alpha'$ is a free parameter which was tuned according to the MAHs of dark matter haloes (seperated at different mass bins) extracted from simulations. Although individual MAHs may deviate from this evolution at specific eras (e.g. at the time of a major merger), the model provided an {\it average} of the simulated MAH trajectories at the different mass bins.  The results of \citet{Wechsler2002} are represented by the magenta lines of Fig. \ref{figEpicG0}. Besides that this model has been pivotal there are some limitations, e.g., it can follow only isolated haloes and is not equipped to capture the effects of subhalo accretion onto larger haloes. Later, \citet{Correa2015} demonstrated that this functional form can only describe succesfully the MAHs at high-redshifts ($z>1$) since their growths are halted by the later accelerated expansion of the Universe in the dark energy dominated era.

\item \citet{vandenBosch2002} using the \citet{PS1974} formalism (PS) and N-body simulations found that their simulated MAHs are well fitted by the following form:
\begin{eqnarray}
\label{eq:Van}
{\log_{10}(\frac{M_h(z)}{M_0}) = -0.301 [\frac{\log_{10}(1+z)}{\log_{10}(1+z_f)}]^v},
\end{eqnarray}
 where $z_{f}$ and $v$ are free fitting parameters which depend on $M_0$ and cosmology. The parameter $z_{f}$ corresponds to the formation redshift defined as $M_h(z_{f}) = M_0/2$. The results of \citet{vandenBosch2002} are represented by the blue dotted line of Fig. \ref{figEpicG0}. Although there was overall a good agreement between PS theory and simulations, there still were some differences (for example the haloes in the simulations started forming earlier). Based on the more advanced Bolshoi simulation, \citet{Van2014} provided a better modeling of the MAHs and potential well growth histories of dark matter haloes. The results of \citet{Van2014} are represented by the black open circles of Fig. \ref{figEpicG0}.

\item
\citet{Zhao2009} used N-body simulations of a variety of cosmologies ($\Lambda$CDM, scale-free, standard CDM, open CDM) to study the MAHs and redshift dependence of concentrations of dark matter haloes. The authors find that there is significant disagreement between empirical models in the literature \citep{Wechsler2002,McBride2009} \footnote{According to \citet{Zhao2009} the \citet{Wechsler2002} and \citet{Zhao2009} models have very different asymptotic behaviors at high redshift because the \citet{Wechsler2002} halo mass is an exponential function of z while \citet{Zhao2009} is a power-law function of z.} and their simulations. We note that Fig. \ref{figEpicG0} indeed agrees with this picture of discrepancy between different authors. The different models converge at lower redshifts but especially at high redshifts (z $ >$ 2) differences of a factor of $\sim 3$ are visible. The effect is more noticeable (even at low redshifts) for the MAHs of haloes with $M_0 > 10^{13} \, M_{\odot}$. \citet{Zhao2009} demonstrated that the simulated mass accretion rate of a halo is tightly correlated to its M(z), redshift, parameters of the adopted cosmology and the initial density fluctuation spectrum.  According to these correlations, the authors developed their {\it empirical} model for both the mass accretion histories and the concentration evolution histories of dark matter haloes. The authors suggest that their models are accurate up to the limit of halo mass that is traced to about 0.0005 times $M_{0}$. The data of the ``Accurate Halo Evolution Calculator'' of \citet{Zhao2009} are public http://202.127.29.4/dhzhao/ and provide the average MAH of a halo given its $M_{0}$. We describe the MAHs of \citet{Zhao2009} in the top right panel of Fig. \ref{figEpicG0}  in a $\log_{10}$(M) - $\log_{10}$(1+z) scale (which is useful to visualize the results at high redshifts, $z>2.5$ - black vertical solid line), while the rest of the panels describe the same model at a linear mass-age scale (which is useful to visualize the results at $z < 2.5)$. The definition of mass we adopt for the \citet{Zhao2009} model in our work is  $M_{200c}$ i.e. the corresponding mass within the Radius $R_{200c}$ enclosing a mean density 200 times the critical value  $\rho_{crit}$. We note the very good consistency between \citet{Van2014} and \citet{Zhao2009}.

\item \citet{Tasitsimi2004} and \citet{McBride2009} suggested that simulated MAHs are well fitted when an additional factor of $(1+z)^{\beta'}$ is added to the \citet{Wechsler2002} simple exponential model, yielding MAHs of the form:
\begin{eqnarray}
\label{Tasi}
{ M_h(z) = M_0 \times (1+z)^{\beta'} \times e^{-\gamma'  z}},
\end{eqnarray}
where the $(1+z)^{\beta'}$ factor is necessary because the growth of the density perturbations is halted in the dark energy dominated era. \citet{Correa2015} constructed an analytic model relying on the PS formalism and the above functional form. In a later study, \citet{Correa2015b} reported some discrepancies of these results with respect the EAGLE cosmological simulations and their semi-analytic model for the MAHs of massive $ M_0> 10^{14} \, M_{\odot}$ and low-mass $ M_0 < 10^{9} \, M_{\odot}$ objects (Fig. 10 of their work) but demonstrated that the parameterization remains as in Eq. \ref{Tasi} and provided further details about the cosmic/average halo growth history. The parameters $\beta'$ and $\gamma'$ are correlated with concentration and linear power spectrum. The results of \citet{Correa2015b} are represented by the red dashed line of Fig. \ref{figEpicG0}.  The definition of Mass we use is once again $M_{200c}$ i.e. the corresponding mass within the Radius enclosing a mean density of 200 times the critical value  $\rho_{crit}$ (same definition as \citet{Zhao2009}). We note the very good consistency between \citet{Wechsler2002} and \citet{Correa2015b} at low redshifts ($z < 1$). 


\end{itemize}


\section{Model of the MAH of dark matter haloes with a Gamma formalism}
\label{sec_model}
 

\begin{figure*}
  \vspace{1.0cm}
\centering
\includegraphics[scale=0.45]{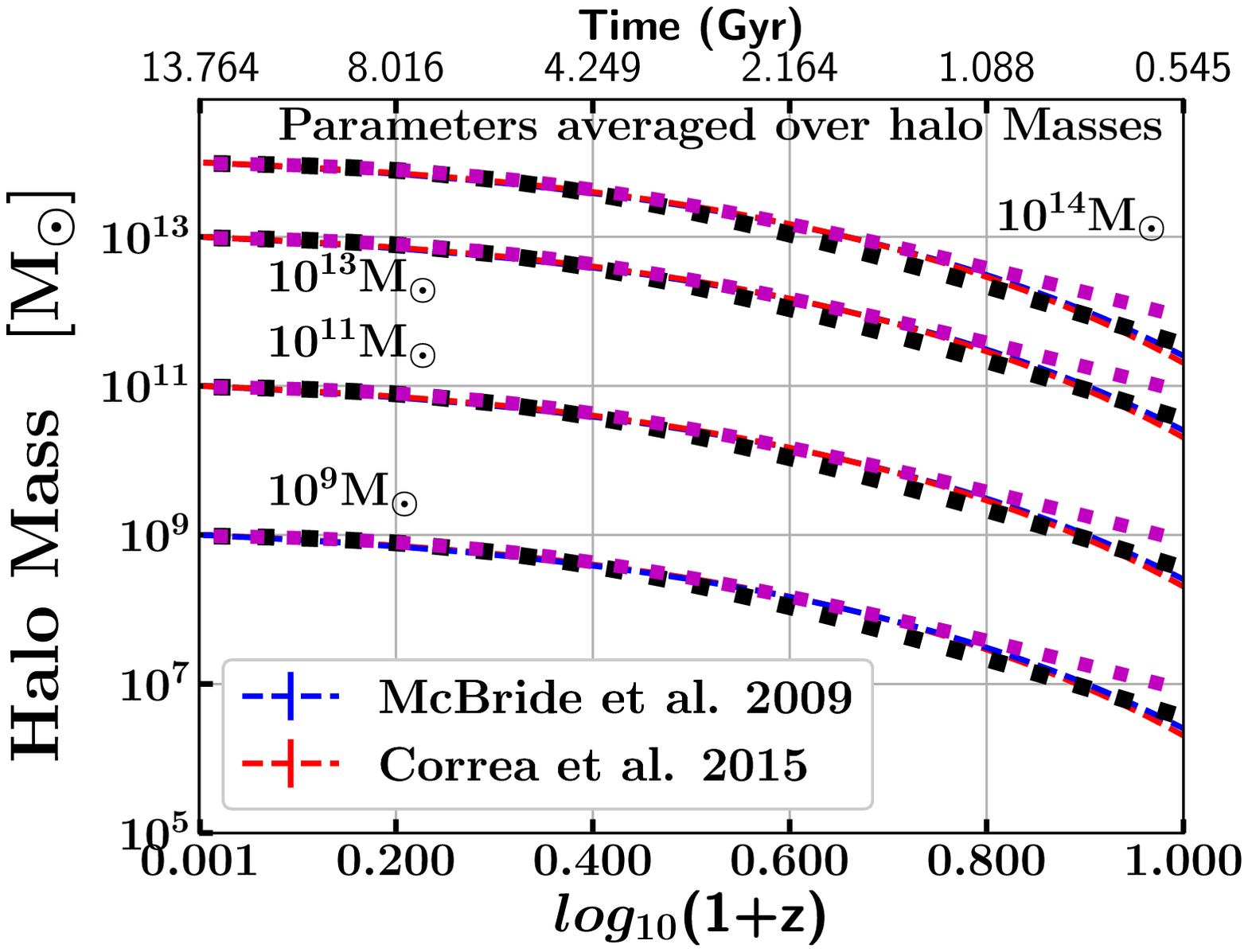}
\includegraphics[scale=0.45]{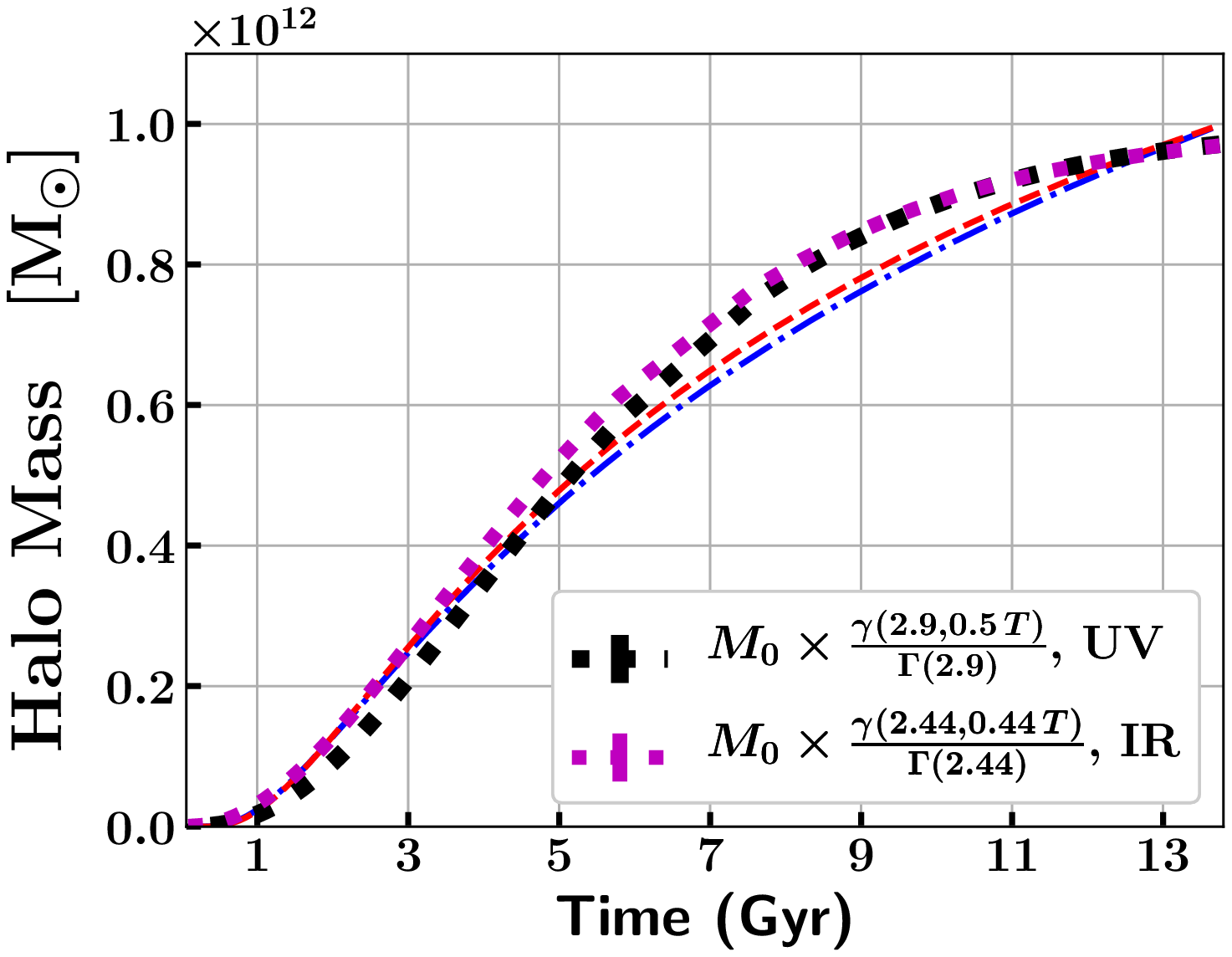}
\includegraphics[scale=0.45]{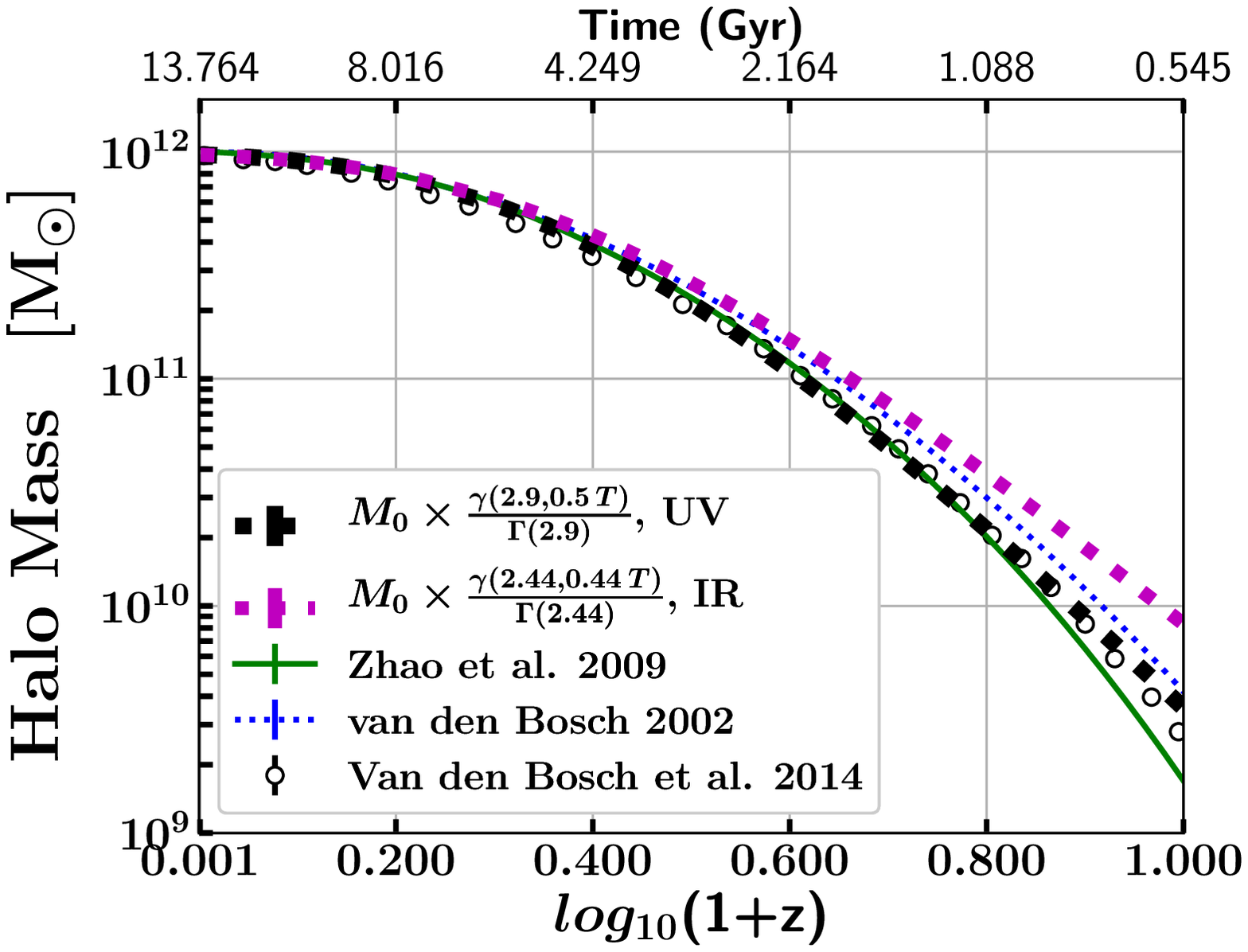}
\includegraphics[scale=0.45]{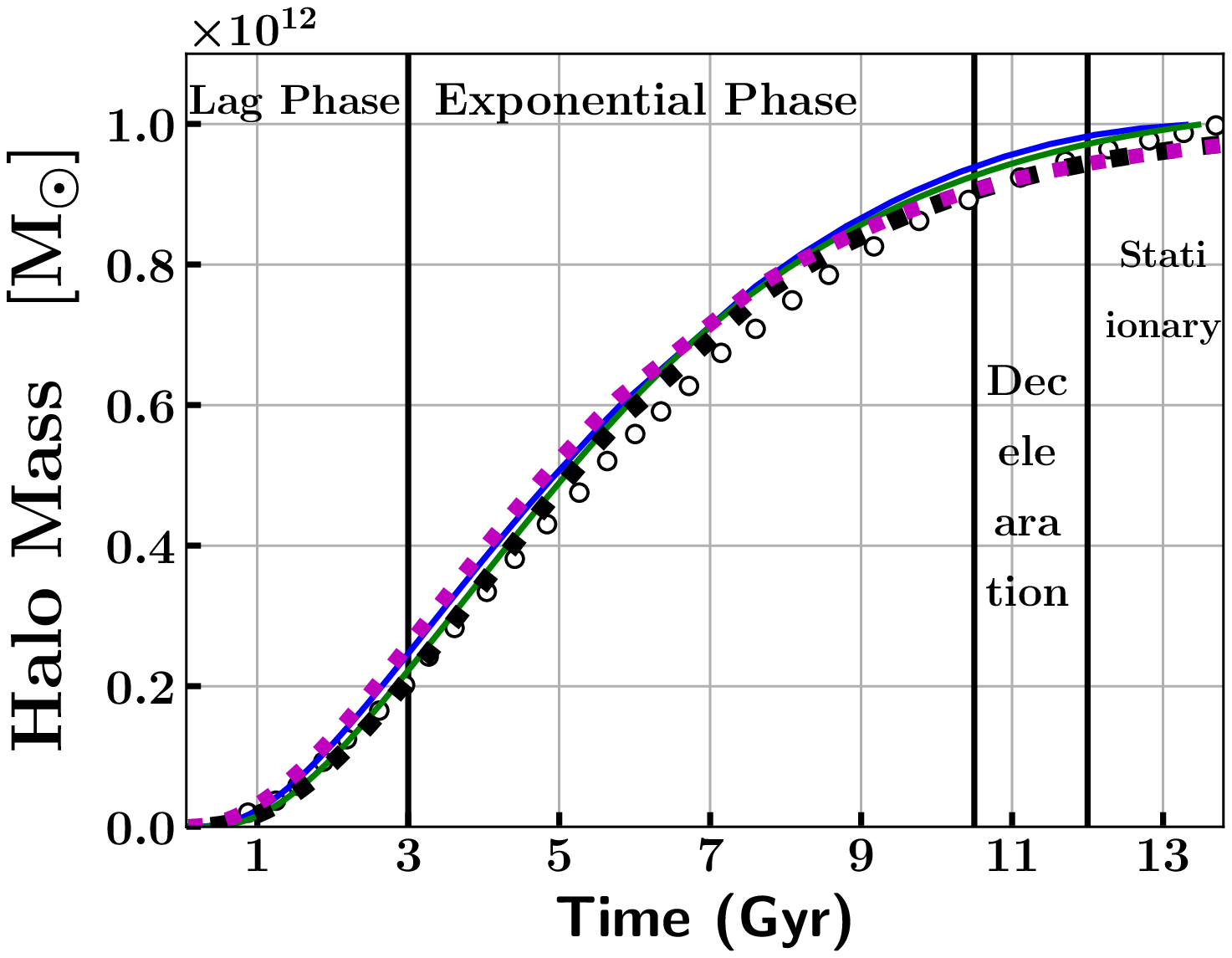}
\vspace{-0.30cm}
\vspace{-0.30cm}
\caption{Top panels: the growth of dark matter haloes at $z \sim 0-9$ derived from the observed UV CSFRD Gamma formalism \citep{Katsianis2021b} with parameters $< \alpha_{\star} > = 2.9$, $< \beta_{\star} > = 0.5$ (black dashed line) and the observed IR CSFRD  with parameters $< \alpha_{\star} > = 2.44$, $< \beta_{\star} > = 0.44$ (magenta dashed line). Alongside the ``average'' simulated MAHs for the $M_h(z) = M_0 \times (1+z)^{\alpha'} \times e^{-\beta' \, z}$ parameterization adopting $<\alpha'> = 0.24$, $<\beta'> = -0.75$ \citep[][red dashed line]{Correa2015b} and $<\alpha'> = 0.10$, $<\beta'> = -0.69$ \citep[][dotted blue line]{McBride2009}. We note that the parameters are averaged for all halo types/masses and are obtained by \citet{Correa2015b} and \citet{McBride2009}, respectively. Bottom panels: the growth of dark matter haloes corresponds to the UV CSFRD parameters alongside the evolution derived from \citet[][Blue dotted line]{vandenBosch2002},  \citet[][green solid line]{Zhao2009} and \citet[][open circles]{Van2014} for haloes that at $z = 0$ have masses of $M_0 = 1.0 \times 10^{12} M_{\odot}$.}
\label{figEpicG1}
\end{figure*}

\subsection{Dark matter halo Growth, Cosmic star formation rate density and $\Gamma$ growth}
\label{averages}

Following the \citet{Katsianis2021b} equilibrium model, the SFR is related to the inflowing baryonic matter as follows:
\begin{eqnarray}
\label{eq:PGgast01}
{\rm  SFR}(M_h, T) = \frac{f_{gal}(M_h,T) \times f_{b}}{1+n-R} \times \frac{dM_h(T)}{dT}.
\end{eqnarray}
In the above, the inflowing baryonic gas is assumed to scale with the DM halo growth ($\frac{dM_h(T)}{dT}$). Here $f_{b}$ is the cosmic baryon fraction and $f_{gal}(M_h,T)$ is the fraction of the incoming baryons that are able to penetrate and reach the galaxy and be used for star formation \citep[e.g.][]{Bouche2010,Dave2012,PengMaolino2014,Dekel2014}, $n$ is the wind mass loading factor and $R$ is the return fraction. We note that this form is equivalent with other models in the literature \citep[e.g.][]{Mutch2013, Tacchella2018, Maniyar2021}. \citet{Katsianis2021b} have demonstrated that the CSFRD of the Universe can be described by a form that resembles a Gamma, $\Gamma(\alpha, \beta)$ form,
\begin{eqnarray}
\label{eq:CSFRD}
\begin{split}
{\rm CSFRD} =&   \frac{ \bar{f}_{gal} \times f_{b} \times \Omega_{ 0} \times \rho_{crit}}{(1+n-R)} \,  \times \\ & \frac{\beta_{\star}^{\alpha_{\star}}}{\Gamma(\alpha_{\star})} \, \times \, T^{\alpha_{\star}-1} \times e^{-\beta_{\star} \, T }  \, { \frac{M_{\odot} \, Gyr^{-1} }{Mpc^3}}, 
\end{split}
\end{eqnarray}
where $\rho_{crit}$ is the critical density of the universe, ${\bar {f}_{gal}}$ is the averaged star formation efficiency from the total gas in the Universe. $T$ is time in $Gyr$ and $\Omega_{0}$ is the matter density parameter at $z \sim 0 $, $\alpha_{\star}$ sets how quickly the SFR rises at the early accretion/merger era, and $1/\beta_{\star}$ is the star formation history timescale\footnote{$\tau_{\star} = 1/\beta_{\star}$ is defined as star formation timescale \citep{Lee2010,Baeuer2013,Hagen2016,Garcia2019,Curtis2021} or star formation history timescale \citep{Weinberg2017} and regulates the delay of the maximum star formation rate (SFR) and the steepness of its decay \citep{Kim2010,Katsianis2014}.} (for more information on the above relations and their physical meaning we refer the reader to section 5 of \citet{Katsianis2021b}\footnote{In short we suggested that the {\it observed} form of the CSFRD is the result of 1) a gas accretion era ($T^{\alpha_{\star}}$) and a depletion era ($e^{-\beta_{\star}  T}$), 2) all the available gas eligible for SF will be consumed at $t = \infty$. 3) There is an equilibrium between infalling gas, outflows and star formation.}). {\it If we associate the star formation with gas supply, in general we can relate $\alpha_{\star}$ with how fast the gas is accreted, and relate $\beta_{\star}$ with how efficient gas is turned into stars or is depleted.}

We assume that the SFH for galaxies in an individual halo can also be described with this functional form. By replacing $\Omega_{0} \times \rho_{crit}$ with $M_{h, final}$ (The maximum final mass of the halo at $T =\infty$), we can simply write the SFH in a halo as:
\begin{eqnarray}
\label{eq:PGgastconsumption0}
\begin{split}
{\rm SFR}(M_h, T) & =  \frac{ f_{gal}(M_h, T) \times f_{b} \times M_{h, final}}{(1+n-R)} \, \times \\
&  \frac{\beta_{h}^{\alpha_{h}}}{\Gamma(\alpha_{h})} \, \times \, T^{\alpha_{h}-1} \times e^{-\beta_{h} \, T }  \, M_{\odot} \, Gyr^{-1}\,.
\end{split}
\end{eqnarray}
$M_{h, final}$ can be written with respect $M_{0}$ as $M_{h, final} = \frac{M_{0}}{f_{0}}$ where $f_{0}$ is the fraction of how much mass the halo has attained at $z = 0$ with respect the $M_{h, final}$ value. We note that we changed the labels of $\alpha_{\star}$ and $\beta_{\star}$ to $\alpha_{h}$ and $\beta_{h}$, respectively, since we will see, in section \ref{sec_param}, that these parameters depend on the halo $M_0$. So we will now focus on halo mass growth and not galaxy stellar mass growth. We label $\tau_{h} = 1/\beta_h$ as the dark matter halo depletion timescale and $\alpha_h$ sets how quickly the halo mass rise attributed to accretion/merger processes. The above can also be understood in the framework of the depletion region, the region at the outskirts of haloes where matter is being depleted for growing haloes \citep{Fong2021,Fong2022}.

By combining equations \ref{eq:PGgast01} and \ref{eq:PGgastconsumption0}, we can trace the halo mass growth  rate as follows:
\begin{eqnarray}
\label{eq:PGgastconsumption00}
{\frac{dM_h(T)}{dT} =  M_{final}\, \times  \frac{\beta_{h}^{\alpha_{h}}}{\Gamma(\alpha_{h})} \,  T^{\alpha_{h}-1}  e^{-\beta_{h} \, T }  \, { M_{\odot} \, Gyr^{-1}}}.
\end{eqnarray}
We integrate equation \ref{eq:PGgastconsumption00} to obtain the following simple $\Gamma$ MAH model:
\begin{eqnarray}
\label{eq:MAHmodel}
\begin{split}
{M_h(T)} &=&  M_{final} \, \times \frac{\gamma(\alpha_{h}, \beta_{h} T)}{\Gamma(\alpha_{h})}\\
&=&  \frac{M_{0}}{f_{0}} \, \times \frac{\gamma(\alpha_{h}, \beta_{h} T)}{\Gamma(\alpha_{h})},
\end{split}
\end{eqnarray}
which is just equivalent to Eq. \ref{eq:GAMMAmodel}. In short we saw the justification for a $\Gamma$ model for the MAHs which originates from our {\it observations} on CSFRD \citep{Katsianis2021b} and a simple equilibrium model.

Before we proceed to use this functional form to fit the related parameters for haloes of different masses from MAHs extracted from simulations and models, it is interesting to simply see the ``average/cosmological'' MAH description of dark matter haloes derived from the CSFRD. 
According to equation \ref{eq:CSFRD}, if we assume ${\bar {f}_{gal}}$ is a constant value, then the CSFRD is just proportional to the average mass accretion and consumption in dark matter haloes. Using the parameters $\alpha_{\star}$ and $\beta_{\star}$ constrained through the CSFRD which represents the average SFR in the Universe and assuming $f_{0} = 1$ (we will see in the section \ref{final} that the later is a sensible assumption on ``average'' and for haloes with $M_0 < 10^{13}$) we are able to obtain an average cosmic MAH in dark matter haloes. We present the resulting average cosmic MAH in Fig. \ref{figEpicG1} as a black dotted line, where $ {\bf < \alpha_{\star} > = 2.9}$ and ${\bf <\beta_{\star}> = 0.5}$ are chosen for the case of the ${\rm CSFRD}_{UV,corr}$\footnote{${\rm CSFRD}_{UV,corr} \approx  {\bf 0.50} \times \frac{{\bf 0.50^{2.9}}}{\Gamma({\bf 2.9})} \times T^{{\bf 2.9}-1} \times e^{-{\bf 0.50}\, T} \, { \frac{M_{\odot} \, yr^{-1} }{Mpc^3}}$} (the case for ${\rm CSFRD}_{UV+IR}$\footnote{${\rm CSFRD}_{UV+IR} (T) \approx  \frac{{\bf 0.44^{2.44}}}{\Gamma({\bf 2.44)}} \times T^{{\bf 2.44}-1} \times e^{-{\bf 0.44} \, T } \, \, { \frac{M_{\odot} \,yr^{-1} }{Mpc^3}}$} are ${\bf <\alpha_{\star}> = 2.44}$ and ${\bf <\beta_{\star}> = 0.44}$) \citep{Katsianis2021b}.

In the top left panel of Fig. \ref{figEpicG1}, we compare our results with the predictions of the semi-analytical model of \citet{Correa2015b} for the ``average'' MAH of dark matter haloes in a $\log_{10}(M)$ - $\log_{10}$(1+z) scale. \citet{Correa2015b} noted that the parameters $\beta'$ and $\gamma'$ in  their model parameterization  (equation \ref{Tasi}) which describe their simulated/theoretical MAHs change very slightly between halo masses of $M_0 = 10^{9}$ and $M_0 = 10^{14}$ $M_{\odot}$ for the EAGLE cosmological simulations (only change by a factor of 2-3). Thus, they provided an approximation for the mean MAH of a halo, by averaging $\beta'$ and $\gamma'$ over all different halo masses $M_0$. The average parameters were calculated as $<\beta'> = 0.24$, $<\gamma'> = -0.75$ \citep{Correa2015b} and the resulting MAHs are represented by the dashed red lines in the top left panel of Fig. \ref{figEpicG1}. Our results which rely on the {\it observed} UV CSFRD (which are favored with respect the IR CSFRD in \citet{Katsianis2021b}\footnote{IR SFRs at high redshifts are possibly overestimated according to simulations/radiative transfer and advanced SED fitting modelling \citep{Hayward2014,Katsianis2015,Martis2019,Leja2020,Katsianis2020}}) and Eq. \ref{eq:MAHmodel} are in good agreement with their model. In addition, we compare our results with the average simulated MAHs of \citet{McBride2009} ($<\beta'> = 0.10$, $<\gamma'> = -0.69$ - dotted blue line) and once again we find a good agreement. The above consistency clearly demonstrates the close connection between the cosmic/average SFH and the average/cosmic MAH. In the top right panel of Eq. \ref{eq:MAHmodel} we make the same comparison in a linear(M) - age scale. We see that there are some small differences of how our average MAH model and that of \citet{Correa2015b} flatten out/decelerate.  These small differences would not be visible in a $\log_{10}(M)$-$\log_{10}$(1+z) scale. This is one of the reasons of why in our work we chose to present both the linear(M) - age (to highlight visually the low redshift regime) and $\log_{10}(M)$-$\log_{10}$(1+z) scales (to highlight the high redshift regime).

It is well known that haloes of $M_{0} \sim 10^{12}$ dominate the total CSFRD \citep{Katsianis2017} and thus a strong connection between the cosmic SFH and the MAHs of these objects is expected. In the bottom left ($\log_{10}(M)$-$\log_{10}$(1+z) scale) and bottom right panels (linear(M) - age scale) of the Fig. \ref{figEpicG1} we present the growth of dark matter haloes derived from the CSFRD and its $\Gamma$ parameterization alongside the evolution derived from the models of \citet{Zhao2009}, \citet{vandenBosch2002} and \citet{Van2014} at $M_{0} \sim 10^{12}$. The comparison of the above models with the MAH derived from our UV CSFRD is excellent for both ways/panels the results are portrayed as both the theoretical MAHs and the one derived from our observation flatten out/decelerate   the same way (seen clearer in the right panel). We note once again that the UV CSFRD produces results in better agreement with the theoretical MAHs as IR SFRs are possibly overestimated at high redshifts \citep{Katsianis2021b}.

In conclusion, observations and theory about the average/cosmic MAH and CSFRD at $z \sim 0-9$ are in good agreement with each other. Also haloes with mass $M_{0} \sim 10^{12}$ also braodly follow a $\Gamma$ growth pattern.

\begin{table*}
  \centering
\resizebox{0.80\textwidth}{!}{%
  \begin{tabular}{cccccc}
    \hline \\
    &  & & 
    Parameters of $\Gamma$ model - \citet{Zhao2009} \\
    \hline \hline
    & $M_{0}$ ($M_{\odot}$) & $f_{0}$ & $\alpha_h$  & $\beta_h$ (Gyr$^{-1}$) & $T_h$ (Gyr) \\
    \hline
    & 0.9e+1 & 1.012 $\pm$ 0.004 & 1.551 $\pm$ 0.039 & 0.522 $\pm$ 0.014 &  0.173 $\pm$ 0.018 \\
    & 0.9e+2 &  1.005 $\pm$ 0.003 & 1.637 $\pm$ 0.038 & 0.523 $\pm$ 0.013 & 0.174 $\pm$ 0.019 \\
    & 0.9e+3  & 1.007 $\pm$ 0.003 & 1.638 $\pm$  0.043 & 0.518 $\pm$ 0.014  & 0.172 $\pm$ 0.021 \\
    & 0.9e+4 & 1.006 $\pm$ 0.003 & 1.676 $\pm$ 0.044 & 0.514 $\pm$ 0.015 & 0.171 $\pm$ 0.020 \\
    & 0.9e+5 & 1.008 $\pm$ 0.003 & 1.692  $\pm$ 0.042 & 0.508 $\pm$ 0.014 & 0.186$\pm$ 0.019  \\
    & 0.9e+6 &  1.005 $\pm $ 0.003 & 1.754 $\pm$ 0.042 & 0.505 $\pm$ 0.012 & 0.193 $\pm$ 0.024 \\
    & 0.9e+7 & 1.001  $\pm$ 0.002 & 1.825 $\pm$ 0.043  & 0.502 $\pm$ 0.012 & 0.185 $\pm$ 0.024 \\
    & 0.9e+8 & 1.002 $\pm$ 0.001  & 1.943  $\pm$ 0.045 & 0.499 $\pm$ 0.010 &  0.187 $\pm$ 0.027 \\
    & 0.9e+9 & 1.001 $\pm$ 0.002  & 2.141 $\pm$ 0.052 & 0.518  $\pm$ 0.011 &  0.186 $\pm$ 0.029 \\
    & 0.9e+10  & 1.000 $\pm$ 0.002 & 2.232  $\pm$ 0.046 &  0.509 $\pm$ 0.010 &  0.184 $\pm$ 0.026 \\
    & 1e+11 & 0.997 $\pm$ 0.002 & 2.451 $\pm$ 0.035 & 0.495 $\pm$ 0.006 &  0.185  $\pm$ 0.021 \\
    & 1e+12 & 0.964 $\pm$ 0.001 & 2.708 $\pm$ 0.046 & 0.470 $\pm$ 0.009 &  0.178 $\pm $ 0.020 \\
    & 1e+13 & 0.911 $\pm$ 0.004 & 3.024 $\pm$ 0.037 & 0.423 $\pm$ 0.006 &  0.172 $\pm$ 0.021 \\
    & 1e14 & 0.801 $\pm$ 0.004 & 3.352 $\pm$ 0.012 &  0.357 $\pm$ 0.003 &  0.168 $\pm$ 0.008 \\
    & 4e14 & 0.564 $\pm$ 0.001 & 3.181  $\pm$ 0.025 & 0.234 $\pm$ 0.002 &  0.165 $\pm$ 0.019 \\
    & 1e15 & 0.356 $\pm$ 0.001 & 3.273 $\pm$ 0.019 & 0.176 $\pm$ 0.002 & 0.167 $\pm$ 0.015 \\
    \hline \\
    &  & & 
    Parameters of $\Gamma$ model - \citet{Correa2015b} \\ 
    \hline \hline
    & 904.5  & 1.007 $\pm$ 0.006 & 1.186 $\pm$ 0.013 & 0.343 $\pm$ 0.005 & 0.132 $\pm$ 0.010 \\
    & 9e+3 & 1.005 $\pm$ 0.006 & 1.209 $\pm$ 0.015 & 0.340 $\pm$ 0.006 & 0.181 $\pm$ 0.008 \\
    & 9e+4 & 1.002 $\pm$ 0.006 & 1.241 $\pm$ 0.019 & 0.337 $\pm$ 0.007 & 0.206 $\pm$ 0.016 \\
    & 9e+5 & 0.999 $\pm$ 0.006 & 1.322 $\pm$ 0.024 & 0.335 $\pm$ 0.006 & 0.199  $\pm$ 0.020 \\
    & 9e+6 &  0.998  $\pm$ 0.006 & 1.369 $\pm$ 0.017 & 0.336 $\pm$ 0.007 & 0.203 $\pm$ 0.008 \\
    & 9e+7 & 0.999 $\pm$ 0.007 & 1.476 $\pm$ 0.022 & 0.343 $\pm$ 0.008 & 0.201 $\pm$ 0.010 \\
    & 8.9e+8 & 0.997  $\pm$ 0.008 & 1.583 $\pm$ 0.027  & 0.338 $\pm$ 0.009  &  0.197 $\pm$ 0.012\\ 
    & 8.67+09 & 0.993 $\pm$ 0.007  & 1.592 $\pm$ 0.044 &  0.335 $\pm$ 0.011 &  0.218 $\pm$ 0.025 \\
    & 2.01e+10 & 0.995 $\pm$ 0.007  & 1.626 $\pm$ 0.057  & 0.321 $\pm$ 0.010 &  0.251 $\pm$ 0.030 \\
    & 8.57e+10 & 0.990 $\pm$ 0.007  & 1.775 $\pm$ 0.044  & 0.334 $\pm$ 0.012 &  0.253 $\pm$ 0.018 \\
    & 7.03e+11 & 0.965 $\pm$ 0.003 & 2.018 $\pm$ 0.082 & 0.326 $\pm$ 0.011 &  0.246 $\pm$ 0.037 \\
    & 6.73e+12 & 0.916 $\pm$ 0.004 & 2.319 $\pm$ 0.071 &  0.332 $\pm$ 0.008 &  0.258 $\pm$ 0.039 \\
    & 6.678e+13 & 0.905 $\pm$ 0.007 & 2.835 $\pm$ 0.070  & 0.365 $\pm$ 0.007 &  0.327  $\pm$ 0.035 \\
    & 1e15 & 0.826 $\pm $ 0.081 & 3.568  $\pm $ 0.089 & 0.384 $\pm$ 0.008 & 0.371 $\pm$ 0.062 \\    
    \hline \hline
  \end{tabular}%
}
\caption{The parameters of $\Gamma$ model (\ref{eq:MAHmodel2}) constrained from the MAHs of the \citet{Zhao2009} and \citet{Correa2015b} models. We include standard deviation errors on the parameter estimates.}
\label{tab:ParamsModels}
\end{table*}

\section{The mass growth of dark matter haloes via a $\Gamma$ formalism}
\label{sec_data}

\begin{figure*}
\centering
\includegraphics[scale=0.42]{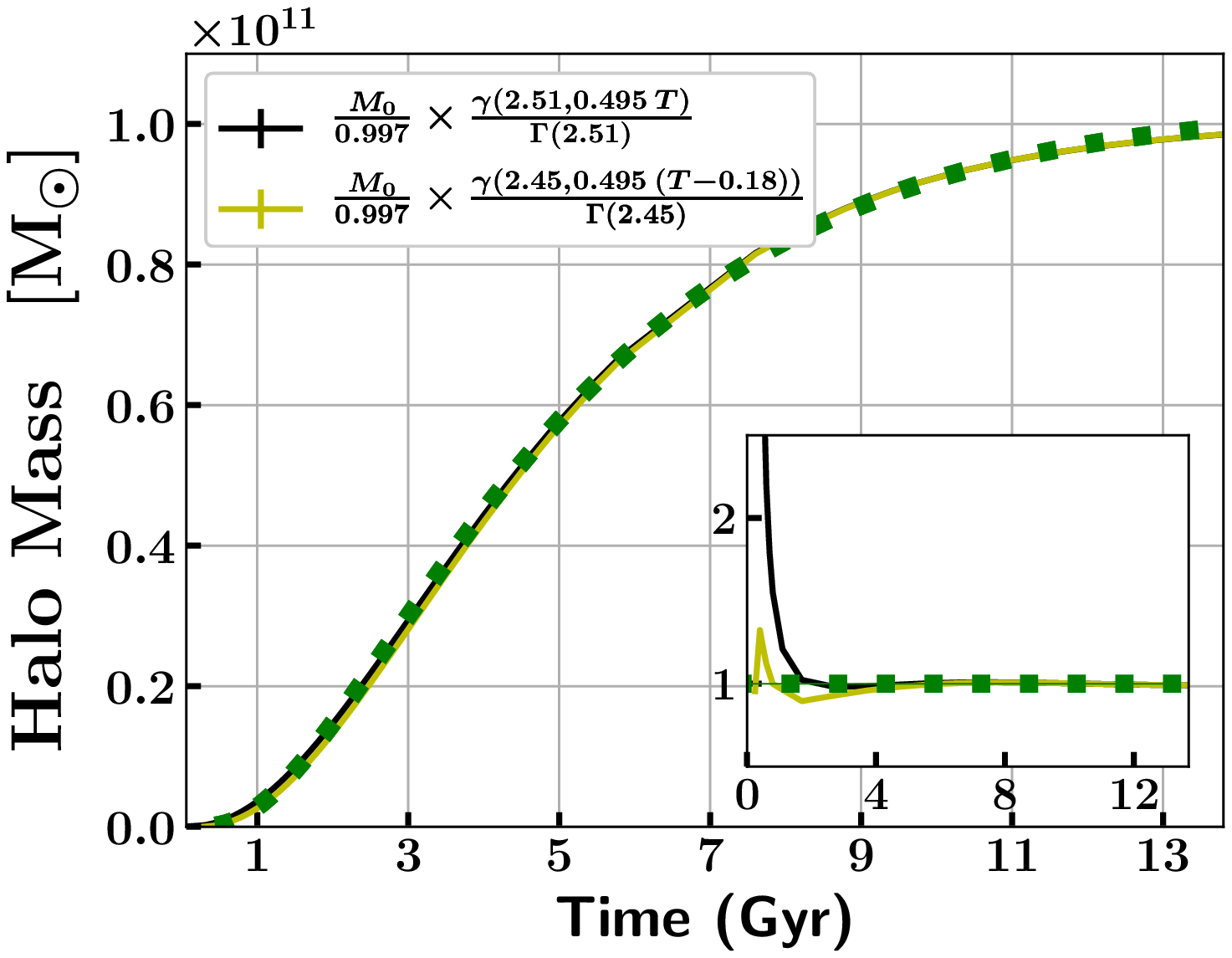}  
\includegraphics[scale=0.42]{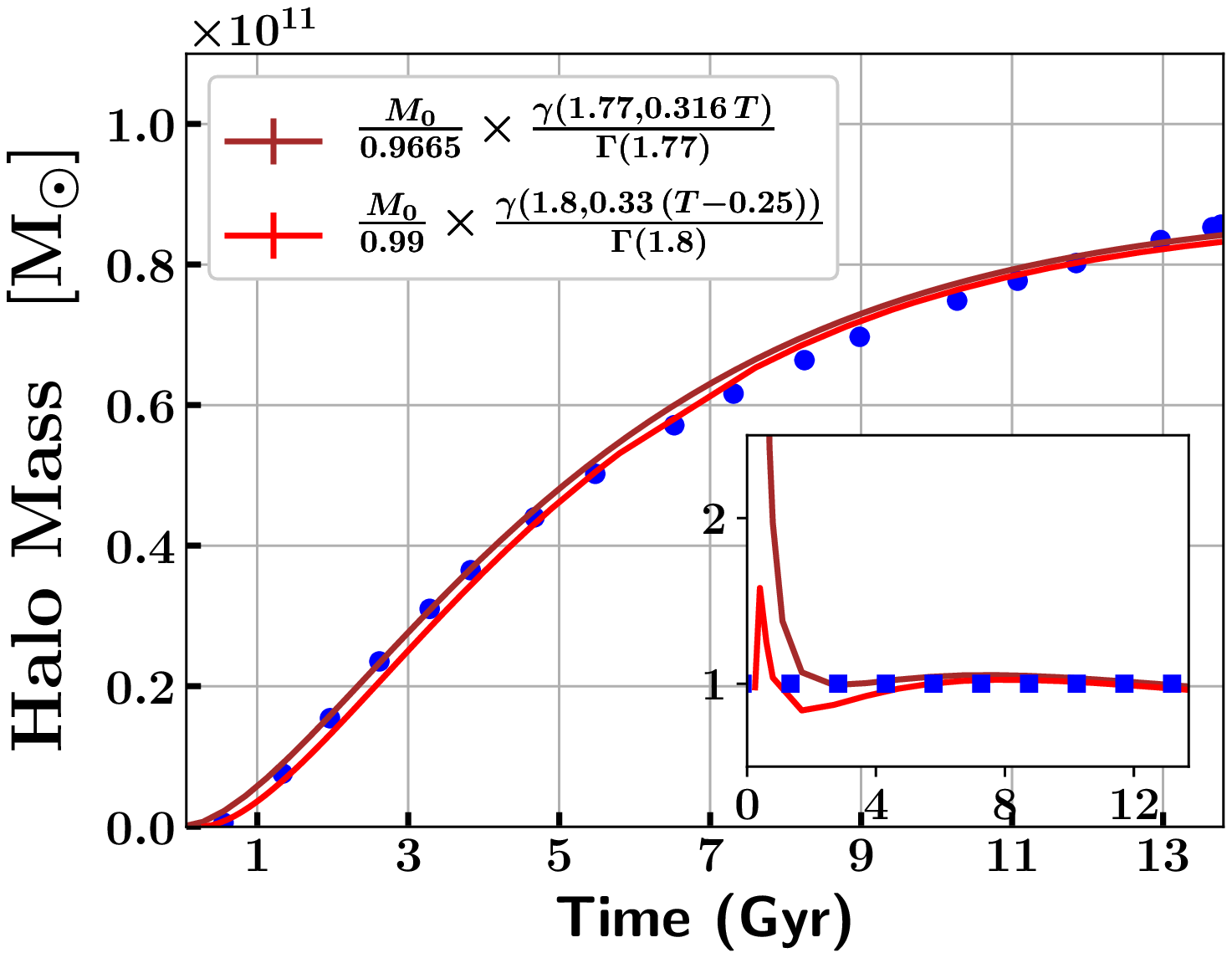}
\includegraphics[scale=0.42]{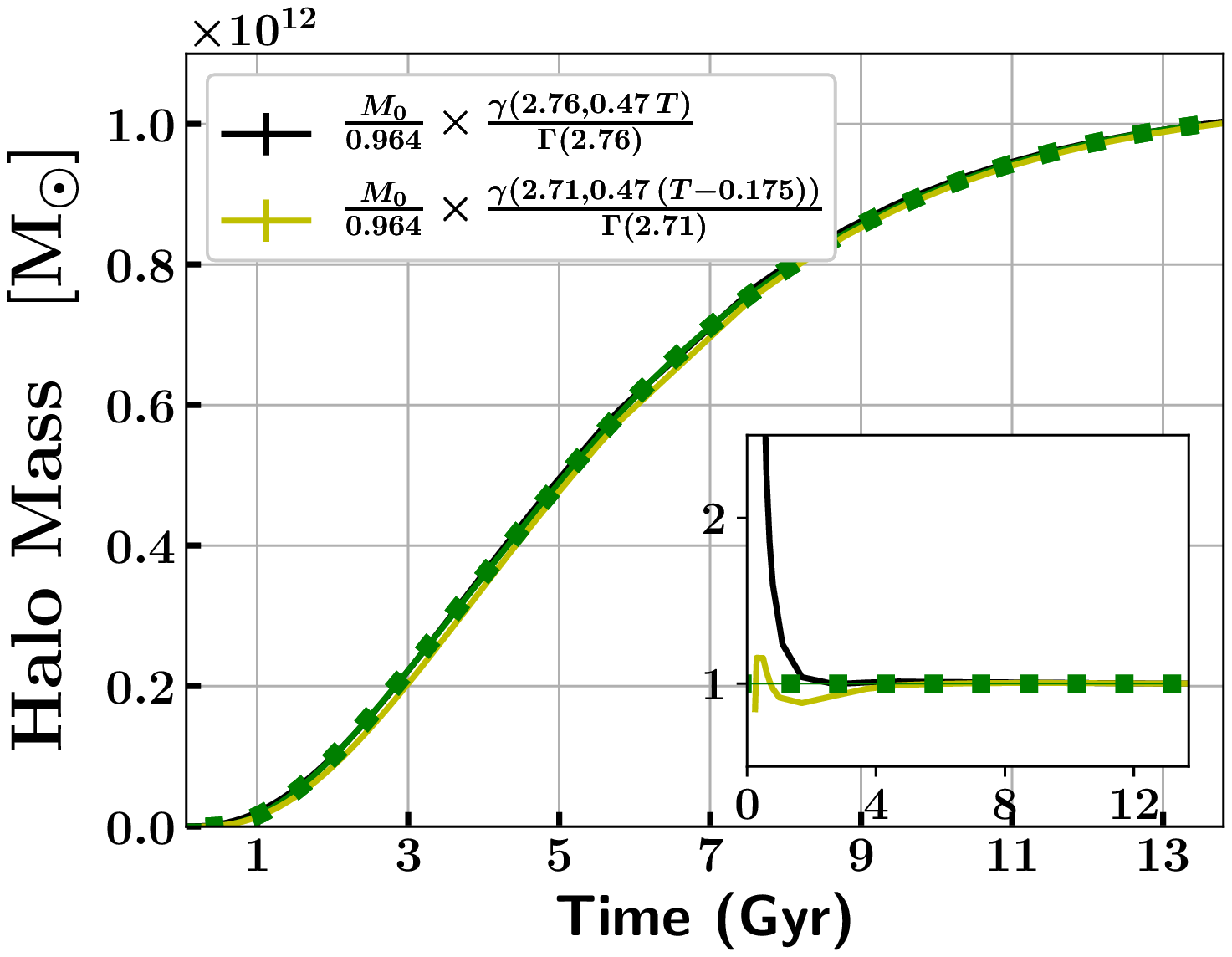}
\includegraphics[scale=0.42]{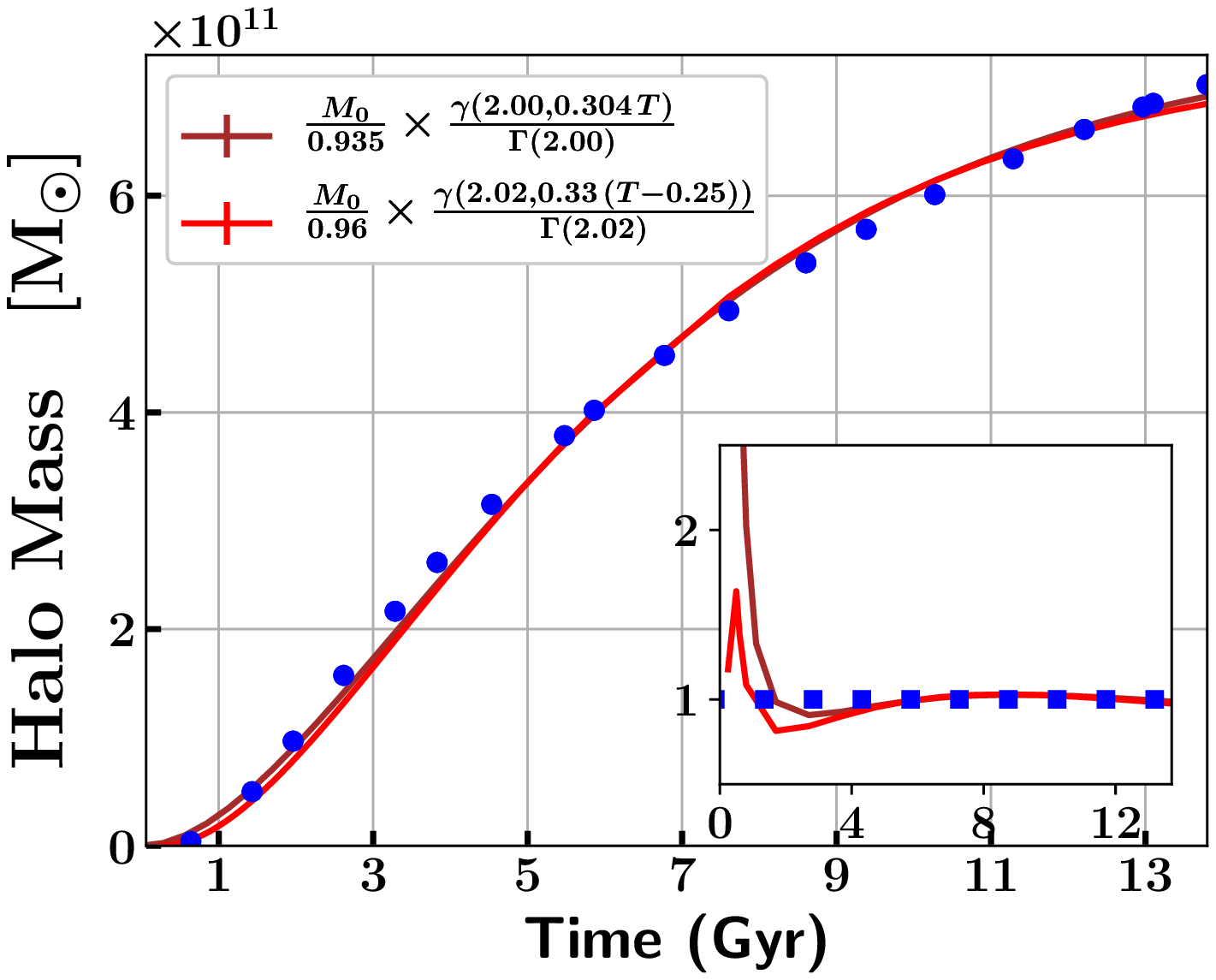}
\includegraphics[scale=0.42]{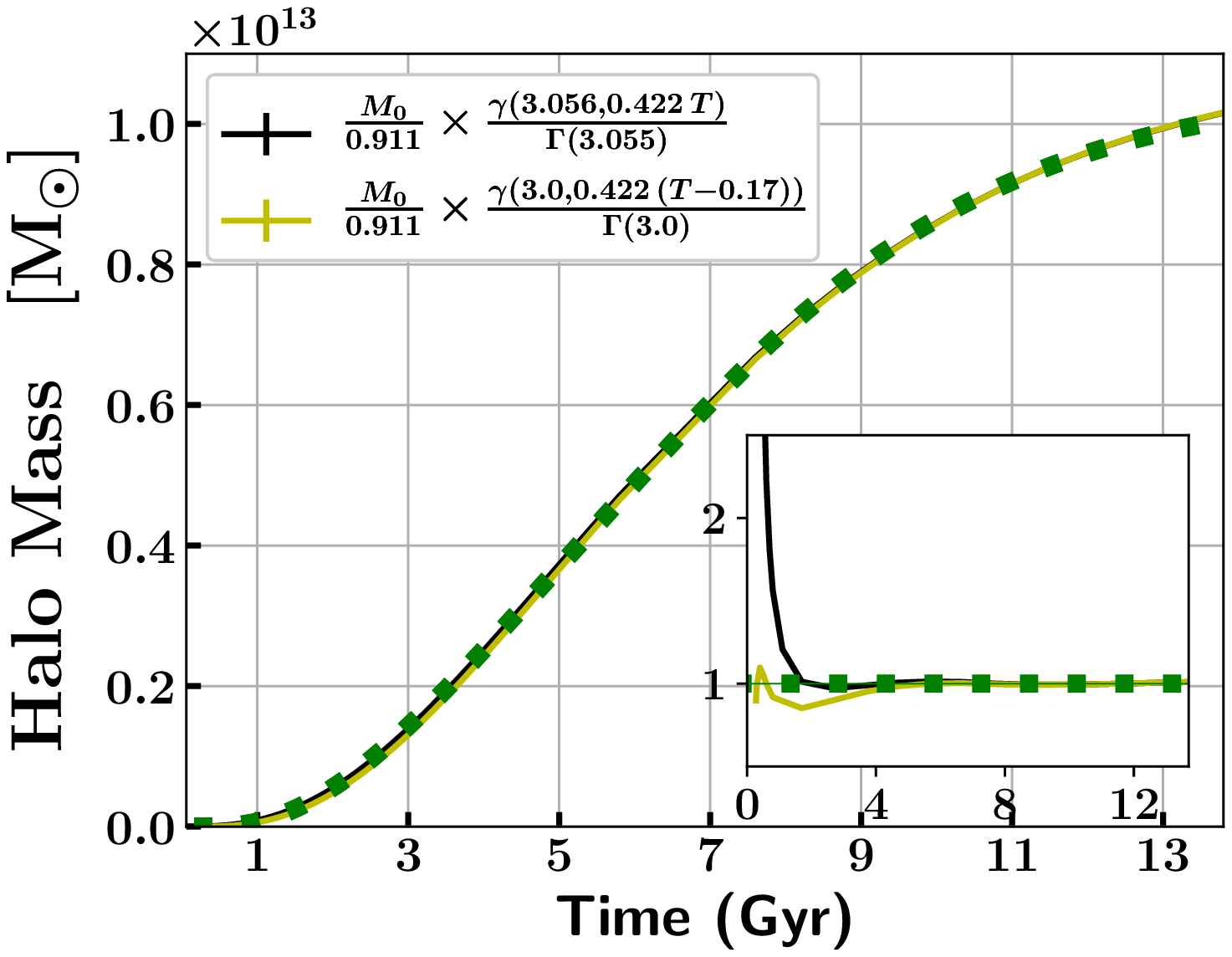}
\includegraphics[scale=0.42]{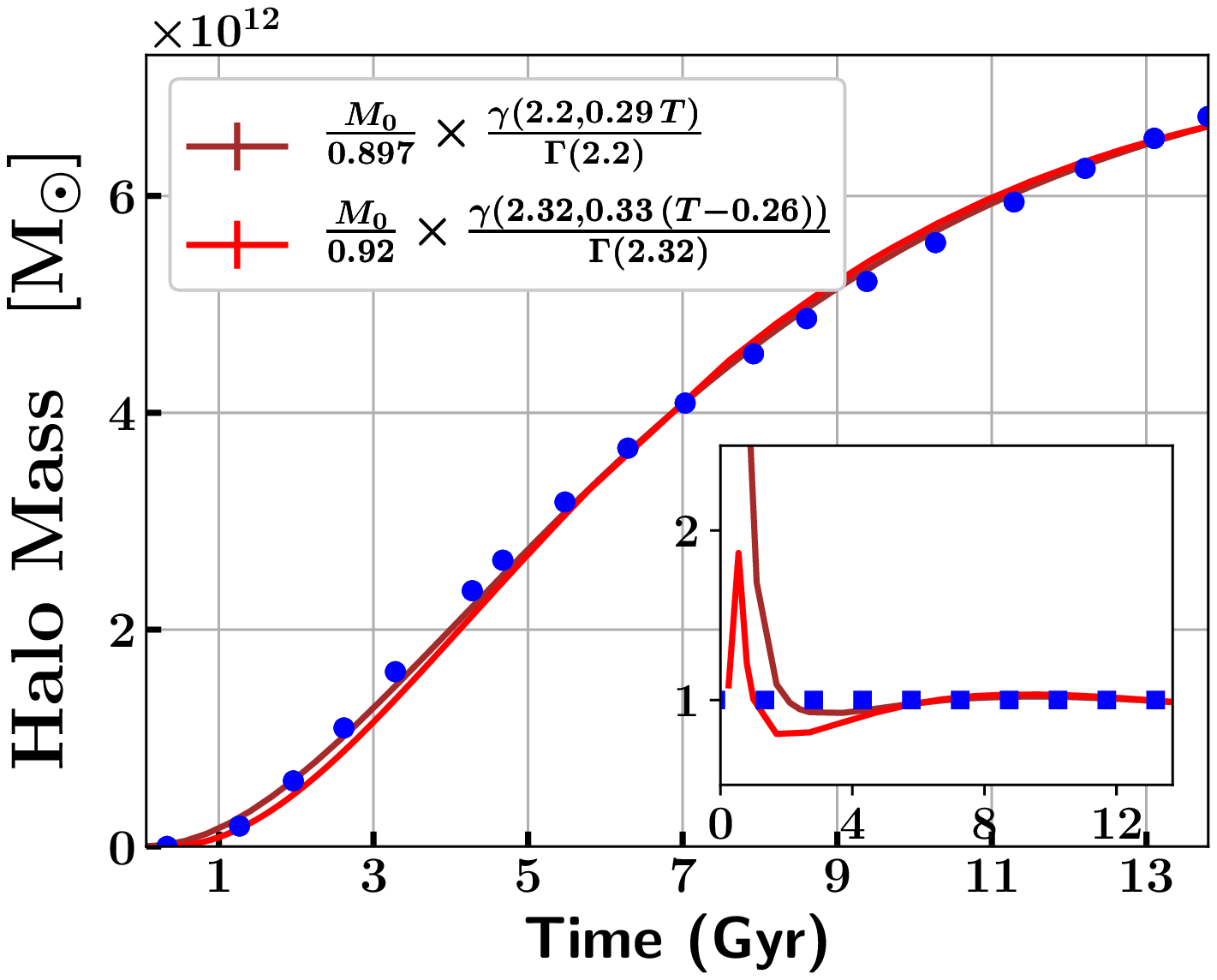}
\includegraphics[scale=0.42]{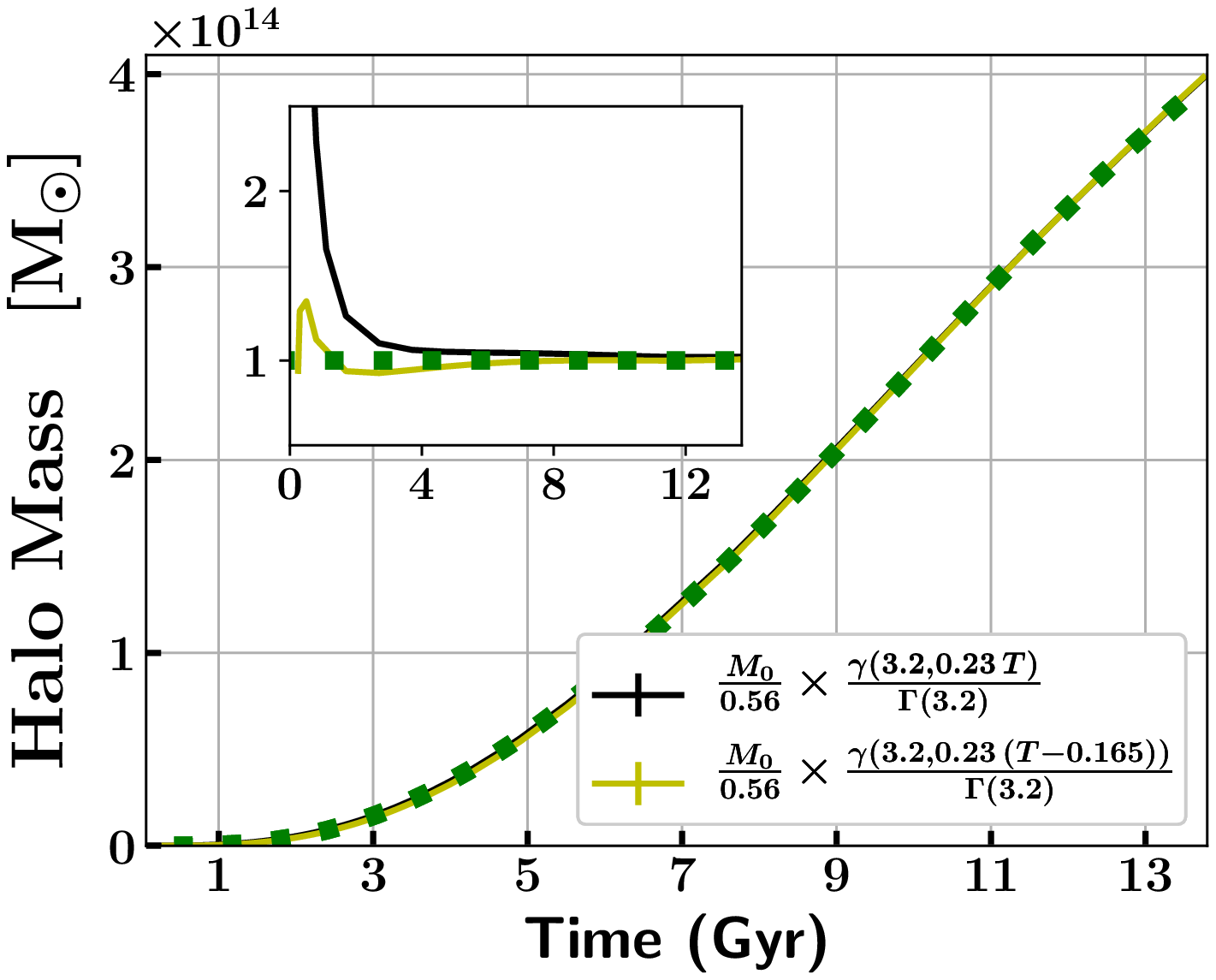}
\includegraphics[scale=0.42]{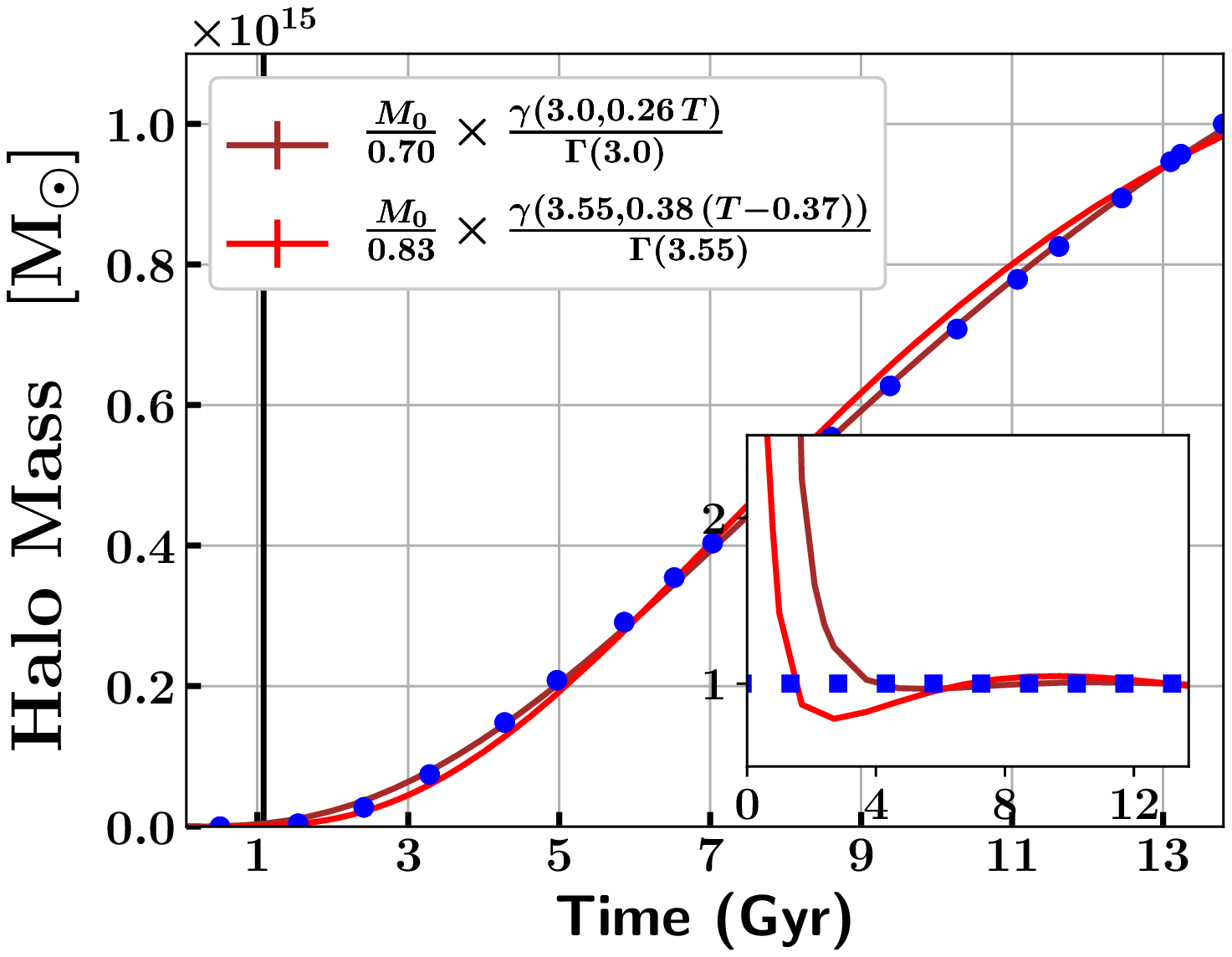}
\vspace{-0.30cm}
\vspace{-0.30cm}
\caption{Left panels: The MAHs obtained from  the \citet{Zhao2009} model (green squares). The black solid line represents a fit to the \citet{Zhao2009} MAHs using the functional form described by Eq. \ref{eq:MAHmodel} with three free parameters. The yellow solid line represents a fit to the \citet{Zhao2009} MAHs using the functional form described by Eq. \ref{eq:MAHmodel2} with four free parameters. Right panels: The MAHs obtained from  the \citet{Correa2015b} model (blue circles). The brown line represents a fit to the \citet{Correa2015b} MAHs using the functional form described by Eq. \ref{eq:MAHmodel} with three free parameters. The red solid line represents a fit to the  \citet{Correa2015b}  MAHs using the functional form described by Eq. \ref{eq:MAHmodel2} with four free parameters.  The importance of the $T_h$ parameter is visible in Fig. \ref{figEpicG4} and early times ($<$ 1 Gyr). }
\label{figEpicG2}
\end{figure*}

\begin{figure*}
  \vspace{1.0cm}
\centering
\includegraphics[scale=0.45]{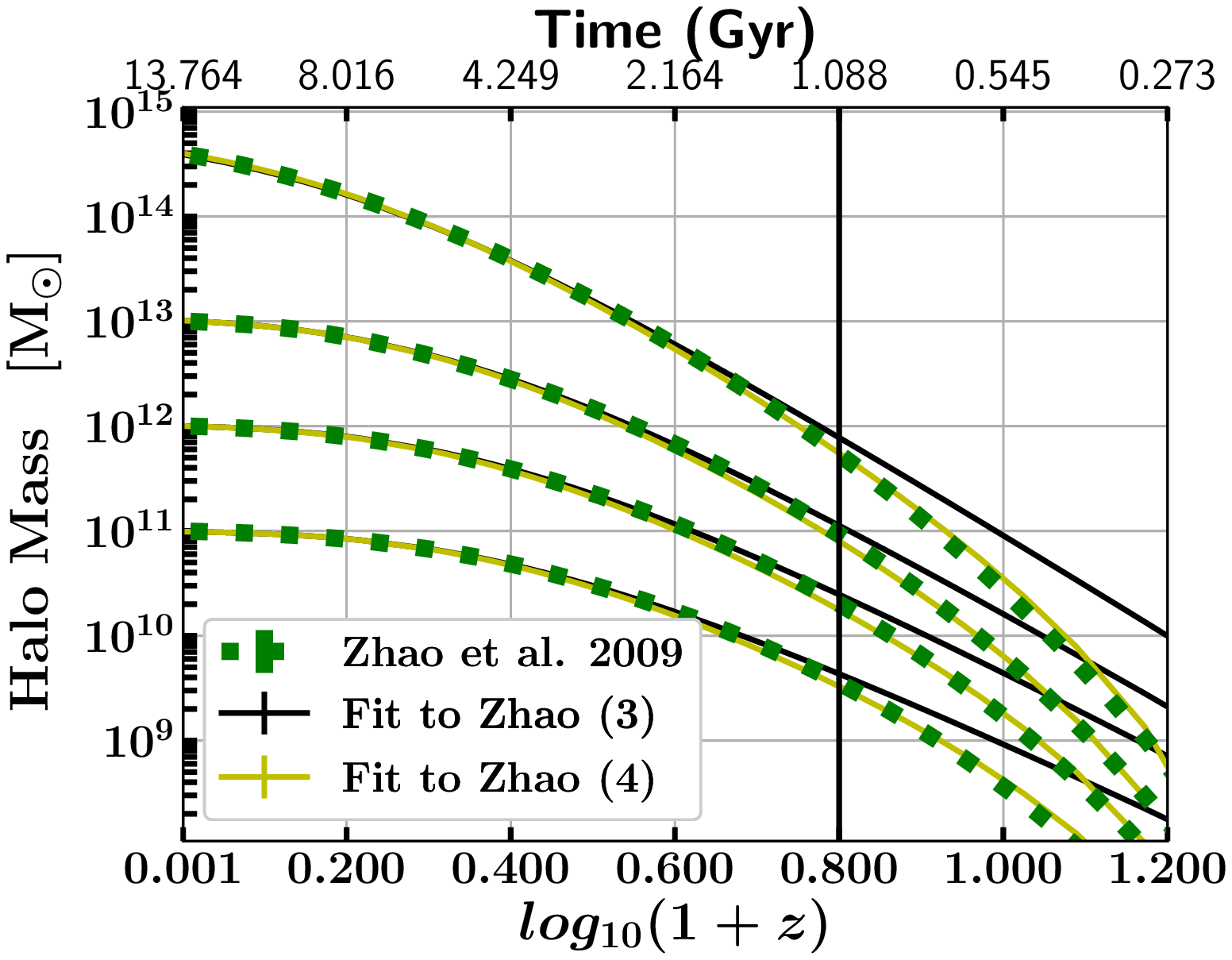}
\includegraphics[scale=0.45]{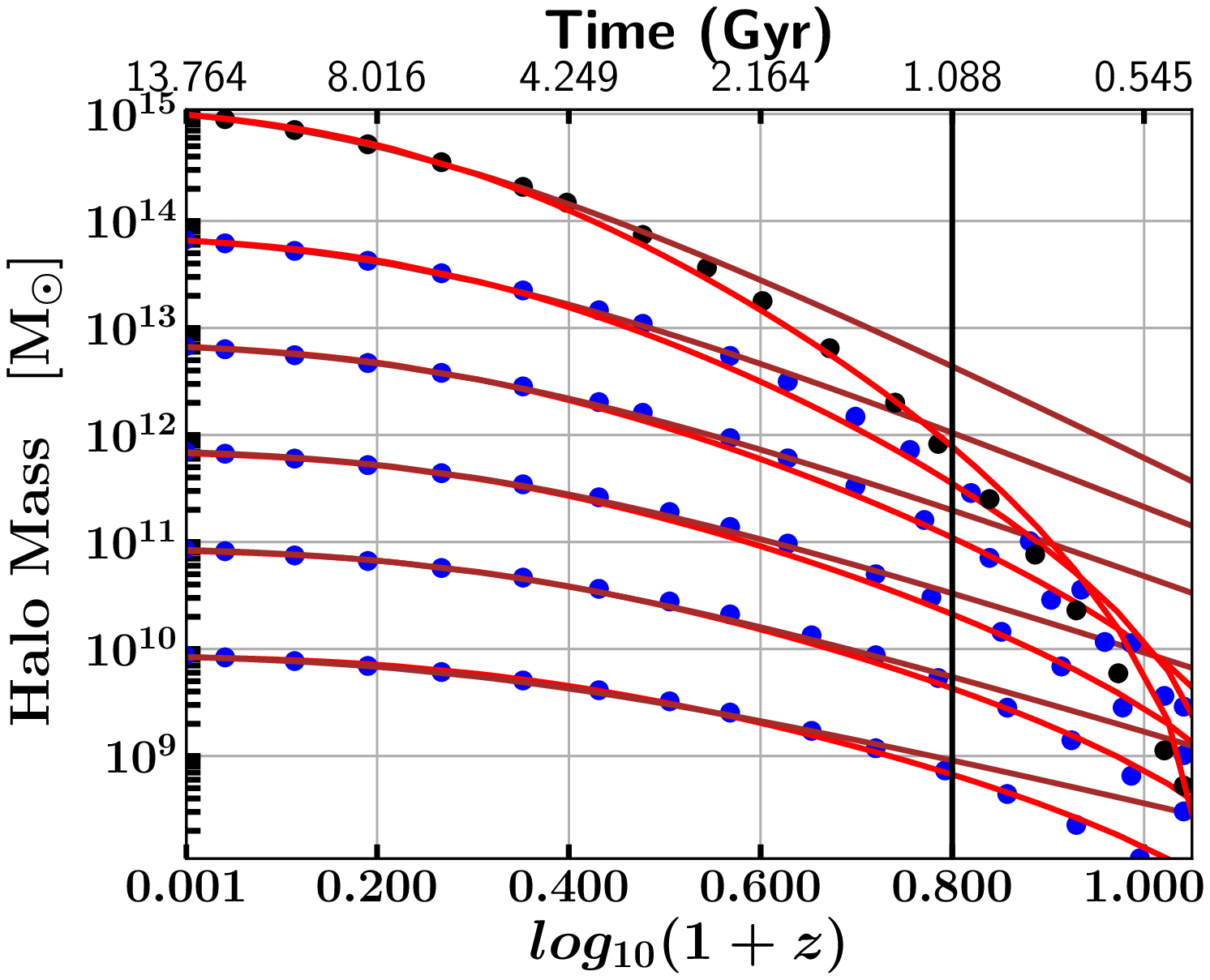}
\includegraphics[scale=0.45]{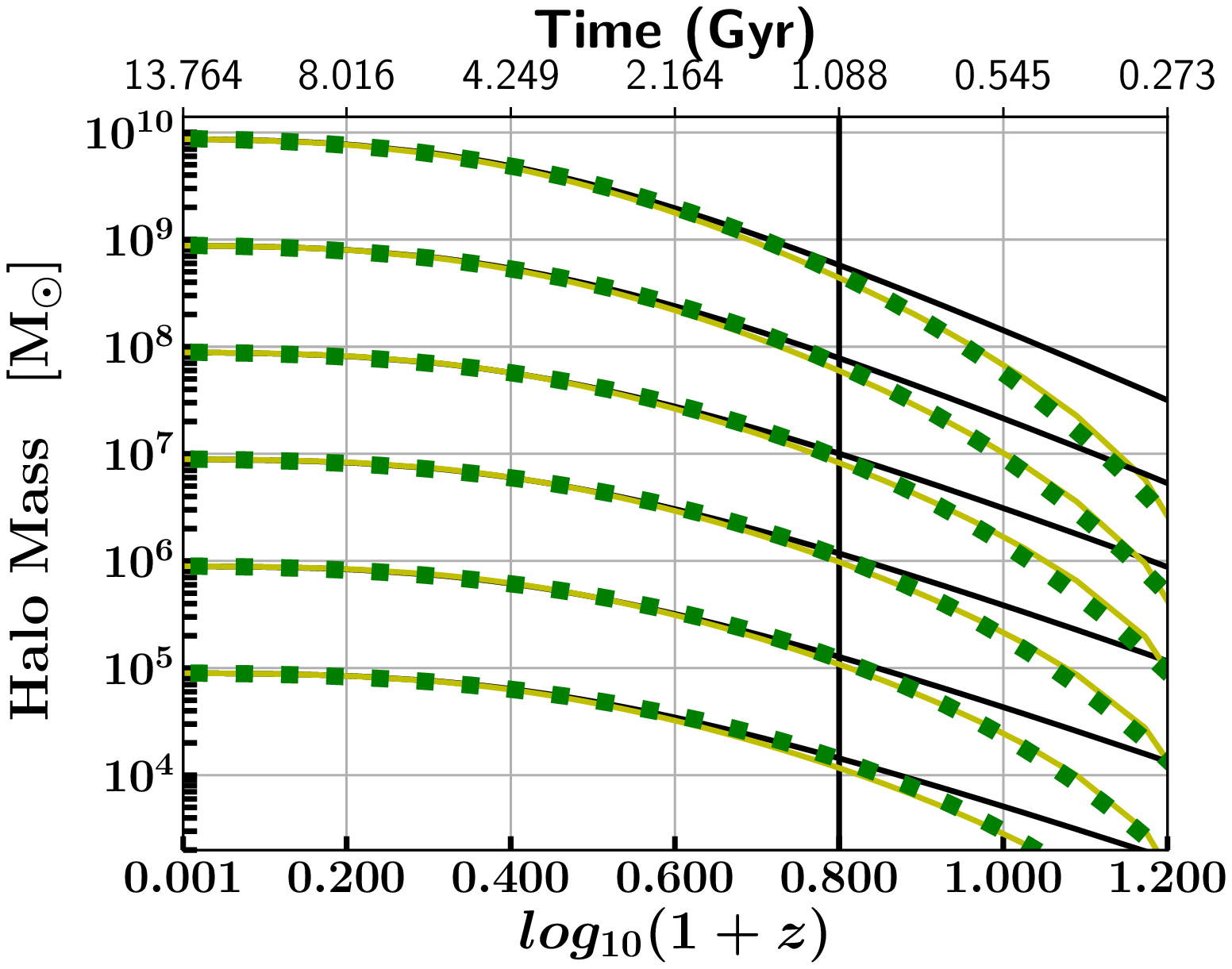}
\includegraphics[scale=0.45]{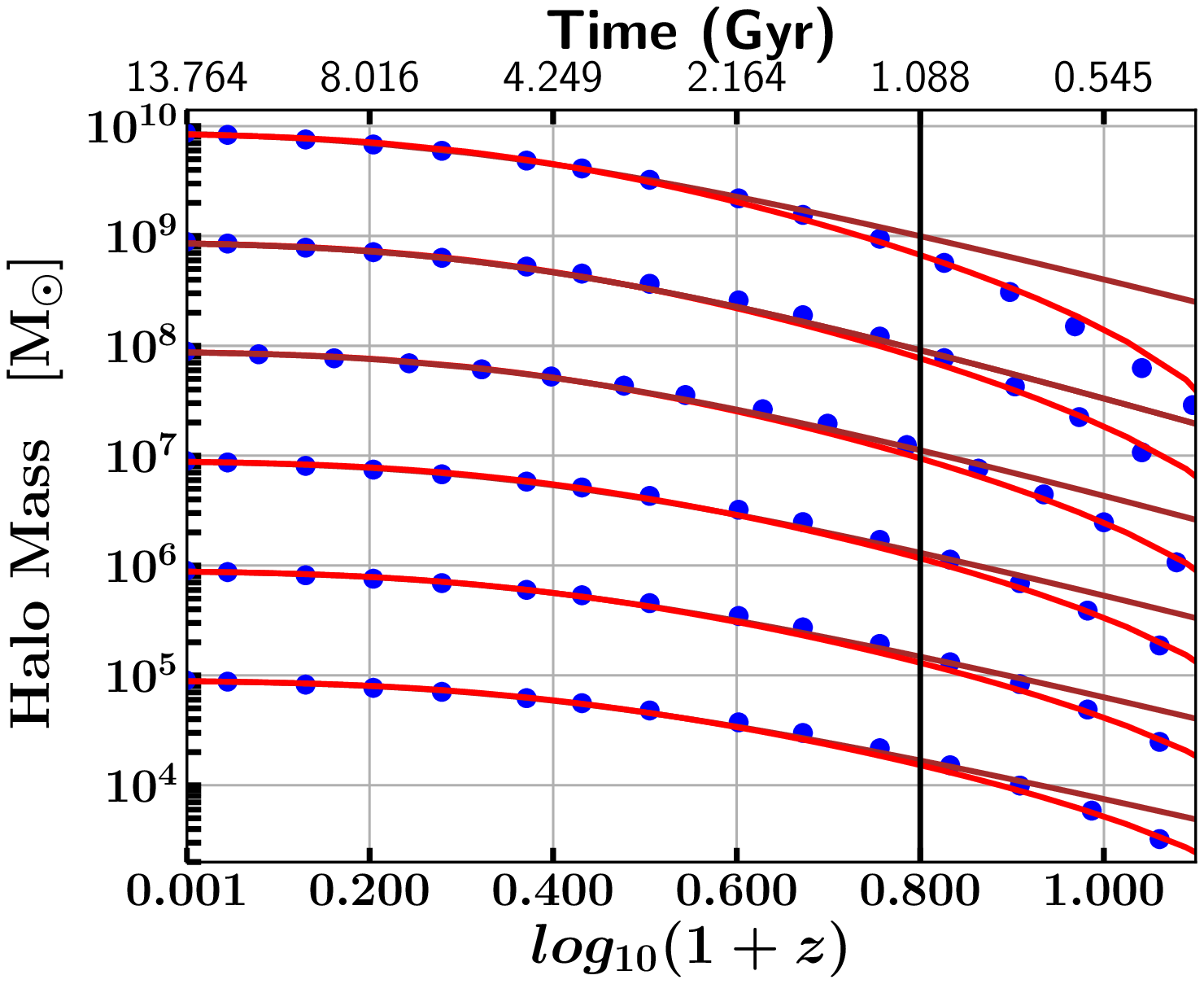}
\includegraphics[scale=0.45]{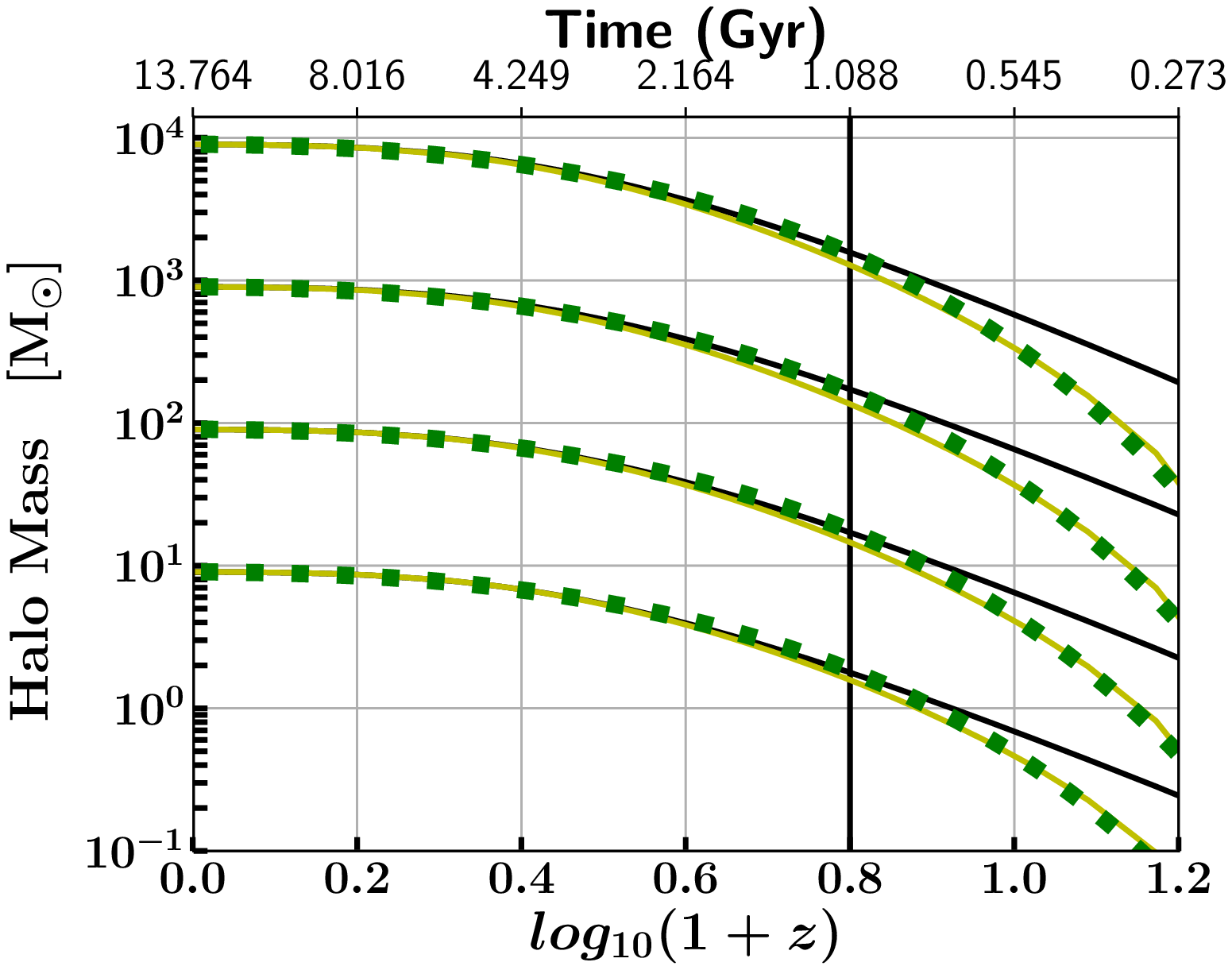}
\includegraphics[scale=0.45]{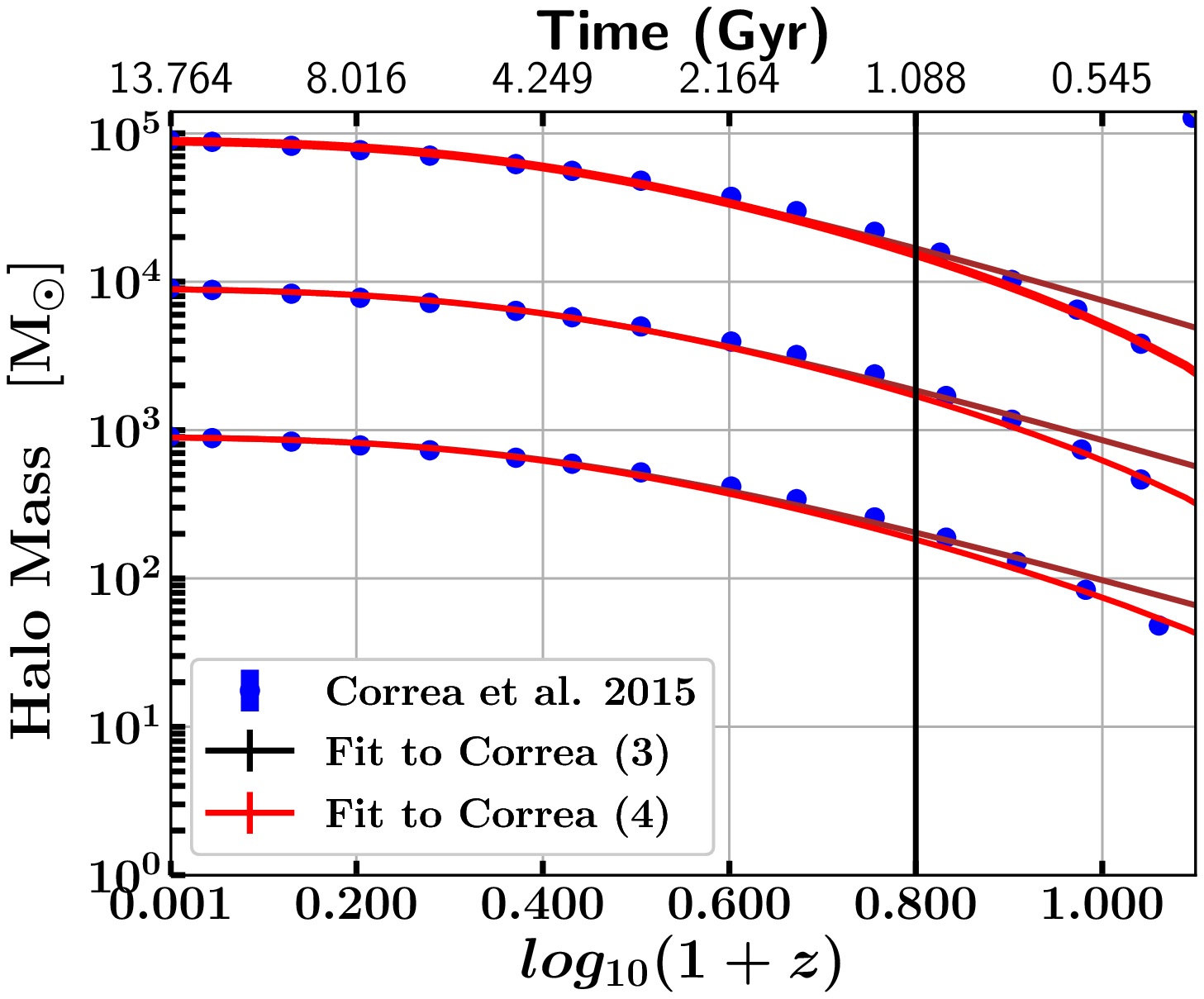}
\vspace{-0.30cm}
\vspace{-0.30cm}
\caption{Same as Fig. \ref{figEpicG2}, including low mass haloes,  but here in a $\log_{10}(M)-\log_{10}(1+z)$ scale.}
\label{figEpicG4}
\end{figure*}

\begin{table*}
  \centering
\resizebox{0.95\textwidth}{!}{%
  \begin{tabular}{cccccc}
    \hline \\
    &  & & 
    Parameters of $\Gamma$ model - \citet{Van2014}/Bolshoi \\
    \hline \hline
    & $M_{0}$ ($M_{\odot}$) & $f_{0}$ & $\alpha_h$  & $\beta_h$ (Gyr$^{-1}$) & $T_h$ (Gyr) \\
    \hline 
& 1e+11 & 0.945 $\pm$ 0.004 & 2.024 $\pm$ 0.026 & 0.345 $\pm$ 0.006 &  0.177 $\pm $  0.036 \\
& 1e+12 & 0.923 $\pm$ 0.003 & 2.352 $\pm$ 0.018 & 0.348 $\pm$  0.004 &  0.175 $\pm$ 0.049 \\
& 1e+13 & 0.866 $\pm$ 0.009 & 2.743 $\pm$ 0.054 &  0.341 $\pm$ 0.010 &  0.166 $\pm$ 0.054 \\
& 1e14 & 0.785 $\pm$ 0.014 & 3.307 $\pm$ 0.041 &  0.337 $\pm$ 0.022 &  0.102 $\pm$ 0.032 \\
& 4e14 & 0.525 $\pm$ 0.007 & 2.856 $\pm$  0.025 & 0.190 $\pm$ 0.004 &  0.224 $\pm$ 0.045 \\
    \hline \\
    &  & & 
    Parameters of $\Gamma$ model - EAGLE \\ 
    \hline \hline \\
& 8.67e+09 & 0.964 $\pm$ 0.017 & 1.422 $\pm$ 0.052 & 0.268 $\pm$ 0.009 &  0.171 $\pm$ 0.039 \\
& 2.01e+10 & 0.952 $\pm$ 0.022 & 1.436 $\pm$ 0.054  & 0.234 $\pm$ 0.012 &  0.182 $\pm$ 0.041 \\
    & 8.57e+10 & 0.900 $\pm$ 0.027 & 1.563 $\pm$ 0.065 & 0.220 $\pm$ 0.008 &  0.185 $\pm$ 0.066 \\
    & 7.03e+11 & 0.892 $\pm$ 0.024 & 1.822 $\pm$ 0.079 & 0.245 $\pm$ 0.009 &  0.185 $\pm$ 0.061 \\
    & 6.73+12 & 0.815 $\pm$ 0.021 & 2.067 $\pm$ 0.051 &  0.227 $\pm$ 0.005 &  0.185 $\pm$ 0.056 \\
& 6.68e+13 & 0.785 $\pm$ 0.023 & 2.718 $\pm$ 0.071 & 0.275 $\pm$ 0.017 &  0.123 $\pm$ 0.034 \\ 
    \hline \hline
  \end{tabular}%
}
\caption{The parameters of $\Gamma$ model (\ref{eq:MAHmodel2}) constrained from the MAHs of EAGLE and Bolshoi simulations. We include the standard deviation errors on the parameters.}
\label{tab:ParamsSims}
\end{table*}

Our objective in this section is to write the MAHs of haloes (derived from cosmological simulations and analytical models) with different $M_{0}$ in the language of the $\Gamma$ formalism (Eq. \ref{eq:GAMMAmodel}), which is inspired by the fractal-power-law growth ($T^{\alpha_{h}-1}$) and mass dependent exhaustion ($e^{-\beta_{h} \times T}$). More information about the motivation of this form and the different phases of evolution related to it (lag phase, exponential phase, deceleration phase and stationary phase) can be found in the appendix \ref{GAMMAMotivation}. We note that our purpose is {\it not} intended to provide a more {\it accurate} MAH model or to precisely reproduce any MAH of the literature (especially since there are some differences between different authors), but a model that is in concordance with the CSFRD model at $z \sim 0-9$. This will allow us to explore the relation between the parameters $\alpha_{\star}$, $\beta_{\star}$ of \citet{Katsianis2021b} and $\alpha_{h}$, $\beta_{h}$ for dark matter haloes, hence to provide a consistent picture for MAHs and SFHs. In addition, at the loss of precisely matching the existing models or simulations we are gaining a physical understanding of MAHs and a connection with other branches of science that $\Gamma$ growths occur.

\subsection{Constraining the parameters for the \citet{Zhao2009} and \citet{Correa2015b} models}
\label{ZhaoBolshoi}

The green squares in the left panels of Fig \ref{figEpicG2} (linear Mass - time scale) and Fig. \ref{figEpicG4} ($\log_{10}(M) - \log_{10}(1+z)$ scale) represent the MAHs of haloes with different $M_{0}$ given by the \citet{Zhao2009} model at the mass range of $M_{0} = 10^{1} - 10^{15} M_{\odot}$. We use the $\Gamma$ form as described by Eq. \ref{eq:MAHmodel} to describe the above evolution. We take $\alpha_h$, $\beta_h$ and $f_{0}$ as free parameters, and fit to the MAHs of haloes with different $M_0$. We find that the parameterization (black solid line of Fig. \ref{figEpicG2}) performs excellent in describing the MAHs of haloes of different masses from the Age of the Universe of $= 1$ Gyr (z = 5.3) to present. However, it does not perform equally well at higher redshifts ($z > 5.3$) where the halo masses derived from the above procedure are typically larger. This limitation can be clearly seen in Fig. \ref{figEpicG4}. The element that is missing to better describe the early evolution is to take into account the fact that galaxies and the galaxy-halo co-evolution does not occur exactly after the Big Bang in the $\Lambda$CDM scenario that we focus on, but instead there is a delay  $T_{h}$, i.e., the final fitting formula is:
\begin{eqnarray}
\label{eq:MAHmodel2}
{M_h(T)} =  \frac{M_{0}}{f_{0}} \, \times \frac{\gamma(\alpha_{h}, \beta_{h} \, \times (T-T_{h}))}{\Gamma(\alpha_{h})}. 
\end{eqnarray}
The above description is represented by the yellow solid lines in the left panels of Fig. \ref{figEpicG2}  and \ref{figEpicG4} and is able to succesfully describe the MAHs of the \citet{Zhao2009} model for $z = 0-9$.  We present the parameters at table \ref{tab:ParamsModels} and include their standard deviation errors. We note that even if there are some values slightly above 1.0 (by 0.001-0.02) for the $f_{0}$ parameter at small haloes we set these values in our analysis at $f_{0}$ = 1.0. Besides that the above choice has a negligible affect to the results we decide in our analysis to set $f_{0}$ = 1.0 for these low mass haloes since the \citet{Zhao2009} model does not describe mass loss and thus values of $f_{0}$ $>$ 1.0 are not physical but just an artefact of the fitting process. 

We follow the same line of logic to describe the MAHs of \citet{Correa2015b} which are represented by the blue filled circles in the right panels of Fig. \ref{figEpicG2}, Fig. \ref{figEpicG4} and table \ref{tab:ParamsModels}. The brown solid line describes our fit using equation \ref{eq:MAHmodel} while the red solid line employs equation \ref{eq:MAHmodel2}. We note that unlike the case of the \citet{Zhao2009} model, which can be described almost perfectly by our $\Gamma$ formalism at z = 0-9, there are some deviations between our best fit and the model of \citet{Correa2015b}. However, these differences are very small for T $>$ 0.5 Gyr (z $<$ 10) and appear very briefly at T $ \sim 0.4-0.5$ Gyr. In addition, we are unable to  describe MAHs with  $M_{0} = 10^{15} M_{\odot}$, where maybe the parameterization \ref{Tasi} employed by  \citet{Correa2015b} and \citet{McBride2009} has its limitations (more discussion in \citet{Behroozi2013}). Furthermore, we are also unable to describe the MAHs of very small haloes of $M_{0} = 10^{0}-10^{3} M_{\odot}$ We conclude that {\it broadly} the MAHs described by the \citet{Correa2015b} model for $M_{0} = 10^{3} - 10^{14} M_{\odot}$ can also be described by $\Gamma$ forms at the redshift range we consider (Fig. \ref{figEpicG4}). 

We stress that both \citet{Zhao2009} and \citet{Correa2015b} models were established using cosmological simulations and involved objects with relatively large halo masses ($M_0 > 10^{9} M_{\odot}$). Thus, their results at lower masses that we present in this work ($M_0 < 10^{9} M_{\odot}$) require excessive extrapolations and are indeed uncertain. In addition, no galaxies form at small haloes, thus there is not direct connection to the CSFRD from which our preferred $\Gamma$ growth is inspired. However, both the \citet{Zhao2009} and \citet{Correa2015b} models are quite sophisticated and they have been tested against zoom-in N-body simulations at this low mass regime performing adequately well. To be more precise the history of dark matter haloes with $M_0 =  10^{1}-10^{9} M_{\odot}$ has been studied using the VVV simulations \citep{Wang2020}/personal communication. The analytical models of \citet{Zhao2009} and \citet{Correa2015b} performed quite well even for these low mass masses. Thus, we find it is interesting to investigate if our $\Gamma$ formalism is adequate to describe the MAHs of these low mass objects and we will be able to constrain better the behavior of the $\alpha_h$ and $\beta_h$ parameters using a large halo mass range ($M_0 =  10^{1}-10^{15} M_{\odot}$). Last, we stress that both the \citet{Zhao2009} and \citet{Correa2015b} models were formulated in order to present the {\it average} trajectory of the MAH of a halo given its $M_{0}$. Thus, our model is also following MAHs at an average sense and individual haloes can deviate from this behavior, especially at the moments of major mergers. Our errors do not reflect this diversity of individual MAHs. We also note that our model is not able to follow halo growth bellow/around the value of $T_{h}$.

\begin{figure*}
\centering
\includegraphics[scale=0.42]{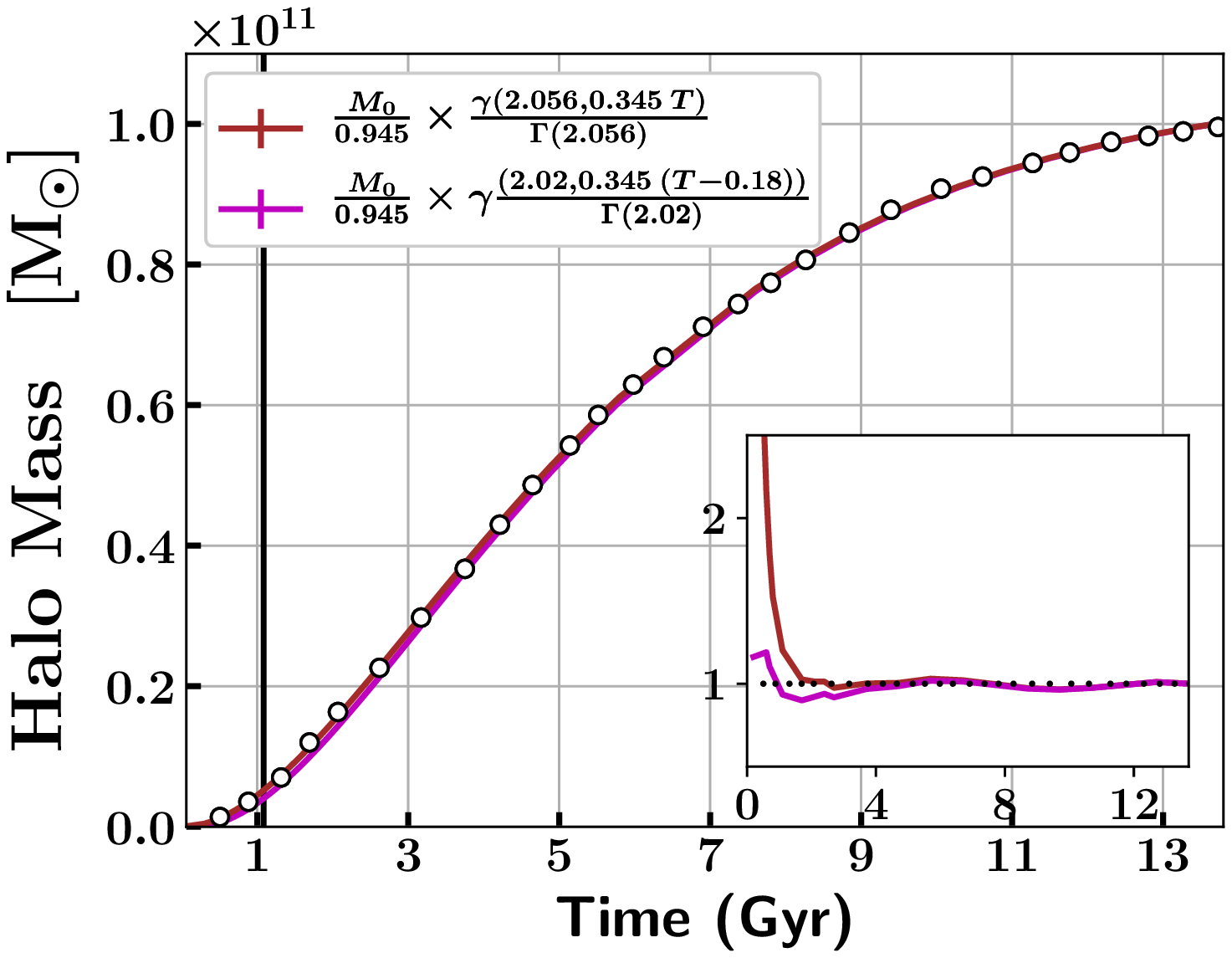}  
\includegraphics[scale=0.42]{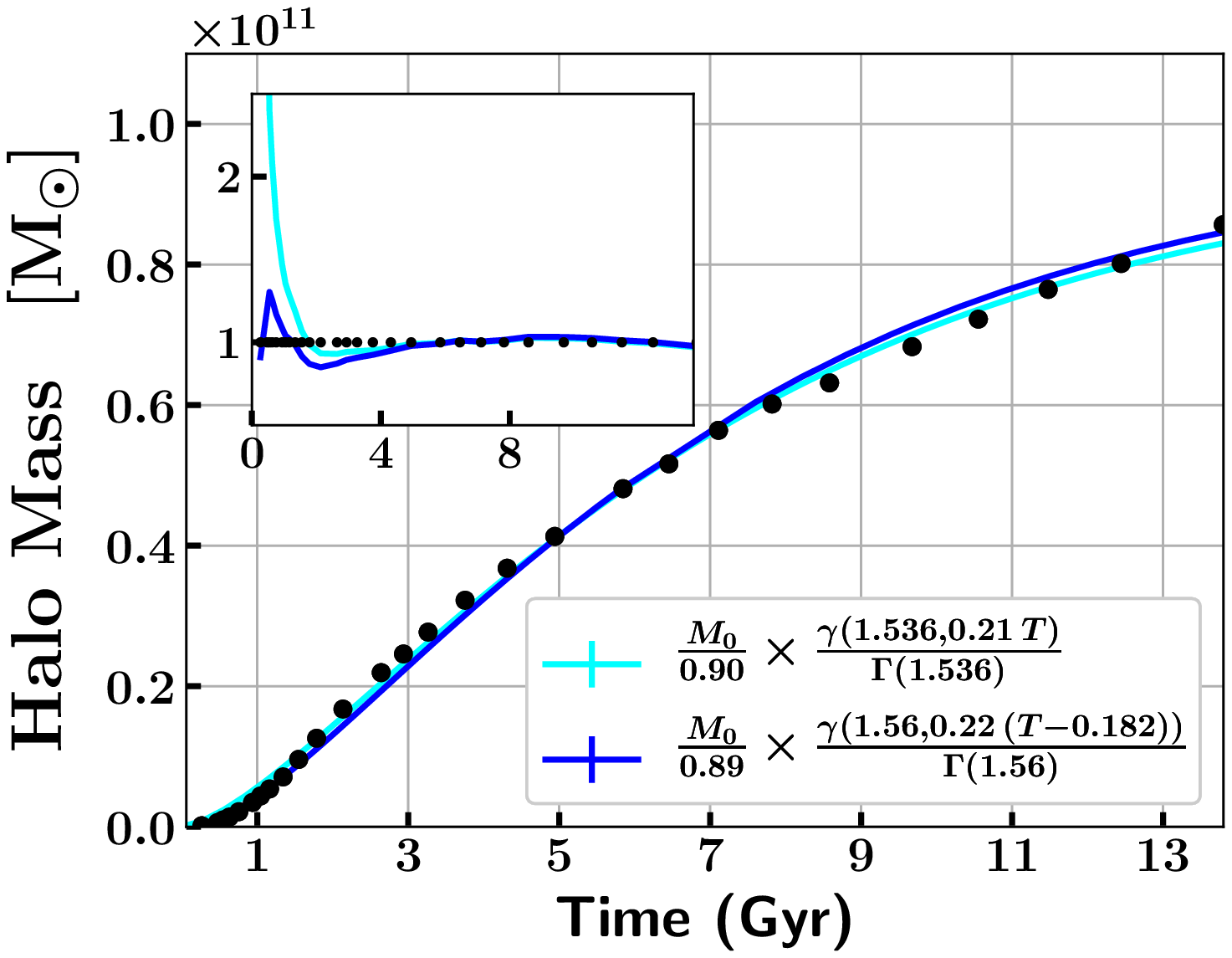}
\includegraphics[scale=0.42]{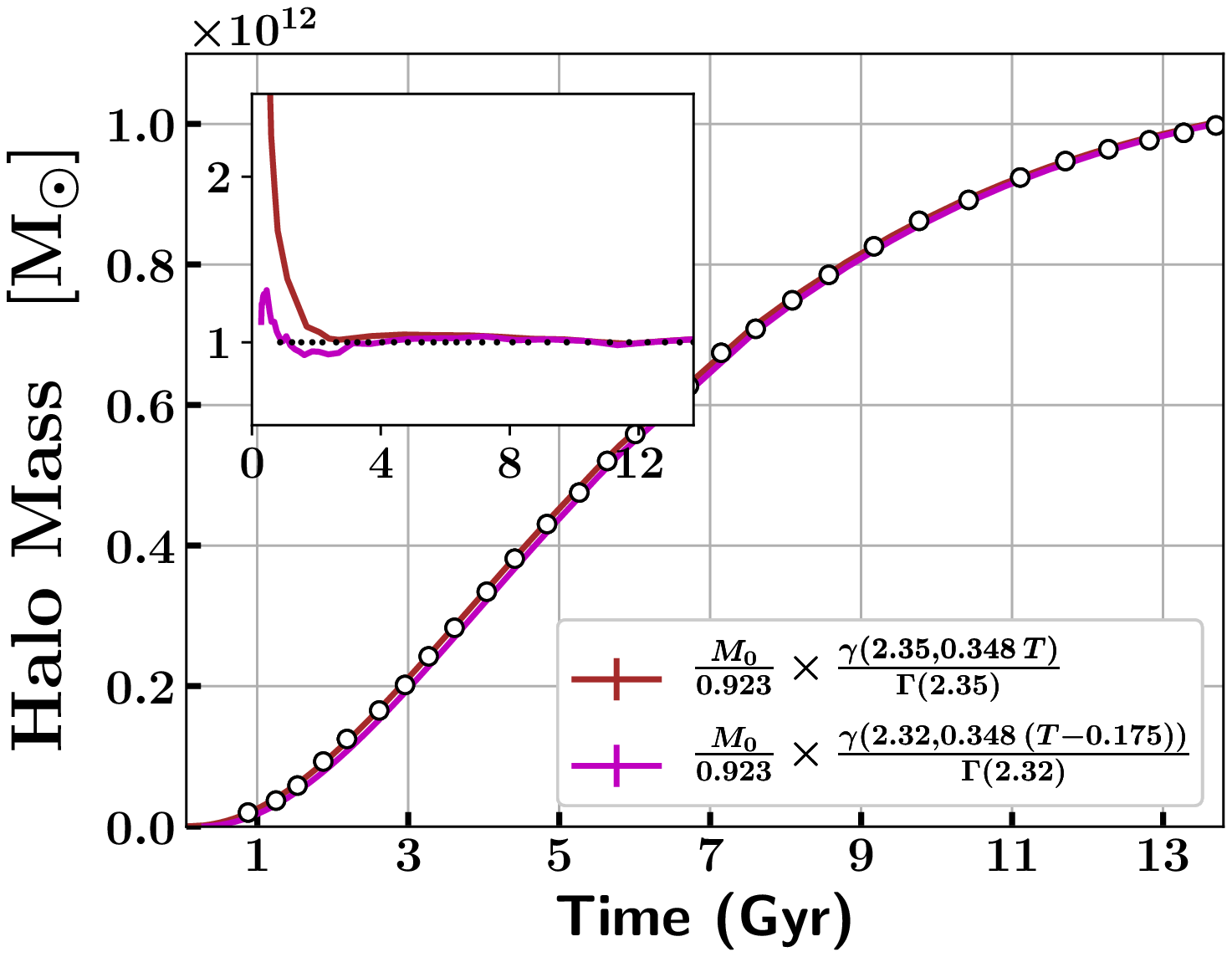}
\includegraphics[scale=0.42]{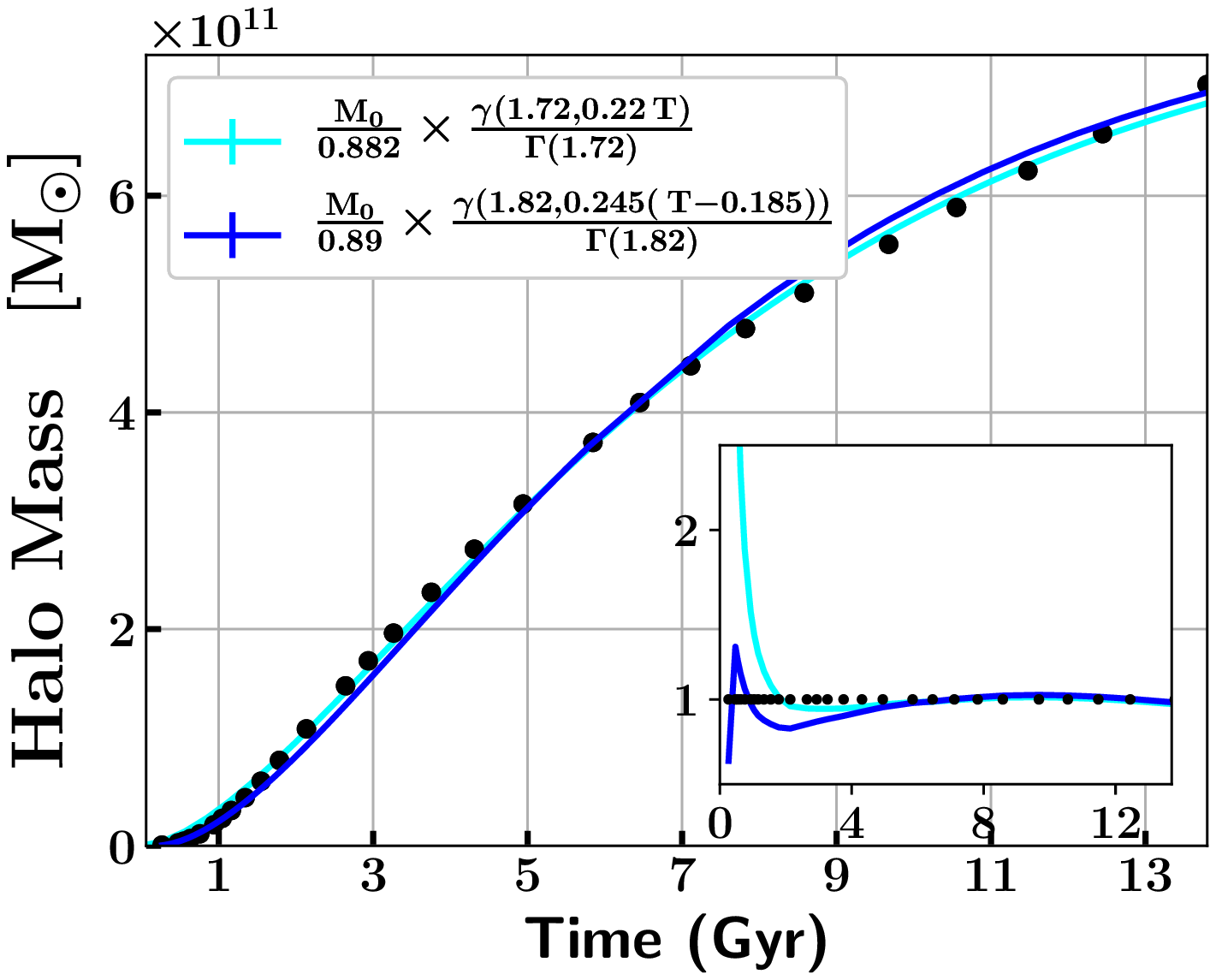}
\includegraphics[scale=0.42]{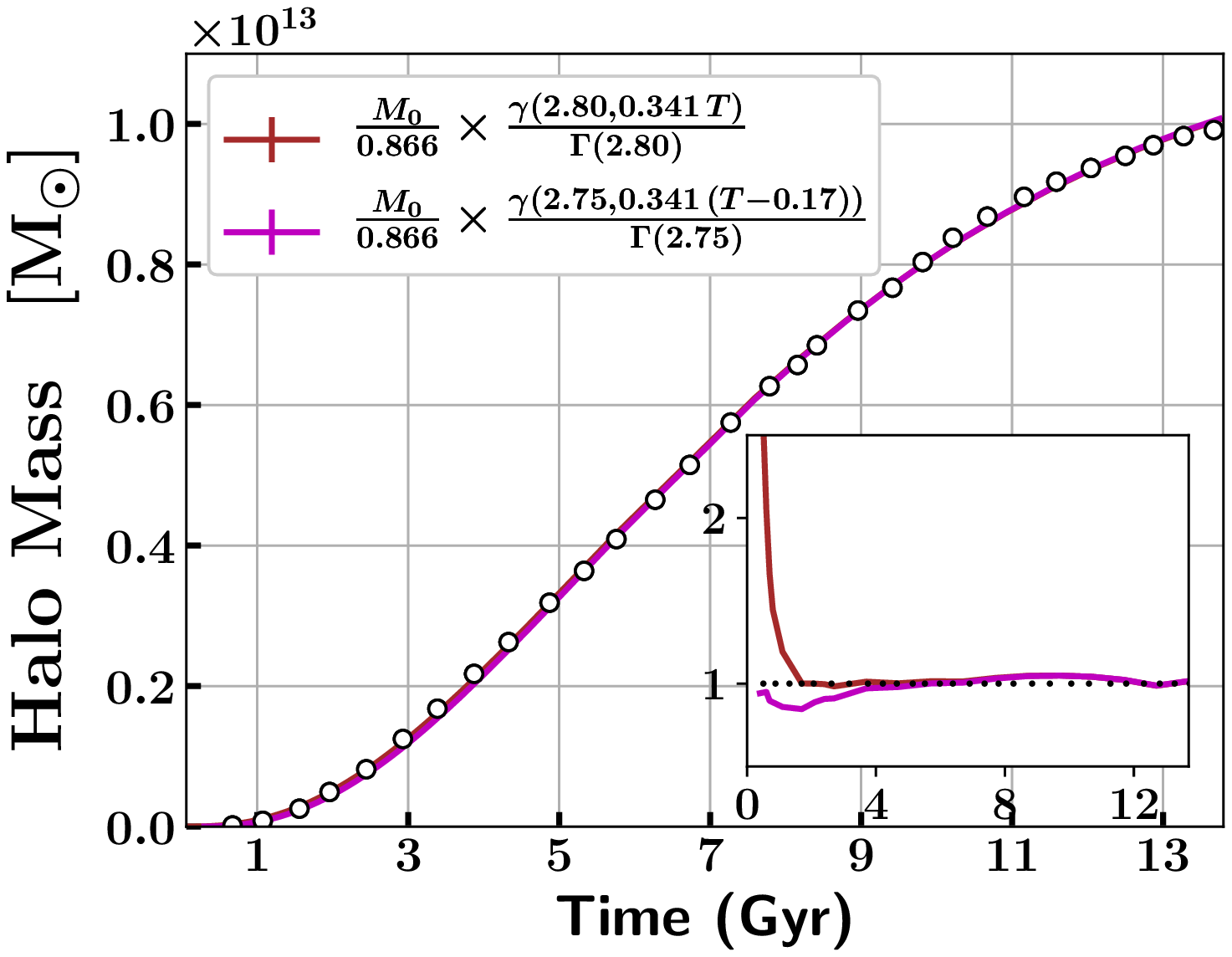}
\includegraphics[scale=0.42]{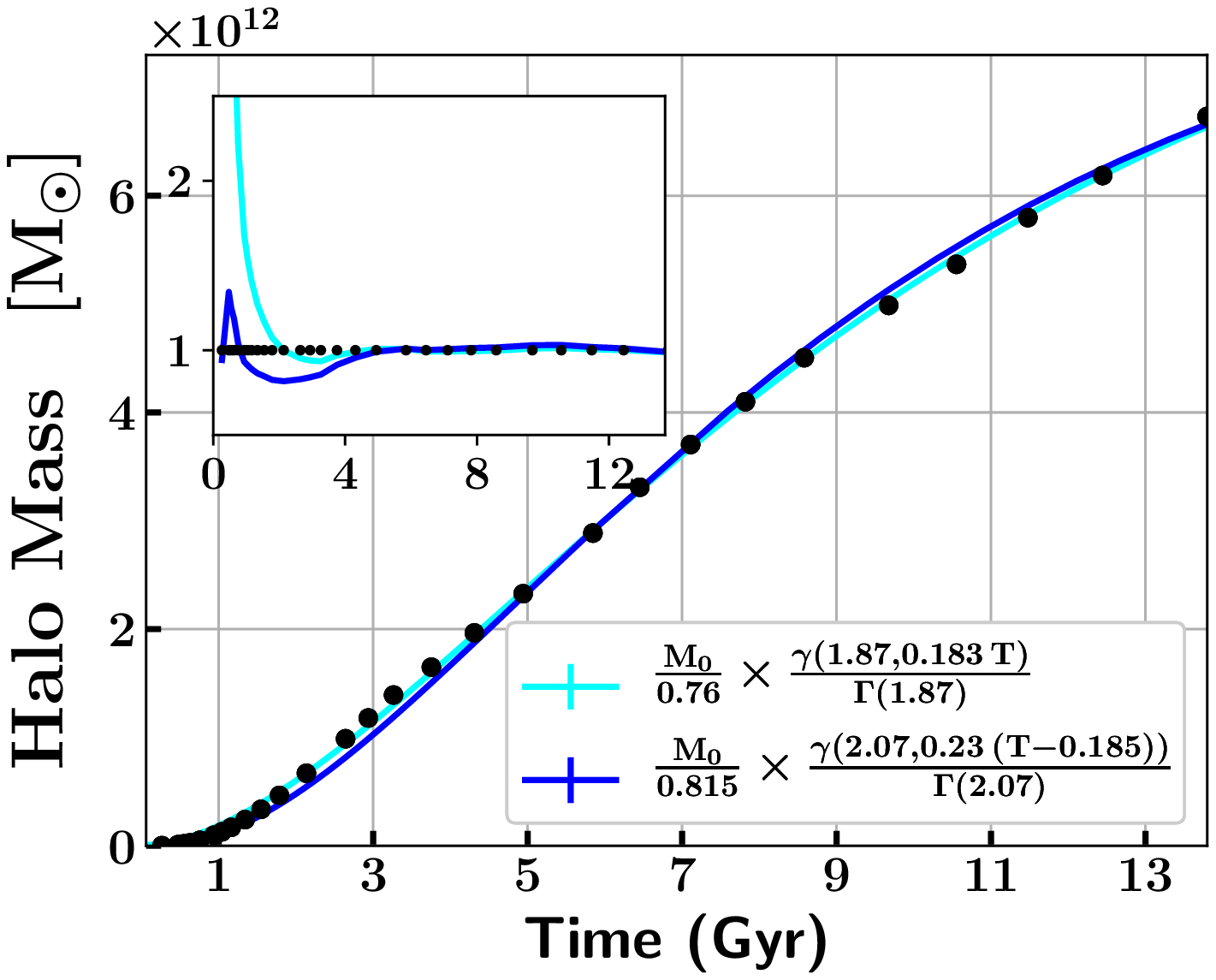}
\includegraphics[scale=0.42]{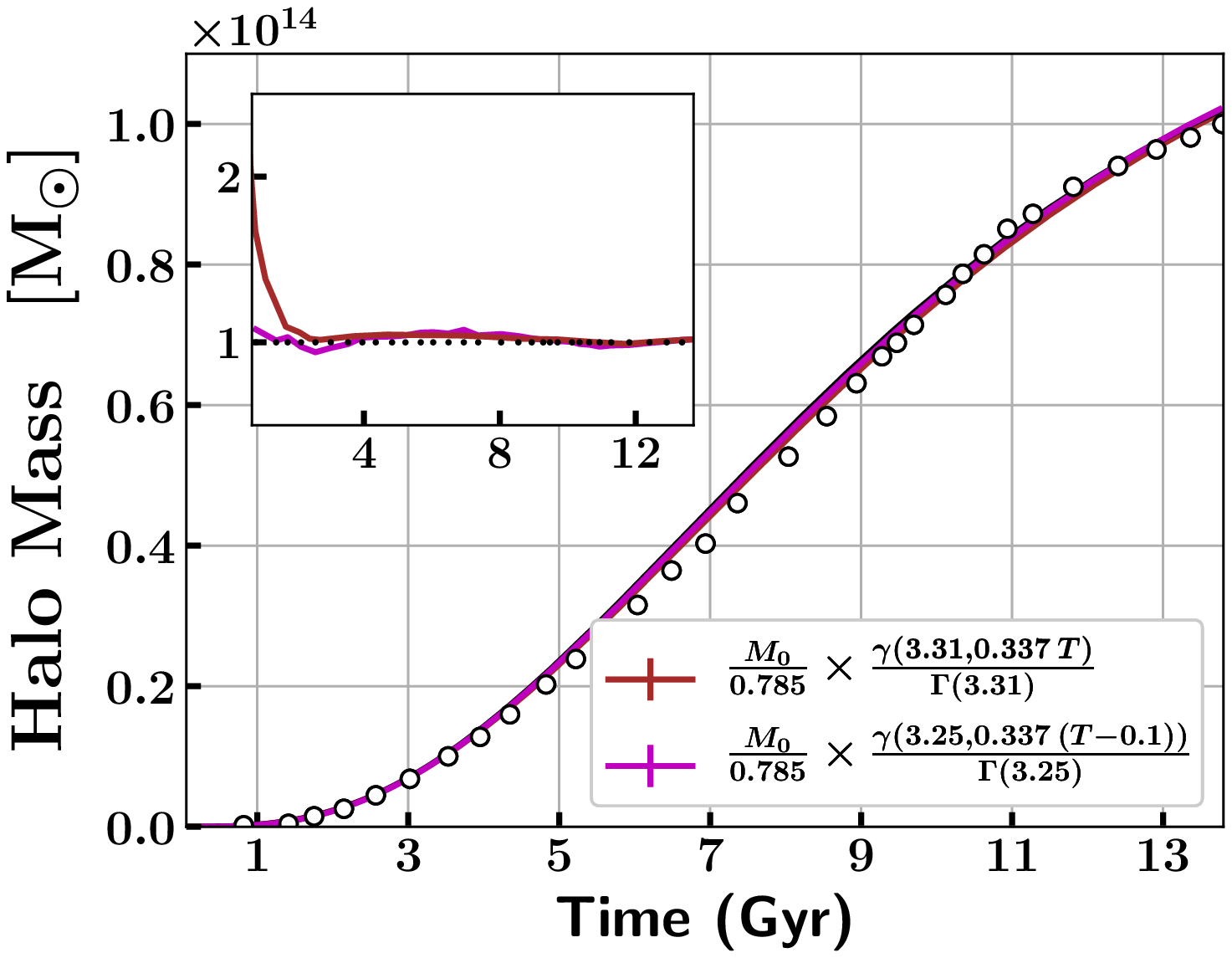}
\includegraphics[scale=0.42]{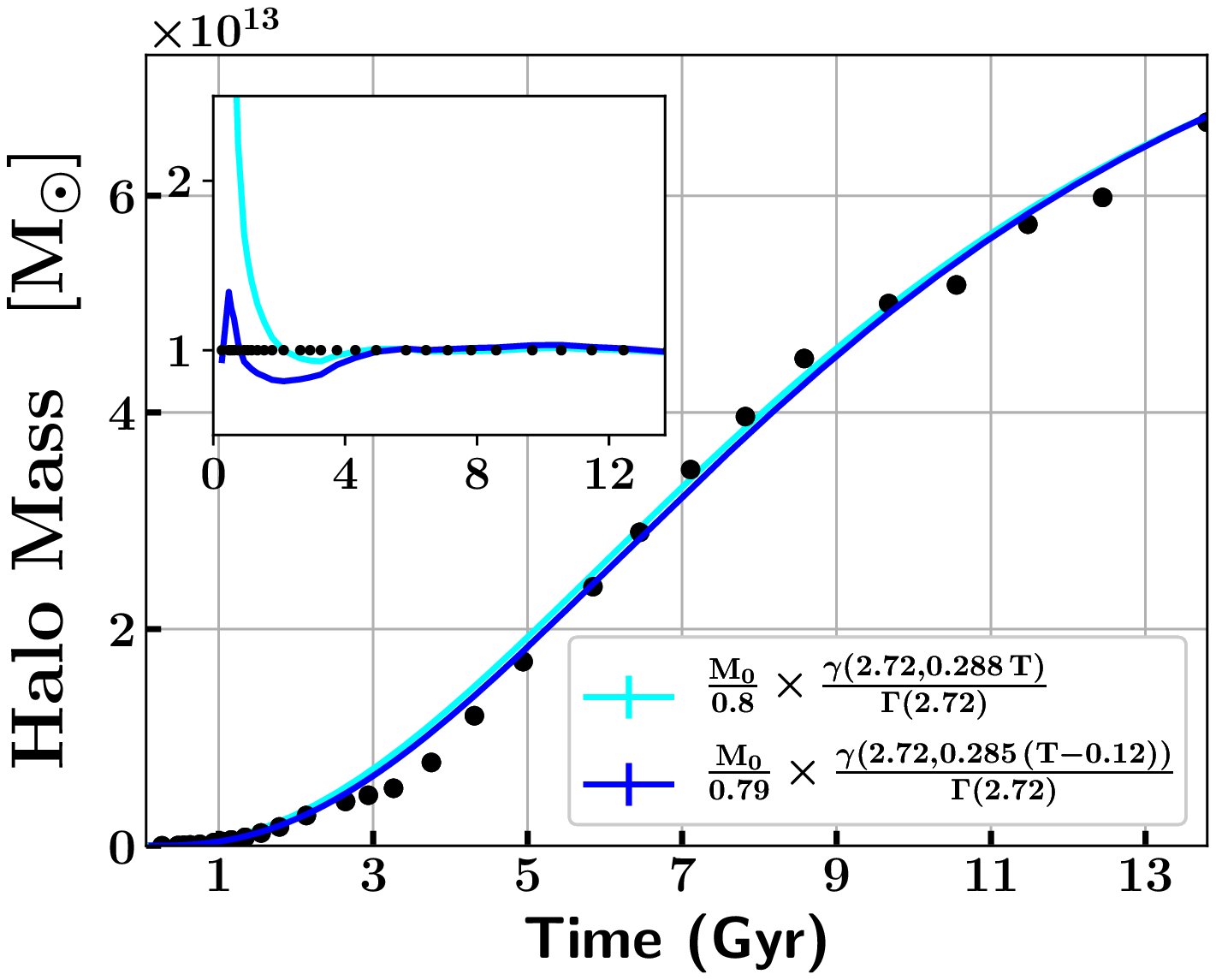}
\vspace{-0.30cm}
\vspace{-0.30cm}
\caption{Left panels: The  MAHs obtained from the \citet{Van2014}-Bolshoi simulations (open circles). The brown solid line represents a fit to the \citet{Van2014} MAHs using the functional form described by Eq. \ref{eq:MAHmodel} with three free parameters. The magenta solid line represents a fit to the \citet{Van2014} MAHs using the functional form described by Eq. \ref{eq:MAHmodel2} with four free parameters.  Right panels: The MAHs obtained from the EAGLE simulations (black filled circles). The cyan solid line represents a fit to the simulated MAHs using the functional form described by Eq. \ref{eq:MAHmodel} with three free parameters. The blue solid line represents a fit to the EAGLE  MAHs using the functional form described by Eq. \ref{eq:MAHmodel2} with four free parameters. The importance of the $T_h$ parameter is only visible in Fig. \ref{figEpicG7} and earlier times $<$ 1 Gyr. }
\label{figEpicG5}
\end{figure*}

\begin{figure*}
  \vspace{1.0cm}
\centering
\includegraphics[scale=0.45]{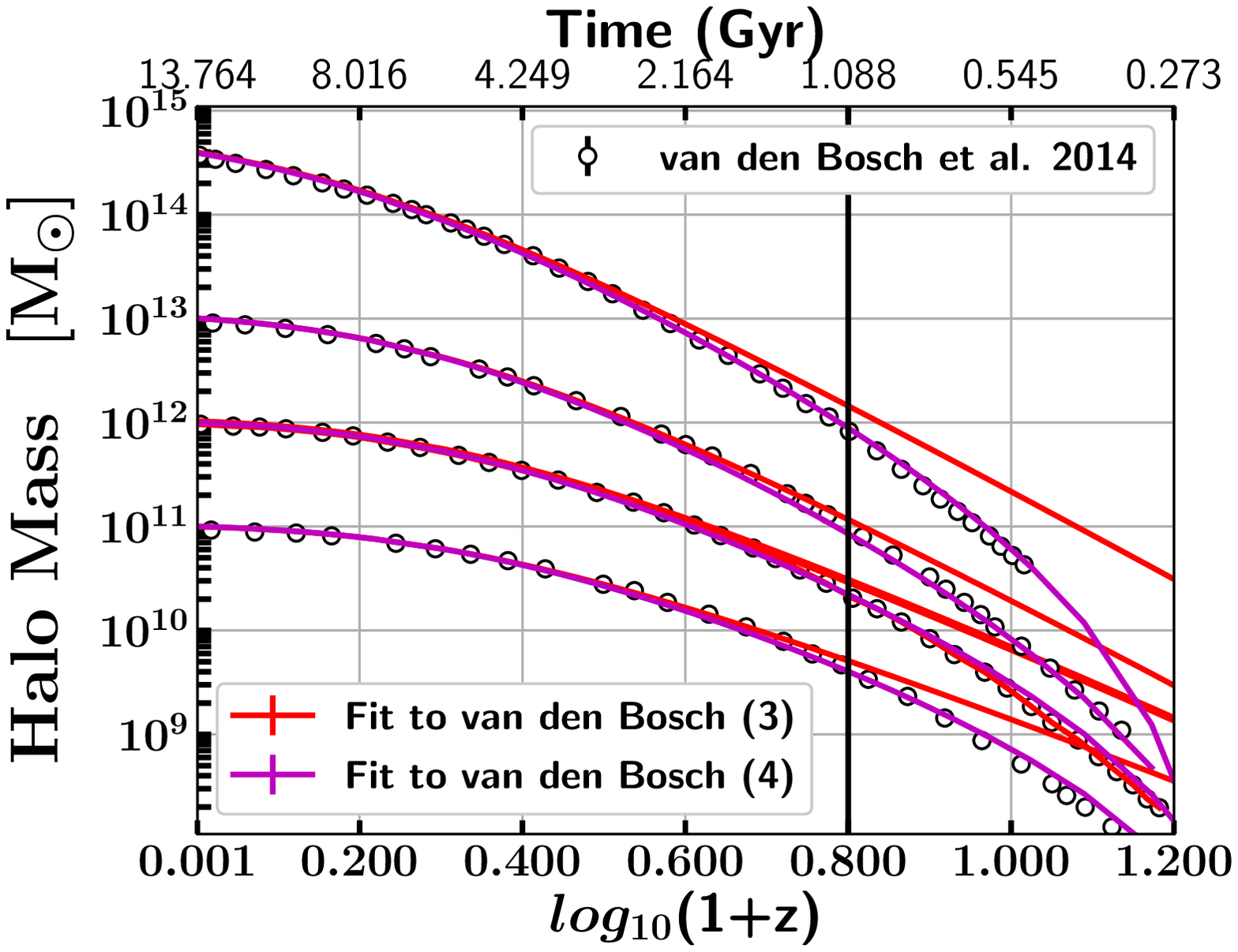}
\includegraphics[scale=0.45]{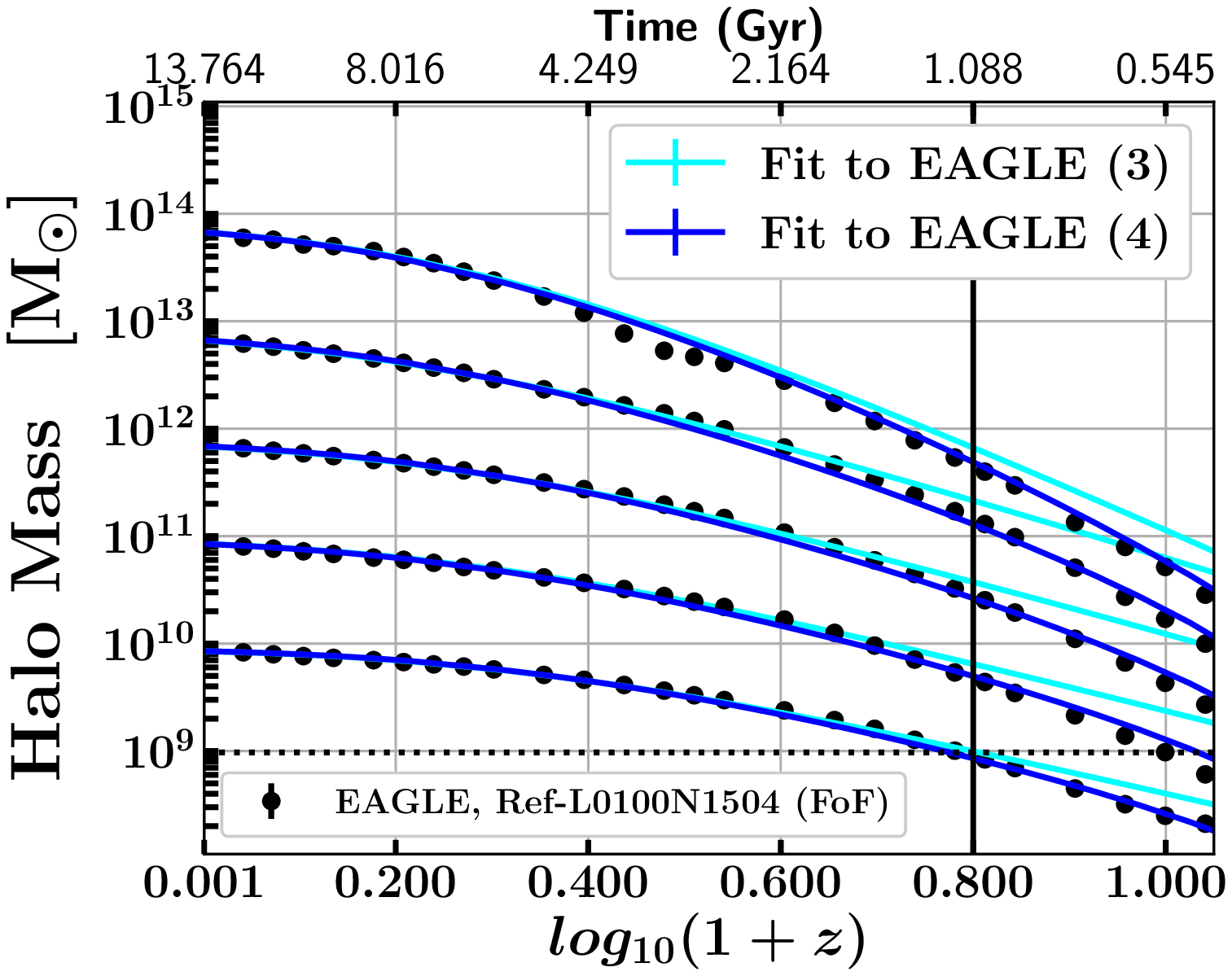}
\vspace{-0.30cm}
\vspace{-0.30cm}
\caption{Same as Fig. \ref{figEpicG5} but in a $\log_{10}(M)-\log_{10}(1+z)$ scale.}
\label{figEpicG7}
\end{figure*}

\subsection{MAH measurements from Simulations}

Besides analytical prescriptions to describe MAHs we decide as well to directly use cosmological simulations. These are unable to probe the low mass haloes due to their low mass resolution. However, we want 1) to investigate if the $\Gamma$ formalism adopted in this work can describe the simulated MAHs and 2) perform a qualitative comparison with the analytical models in section \ref{sec_param}. 

\begin{itemize}

\item
\citet{Van2014} presented a detailed study of how dark matter haloes assemble their mass and grow their potential wells relying on the Bolshoi simulation \citep{Klypin2011}. The publicly available halo catalogs comes from merger trees obtained using the phase-space halo finder ROCKSTAR \citep{Behroozi2013}. The authors presented the results obtained from the Bolshoi simulation, where each line is the average obtained from all haloes in a mass bin that is 0.2 dex wide. We use the same MAHs as in \citet{Van2014}. The results are represented by the open circles in the left panels of Fig. \ref{figEpicG5} (linear Mass - time scale ) and Fig. \ref{figEpicG7} ($\log_{10}(M) - \log_{10}(1+z)$ scale).  The brown solid line describes our fit using equation \ref{eq:MAHmodel} while the magenta solid line employs equation \ref{eq:MAHmodel2}. The parameters can also been found at table \ref{tab:ParamsSims}. We find that our $\Gamma$ form is an excellent parameterization to describe the simulated MAHs found in Bolshoi.

\item
The EAGLE simulations include a detailed merger tree in their user-friendly SQL public catalogue \citep{McAlpine2016}. haloes are identified using the friends-of-friends (FoF) algorithm of \citep{Davis1985}. Then the SUBFIND algorithm is used to define substructure candidates by identifying over-dense regions within the FoF halo \citep{springel2001}. Descendant subhaloes and hence galaxies are identified using a D-Trees algorithm \citep{Jiang2014,Qu2017} which traces subhaloes using the $N_{link}$ most bound particles of any species, identifying the subhalo that contains the majority of these particles as a subhalo's descendant at the next output time. The above trees have been used in numerous studies \citep[e.g.][]{Lagos2018,Katsianis2019}. In this study, we use two simulations of the EAGLE suite. We use the large box reference run labelled as Ref-L0100N1504 (hereafter, L100) in order to follow the MAHs of intermediate mass haloes (with $M_0 = 0.5 \times 10^{12} - 1.0 \times 10^{14} M_{\odot}$) and the highest resolution/smaller box-size run, labelled as Ref-L0025N0752 (hereafter, L25), in order to capture the growth of low mass haloes (with $M_0 = 0.5 \times 10^{10} - 0.5 \times 10^{12} M_{\odot}$). For more information on EAGLE we refer the reader to \citet{Schaye2015} and \citet{Crain2015}. 

We trace a galaxy at $z \sim 0$ and obtain the FoF mass of its host halo. Then using the merger tree we find the main  progenitors of this galaxy at different redshifts and obtain the host halo masses of the progenitors at these different eras. We deem this procedure to be compatible with our approach of tracing the MAH via galaxy growth/SFR detailed in section \ref{sec_model}. Following \citet{Diemer2017} we restrict our analysis for haloes that have never been subhaloes. The MAHs of different haloes with similar $M_0$  can be quite different \citep{vandenBosch2002,Mutch2013}. Thus, in order to quantify the average growth at similar masses we follow a procedure similar to \citet{Wechsler2002} and bin all halo masses among different merger trees with similar $M_0$ and then calculate the average value of all the halo masses at a given output redshift/time. The results from EAGLE are represented by the black filled circles in the right panels of Fig. \ref{figEpicG5} (Linear Mass - Age scale) and Fig. \ref{figEpicG7} ($\log_{10}$ Mass - $\log_{10}$(1+z) scale).  The cyan solid line describes our fit using equation \ref{eq:MAHmodel} while the blue solid line employs equation \ref{eq:MAHmodel2}. We see that once again the $\Gamma$ forms inspired from observations \citep{Katsianis2021b} can also broadly describe the simulated MAHs of EAGLE. The parameters can also been found at table \ref{tab:ParamsSims}.

\end{itemize}

\section{The physical interpretations of the four parameters of the MAH and CSFRD model}
\label{sec_param}

The objective of this section  is to study the dependence of the $f_{0}$ (\ref{final}),  $\beta_h$ (\ref{beta_h}), $\alpha_h$ (\ref{alpha_h}) and $T_{h}$ (\ref{TGr}) parameters on $M_{0}$ among the different models and simulations described in Fig. \ref{figEpicG2}, Fig. \ref{figEpicG4}, Fig. \ref{figEpicG5} and Fig. \ref{figEpicG7}. This will allow us to broadly describe the MAHs of haloes with different $M_{0}$ and gain further insight on the relation between CSFRD and the average MAH of haloes.

\subsection{The $f_{0}$ parameter}
\label{final}

\begin{figure*}
\centering
\includegraphics[scale=0.5]{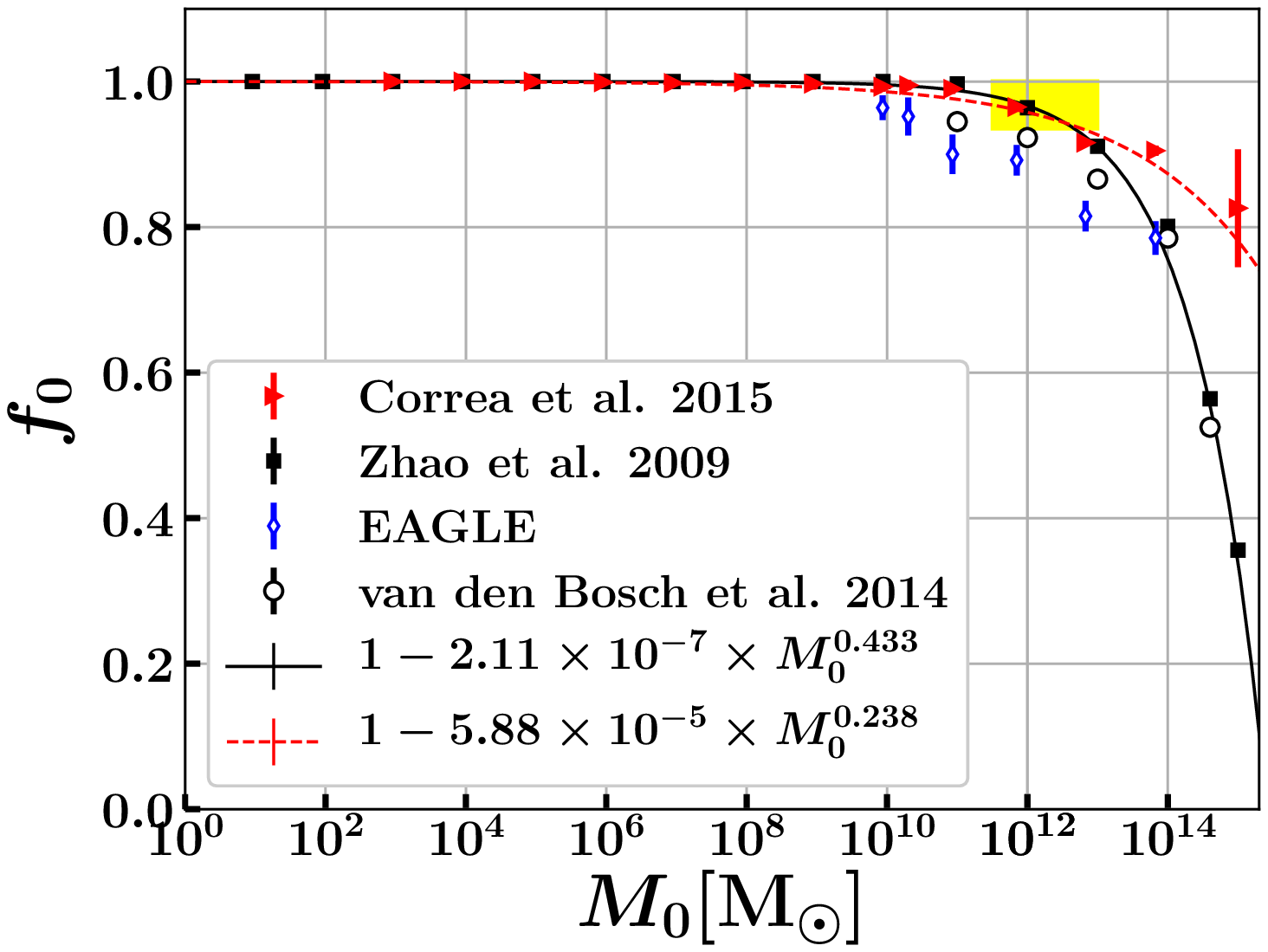}
\includegraphics[scale=0.5]{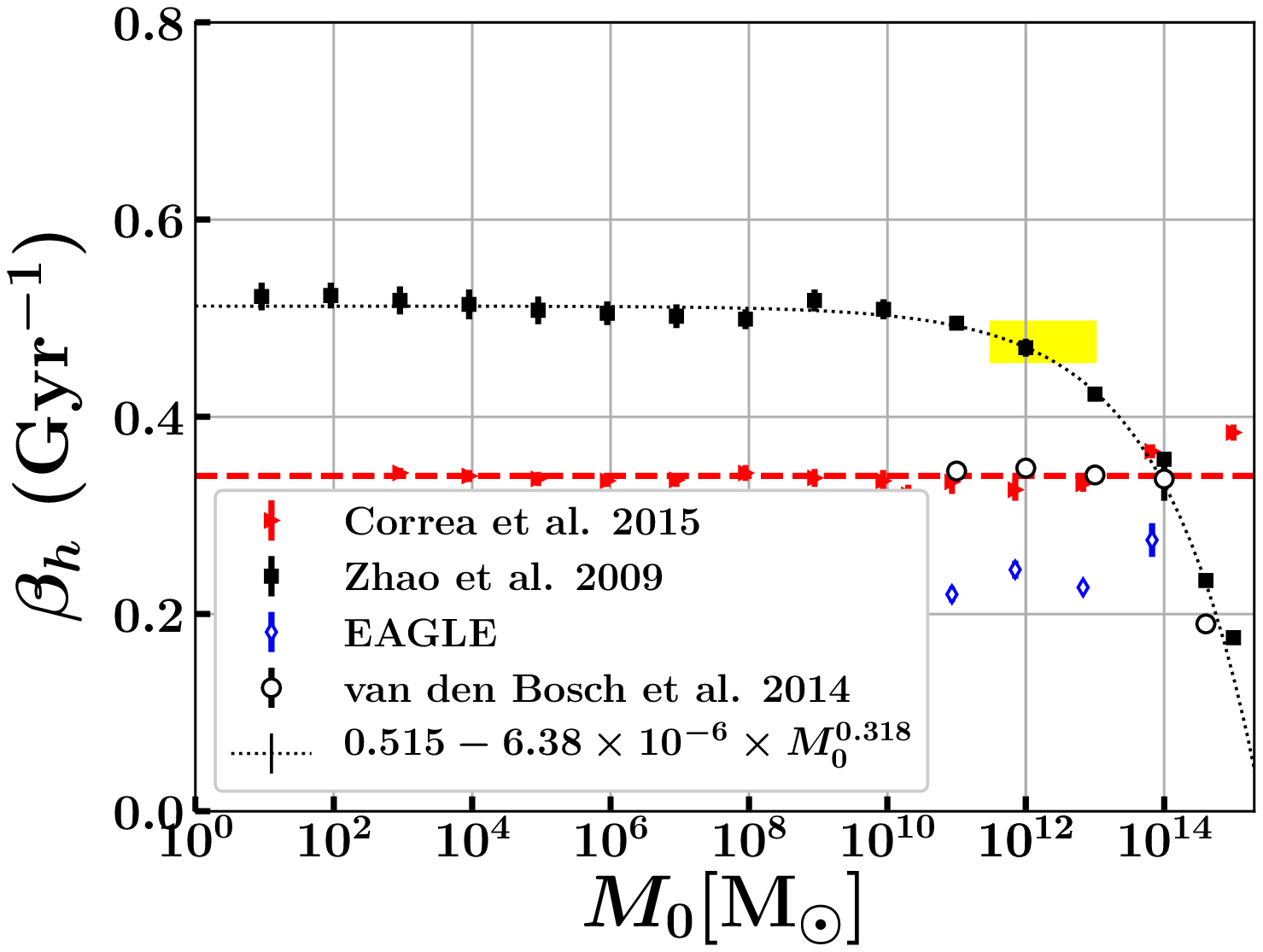}
\includegraphics[scale=0.5]{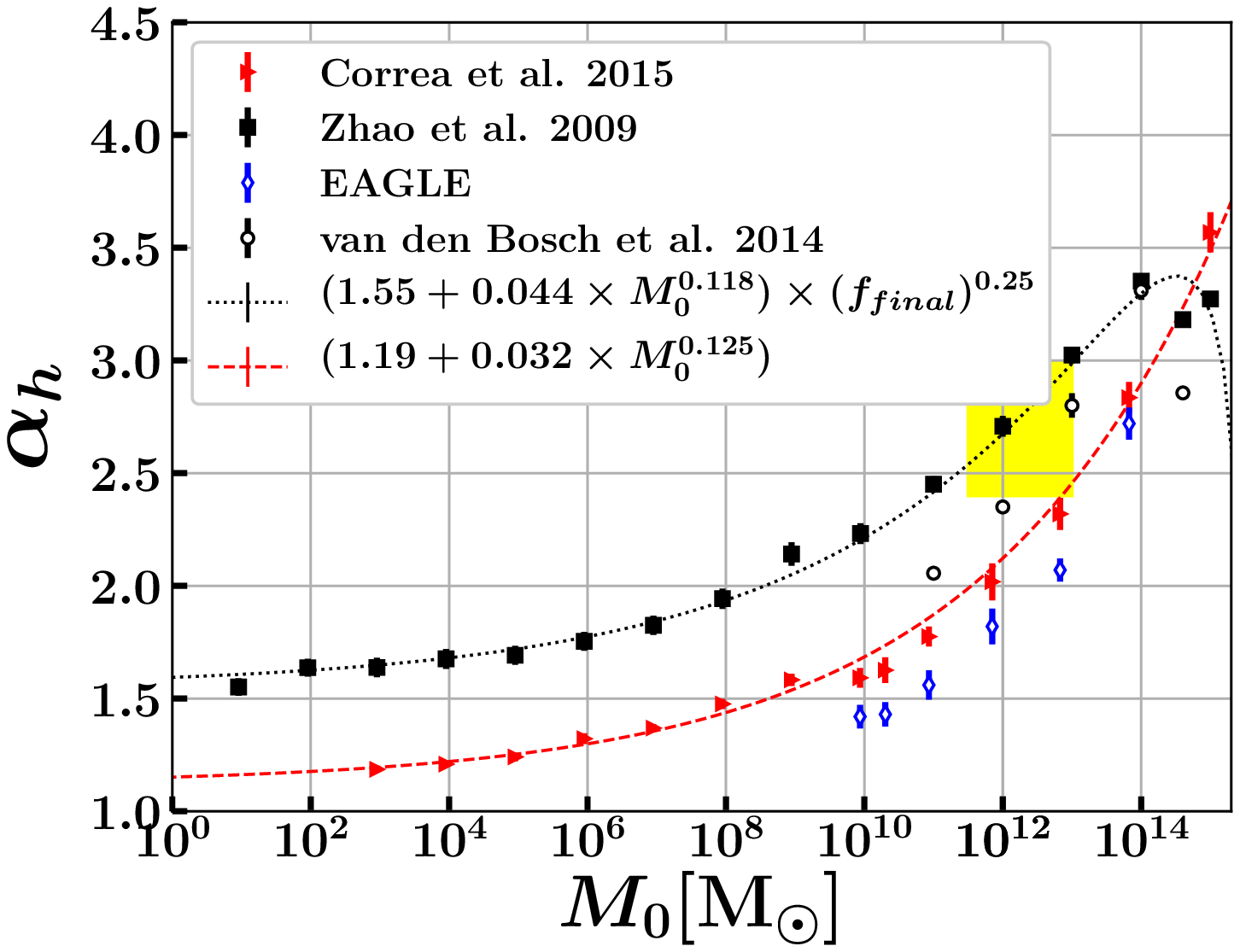}
\includegraphics[scale=0.5]{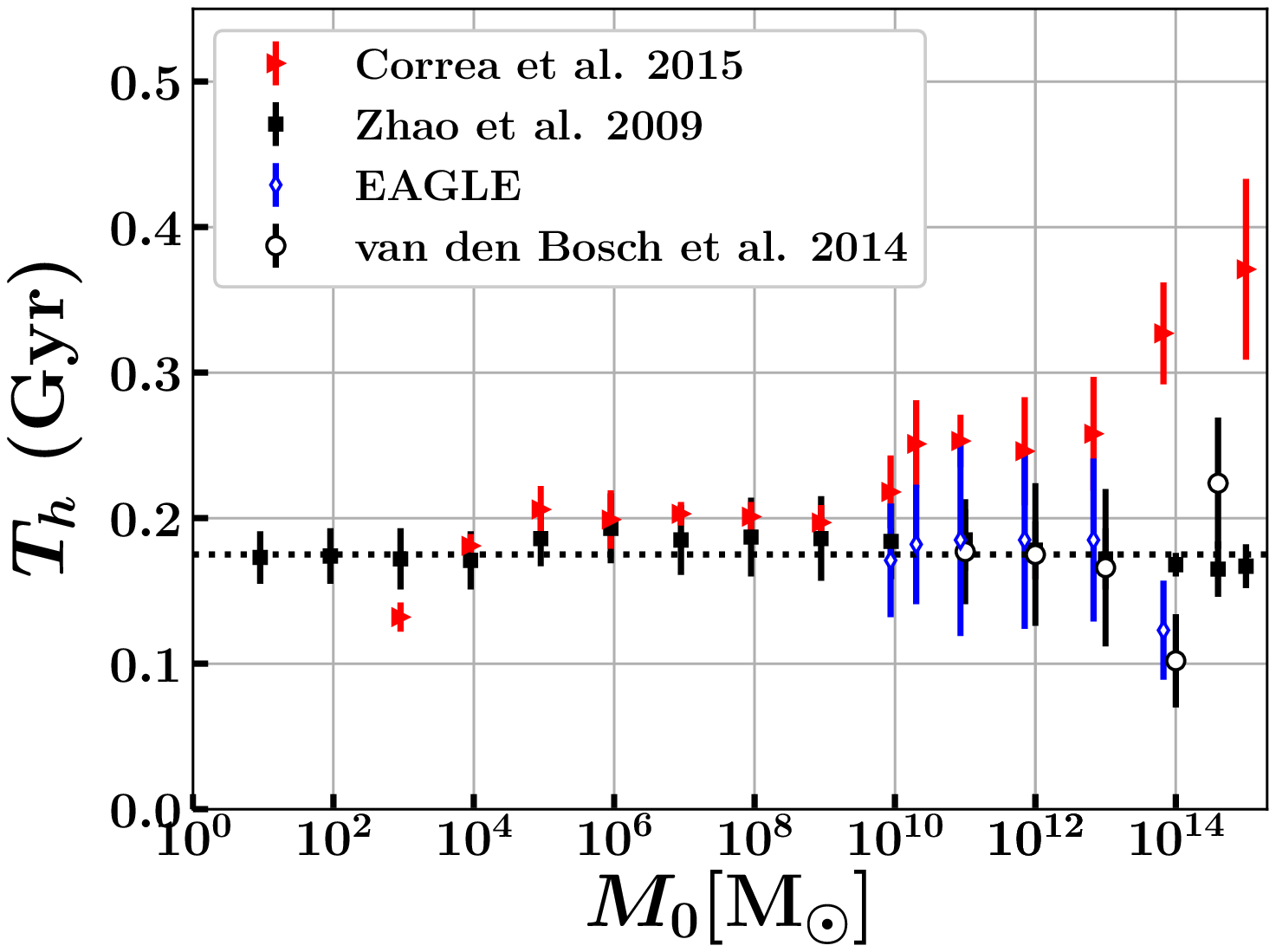}
\vspace{-0.30cm}
\vspace{-0.30cm}
\caption{Top left panel: The $f_{0}$ parameter at the mass range $M(0) = 10^{1}-10^{15} M_{\odot}$ found for the EAGLE (blue open diamonds)/Bolshoi (black open circles) simulations and the \citet{Zhao2009}/\citet{Correa2015b} models black dotted line/red dashed line. Top right panel: The $\beta_h$ parameter. Left bottom panel: The $\alpha_h$ parameter. Right bottom panel: $T_{h}$ parameter. The yellow area represents the parameters obtained for the $10^{11.5}-10^{13} \, M_{\odot}$ from the observations of the CSFRD. For the case of $f_{0}$ the CSFRD would suggest values of $\sim $ 1, the $\beta_h$ parameter between 0.44 Gyr$^{-1}$ (IR) to 0.5 Gyr$^{-1}$ (UV) while the $alpha_h$ parameter between 2.44 (IR) to 2.9 (UV).}
\label{figEpicG454}
\end{figure*}

We remind that the $f_{0}$ parameter describes how much mass a halo has attained at present time ($z = 0$) with respect the final mass at $\infty$ ($f_{0} = M_{0}/M_{final}$). We note this parameter does not have units. In the top left panel of Fig. \ref{figEpicG454} we plot the dependence of the $f_{0}$ parameter on $M_0$ within the range of $M_0 = 10^{1} -  10^{15} M_{\odot}$. It is clear that all models and simulations agree qualitative on the fact that more massive haloes assemble their mass later, consistent with hierarchical structure formation and with numerous previous findings \citep{Fakhouri2010,Yang2011,Van2014}. Via our parameterization and the $f_{0}$ parameter we can quantify this effect.

For the \citet{Zhao2009} model (black dotted line) we find:

\begin{eqnarray}
\label{ffinal}
f_{0} = 1-2.11 \times 10^{-7} \times M_{0}^{0.4323 \pm 0.007}.
\end{eqnarray}

For the \citet{Correa2015b} model (red dashed line) we find:

\begin{eqnarray}
\label{ffinal2}
f_{0} = 1-5.88 \times 10^{-5} \times M_{0}^{0.2378 \pm 0.031}.
\end{eqnarray}
  
Both behaviors can be parameterized via a power-law, $f_{0} = 1- c \times M_{0}^{d}$ and agree that a) on {\it average} low mass haloes between $M_0 = 10^{1} -  10^{10} M_{\odot}$ have attained most of their final mass ($f_{0} = 1$) already (i.e. they have entered their stationary phase in terms of our $\Gamma$ formalism, appendix \ref{GAMMAMotivation}). b) haloes between $M_0 = 10^{10} - 10^{13} M_{\odot}$ have attained on average more than $95\%$ of their final mass (i.e., they have entered the deceleration phase and soon will stop growing according to our $\Gamma$ form). This is the reason why the parameterization of \citet{Katsianis2021b} for the CSFRD - Average MAH assumed $f_{0} = 1$ (subsection \ref{averages}) and the underlying reason why the star formation of the Universe, which is dominated mostly from haloes of $M_0 = 10^{11.5} - 10^{13} M_{\odot}$, has already entered a deceleration-stationary phase. 

Besides the qualitative agreement between equations \ref{ffinal} and \ref{ffinal2} it is important to note that the $f_{0}$ parameter of high $M_{0}$ haloes of the \citet{Zhao2009} model is significantly lower with respect to the one from \citet{Correa2015b} (for example a halo with $M_{0} = 10^{15} \, M_{\odot}$ has attained only $40 \, \%$ of its final mass in the \citet{Zhao2009} model but already $80 \, \%$ in the \citet{Correa2015b} model). Our observations for the CSFRD \citep{Katsianis2021b} and average MAH (subsection \ref{averages}), which both agree on a $<f_{0}>$ value of $\sim 1.0$ (yellow shaded area), cannot disentangle which or our descriptions represents better reality (as both models broadly agree that $f_{0} > 0.95$ for $M_0 = 10^{11.5}-10^{13} M_{\odot}$). However, we have to note that both the Bolshoi simulations (open black circles) and the EAGLE simulations (open blue diamonds) are closer to the $f_{0}$ results obtained from the \citet{Zhao2009} model. In a future study we will look into the larger scale mass reservoir which can drive the further growth of these massive haloes, in the framework of halo boundary and depletion radius recently proposed in \citet{Fong2021}.

\subsection{The $\beta_h$ parameter}
\label{beta_h}

We remind that the parameter $\beta_{h}$ (which has units of inverse time) determines the rate ($e^{-\beta_{h} \, \times T}$) at which halo growth decelerates and describes the mass dependent depletion of the available resources for halo growth  due to the current halo growth (more information can be found in appendix \ref{GAMMAMotivation} for the further physical interpretation of the form). At the same time its inverse $\tau_h = 1/\beta_{h}$ gives us the timescale at which this process occurs. In the top right panel of Fig. \ref{figEpicG454} we plot the dependence of the $\beta_{h}$ parameter on $M_0$ within the range of $M_0 = 10^{1} -  10^{15} M_{\odot}$. For the \citet{Zhao2009} model (black dotted line) we find :

\begin{eqnarray}
\label{betahh}
\beta_h(M_0) = 0.515-6.38 \times 10^{-6} \,  M_{0}^{0.318 \pm 0.013} Gyr^{-1}.
\end{eqnarray}
The value of $0.515$ $Gyr^{-1}$  represents the maximum $\beta_h$, which is achieved from low mass haloes between $M_0 = 10^{1}-10^{11.5} M_{\odot}$. The cosmic CSFRD gives us broadly the parameters of haloes between $M_0 = 10^{11.5}-10^{13} M_{\odot}$ $Gyr^{-1}$, since these dominate the CSFRD \citep{Katsianis2017}\footnote{This halo range corresponds to haloes that contribute most of the star formation in the Universe because of the combination of three factors: gas accretion efficiency, star formation efficiency, and halo abundance. This is consistent with the conditional luminosity function model prediction given in \citet{Yang2003}, where they found that the galaxies in the Milky Way sized haloes are the most efficient producers of blue light in the local Universe. It is also consistent with the SFRD model prediction given in \citet{Yang2013}, where they found that haloes with mass $M_0 \sim 10^{12}h^{-1}M_{\odot}$ at redshift $z\sim 1$ provides the most contributions of the SFRD in the Universe (more information in their Fig. 8).}, and suggest values of $<\beta_{h}> = <\beta_{\star}> = 0.44-0.5$ Gyr$^{-1}$ (section \ref{sec_model}) represented by the yellow shaded area \footnote{The yellow area is obtained by the UV and IR observations of \citep{Katsianis2021b}. The  $<\beta_{\star}>$ is 0.44 Gyr$^{-1}$ according to IR data and 0.5 Gyr$^{-1}$ according to UV data, (section \ref{sec_model}).}. This is indeed reflected in Fig. \ref{figEpicG454} for the \citet{Zhao2009} model which has values between 0.4 to 0.5 Gyr$^{-1}$ at this range.  The timescale of the depletion can be written as $<\tau_h> = 1/<\beta_{h}> = 1/0.5 - 1/0.4 = 2 - 2.5$ Gyr.

It is a common practice in the literature to express a timescale of a process with respect the dynamical time (${t}_{\mathrm{dyn}}$\footnote{The ${t}_{\mathrm{dyn}}$ is defined as the ratio of a characteristic size and a characteristic velocity. It is a useful timescale when studying MAHs as it represents the time it takes a particle to reach the apocenter of its first orbit after infall \citet{Diemer2017}. The dynamical time is also used as the finite time it takes for gas to settle into the central galaxy after being accreted onto the halo \citep{Robertson2005}.}) in order to connect it with other galaxy/halo properties and better grasp what aspects are affecting it. For example, \citet{Davies2016} pointed out that the alignment process in galaxies typically has a timescale of $<\tau_{aligment}> = 1/\epsilon \times {t}_{\mathrm{dyn}}$ where $\epsilon$ is the ellipticity of the potential. We remind that the dynamical time can be written as \citep{Robertson2005,Solanes2018}:
\begin{eqnarray}
\label{Dynamical}
{t}_{\mathrm{dyn}} (z) = \frac{R_{Halo}}{V_{Circ}} = 2  \times (1+z)^{-3/2} \, {\rm Gyr},
\end{eqnarray}
and thus from equation \ref{Dynamical} the {\it dynamical timescale at z = 0 is ${t}_{\mathrm{dyn}} (0) = 2$ Gyr}. We follow this line of logic and write our average/cosmic depletion timescale $<\tau_h> = 2$ Gyr from our observations and the \citet{Zhao2009} model with respect to the ${t}_{\mathrm{dyn}}$. The connection is simple as follows :
\begin{eqnarray}
\label{DynamicalHAlogrowth}
<\tau_h> = 1/<\beta_{h}> = 2 \, {\rm Gyr} = {t}_{\mathrm{dyn}} (0).
\end{eqnarray}
We see that the average depletion timescale $<\tau_h>$ coincides with the dynamical timescale {\it at z = 0}. A question that arises is why the depletion timescale is constant/independent of time and equal to the ${t}_{\mathrm{dyn}} (0) = 2$ Gyr value and not to the ${t}_{\mathrm{dyn}} (z)$ (i.e. how the redshift scaling of equation \ref{Dynamical} was removed)?


\citet{Genzel2015} argued that the {\it molecular gas depletion time} also scales very weakly with cosmic epoch ($t_{depl, Mol} (z) \propto (1+z)^{-0.34}$) besides the fact that a $(1+z)^{-3/2}$ factor would be expected due to the ${t}_{\mathrm{dyn}}(z)$ scaling (equation \ref{Dynamical}). The authors suggested two additional corrections that could leave the molecular parameter almost redshift independent and remove the $(1+z)^{-3/2}$ dependence. 1) the accelerating expansion of the Universe. The Hubble parameter in a $\Lambda$CDM universe changes more slowly at late times and approximates to $H(z) \propto (1+z)^{Co}$ , the average $Co$ for the redshift range z = 0-2.5 is -0.98. The above counteracts the $(1+z)^{-3/2}$ dependence.  2) depletion timescales are also dependent on the concentration parameter which is smaller at high-z than at z = 0. Taken together, these two effects removed the $(1+z)^{-3/2}$ dependence introduced by the dynamical scale for the molecular depletion timescale $t_{depl, Mol}$ resulting in a redshift independent $t_{depl, Mol}$ parameter \citep{Genzel2015}. It seems that the above apply as well to the average depletion timescale $<\tau_h>$ discussed in our work which is also time independent. 

Thus, the depletion timescale of  $<\tau_h> = 1/<\beta_{h}> = 2$ Gyr coincides with other {\it important} characteristic timescales of halo growth (at z = 0) and galaxy formation. Some of those are the Dynamical Friction\footnote{Dynamical friction (DF) time is the interval between entry and merger. DF causes a satellite to lose energy to the background dark matter halo and eventually causes the satellite to sink towards the centre and merge with the central galaxy of the host halo.} \citep{Binney1987,Mo2010}, star formation history \citep{Katsianis2021b}, free-fall of the host halo \citep{Ogiya2019}, and dynamical timescales \citep{Diemer2020,Lsmith2022} have all been found to be $\sim 2$ Gyr.



Going beyond the average $<\beta_{h}>$ and focusing on haloes with different $M_{0}$ we see a power-law relation between $\beta_{h}$ and $M_{0}$ which has an exponent of 0.318 $\pm 0.013$ for the \citet{Zhao2009} model (equation \ref{betahh}). This is very close to 1/3 so the $\beta_h(M_0)$ parameter appears to be related to the  radius of the halo. But should there be a relation of $\beta_{h}$ with $R$/$M_{0}^{1/3}$ in Eq. \ref{betahh}? According to \citet{Zhao2003a}, for any halo and cosmology \citet{Zhao2009}, regardless of $M_{0}$ {\it and regardless if there is an accelerating expansion like in $\Lambda$CDM}, the growth history of haloes is such that the early phase is characterized by rapid halo growth dominated by major mergers, which reconfigure and deepen the gravitational potential wells, forming the core structure. The late phase is characterized by much slower growth predominantly driven by accretion of material onto the {\it halo outskirts, R,} with the inner structure and potential being unnaffected. Thus, for the case of the \citet{Zhao2009} model a radius/Mass dependence is not a surprise as after the cores have been formed there will be growth mostly driven at the outer regions. The larger the radius (further from the core) - the slower the resources for growth are consumed. We find that indeed with increasing radius/$M_{0}$ we have a decreasing $\beta_h$ parameter following a power-law relation.

We have to note that we do not find the above mass dependence of the $\beta_h(M_0)$ to $M_{0}$ for the model of \citet{Correa2015b} which seems to imply a mass independent $\beta_h = 0.335$ $Gyr^{-1}$. We stress that the results we obtained from the model of \citet{Correa2015b} are in better agreement both with the EAGLE (blue open diamonds of the top left and middle panels of Fig. \ref{figEpicG454} ) and the Bolshoi (Open black circles of top left and middle panels of Fig. \ref{figEpicG454})  which suggest on average a 30$\%$ lower value for $\beta_h$ obtained from the \citet{Zhao2009} and our average MAH derived from the CSFRD (subsection \ref{averages}). Nevertheless, values of 0.35 to 0.5 Gyr$^{-1}$ should be expected for the $<\beta_h>$ and  $<\beta_\star>$ according to our analysis.

\subsection{The $\alpha_h$ parameter}
\label{alpha_h}

In the bottom left panel of Fig. \ref{figEpicG454} we see the dependence of the $\alpha_h$ parameter with $M_0$ within the range $M_0 = 10^{1}-10^{15} M_{\odot}$. For the case of \citet{Correa2015b} we see a power-law relation of $\alpha_h$ with $M_0$ occurring as follows:
\begin{eqnarray}
\label{alphah}
{\alpha_h(M_0) = 1.19 + 0.032 \times M_0^{0.125 \pm 0.017}}.
\end{eqnarray}
This means that small changes in $\alpha_h$ increase significantly $M_{0}$ (i.e. for $\alpha_h = 1.19$ we have $M_{0} =  1 M_{\odot}$ while for $\alpha_h = 3.2$ we have $M_{0} = 10^{14} M_{\odot}$). This power-law relation between $\alpha_{h}$ and $M_{0}$ is also found for the EAGLE simulations. The \citet{Zhao2009} model and the Bolshoi simulations have also a similar power-law behavior with similar expoenent up to $10^{14} M_{\odot}$ with a turn-over at higher masses described by the $f_{final}$ parameter\footnote{The turn-over of the relation obtained from the \citet{Zhao2009} model can be described intuitively using $f_{0}$ as: ${\alpha_h(M_0) = (1.55 + 0.044 \times M_0^{0.118}) \times f_0^{0.25}}$.}
.

We remind that the $\alpha_h$ parameter is the exponent ($T^{\alpha_h-1}$) in our $\Gamma$ formalism, which describes the early accretion of matter/smaller haloes into the halo boundary. An $\alpha_h = 1$ value implies a growth without major mergers or steps \citep{Katsianis2021b}, but a simple exponential settle down rate of the existing matter of the region. In this scenario as outlined in \citep{Katsianis2021b}, the available matter $M_{final}$ of an isolated system is consumed and just collapses to form a structure. A scenario with $\alpha_h > 1$ instead involves a case in which the system accretes new matter to the halo regime and further merging occurs at different time-steps, especially at early times. This results in the total available matter to form a dark matter halo with a mass rate rising by a factor of $T^{\alpha_h-1}$.

There is a complementary way to see $\alpha_h$ and the power-law growth. A power-law growth behavior could imply also a fractal growth (also see appendix \ref{GAMMAMotivation}) which is the result from a power-law distribution, a self-similar scale free structure or a process that occurs following an exponential distribution  occurrence \citep{Askar2019,Mori2020,ZiffandZiff}. An example is the case of the transmission of COVID-2019 which rate of daily new cases as a function of time, $I(t)$, has also a $\Gamma$ form which is described by a power-law behavior at early times combined by an exponential decline at later times \citep[$I(t) = K \times t^{D-1} \times e^{-t/\tau_0}$]{ZiffandZiff,Vazquez2020}, like Eq. \ref{eq:PGgastconsumption00} which describes halo growth rate. The exponential decline is caused by the fact that people are getting more and more vaccinated, are naturally immunized and they follow social distancing the more the total number of cases rise. On the above example which is caused by a scale-free network the power-law exponent D of $t^{D-1}$ represents the network diameter, or simply put the number of generations a disease has been transmitted from patient zero up to the implementation of the lockdown-social distancing \citep{Medo2020,Vazquez2020}. Thus, our $\alpha_h$ parameter for the growth of haloes can be interpreted similarly as the number of mergers/steps occurring up to the time that the expansion starts to decrease the matter density in the Universe and less halo growth occurs.

The $\alpha_{\star}=2.4-2.9$ (yellow shaded area, IR data suggest 2.4, while UV data suggest 2.9) value obtained from the CSFRD given in \citet{Katsianis2021b} and subsection \ref{averages} is close to the model of \citet{Zhao2009} and the Bolshoi simulations \citep{Van2014}. The \citet{Correa2015b} model is close to the results from the EAGLE simulations and the values of the IR data ($\alpha_{\star}=2.4$).

\subsection{The $T_{h}$ parameter}
\label{TGr}

Our model for MAH is tightly related to the SFHs of galaxies, which galaxies start growing after 150-300 million years after the Big Bang. The growth discussed in our work occurs with a delay $T_{h}$ which is the time interval between the Big Bang and the time the halo embryo is formed and is able to host star formation. We note that this is different than the halo formation time defined usually as $M_{0}/2$. In the bottom panel of Fig. \ref{figEpicG454} we present the $T_h$ parameter with respect $M_{\odot}$ and indeed we find that $T_{h}$ is $\sim 170 $ million years ($0.17$ Gyr) for most haloes in \citet{Zhao2009} model, while there is a $M_{0}$ dependence for our model which relies on \citet{Correa2015b} which suggest values between 100-360 Million years.

A value of 150-360 million years for $T_{h}$ would be appropriate indeed for $\Lambda$CDM. According to numerical simulations, dark matter haloes become sufficiently massive enough to induce star formation at 150-250 million years after the Big Bang, corresponding to the redshift range $z \sim 15-20$ \citep{Villanueva2018,Laporte2021}. From observations, deep imaging with the Hubble and Spitzer Space Telescopes have revealed that when the Universe was 600 million years old, there were already main sequence stars older than 250 million years, which would imply galaxy formation originating before a redshift $z > 14$ and point the dating birth of galaxies at 150-300 Million years after Big Bang \citep{zheng2012,Oesch2018}.

Our $T_{h}$ parameter is in line with the above narrative from both simulations and observational studies. However, we note that in the near future we may be able to detect a large number of high star forming galaxies via the James Webb Space Telescope and this may challenge $\Lambda$CDM and the current standard model of galaxy formation, while our $\Gamma$ formalism will just have to adopt a smaller $T_h$ parameter. Recently there have been unexpected active galaxies that are detected beyond $z > 13$ \citep{Pacucci2022}. For example, HD1 \citep{2022ApJ...929....1H} is a very bright and high star forming, suggesting that active galaxies already existed in the Universe only 300 million years after the Big Bang and galaxy formation occurs much earlier. HD1 is hardly explained with current theoretical models of galaxy formation which would suggest a system with much less star formation.

\section{Conclusions}
\label{sec_concl}

The cosmic star formation rate density (CSFRD) of galaxies and the mass accretion histories of dark matter haloes are two topics of great interest. It is a common practice to build galaxy formation models, which incorporates theoretical (relying on the PS formalism) or simulated (relying on N-body simulations) MAHs, and then follow the star formation prescriptions in dark matter haloes. So far the empirical MAH models are typically modelled as a function of redshift $M_h(z)$.  Motivated by the recent Gamma functional form CSFRD model obtained by \citet{Katsianis2021b}, which is better associated with star formation physics (i.e., gas accretion and star formation time scales), we set out to investigate a MAH model as a function of cosmic time, $M_h(T)$. The consistent formalism of MAHs and SFHs will allow us to better understand the connections between halo mass growth and galaxy growth in dark matter haloes as we explore the parameters needed to describe the above. Our results can be summarized as follows: 

\begin{itemize}
  
\item \citet{Katsianis2021b} demonstrated that the {\it observed} CSFRD has an evolution which resembles a ${\Gamma}$ distribution. In that work we find that the total SFH of galaxies in dark matter haloes can also be described by such a functional form while $\alpha_{\star}$ and $\beta_{\star}$ can be expressed as a function of halo mass today, $M_{0}$. We can obtain the following form for the MAHs of dark matter haloes: $M_h(T) = \frac{M_0}{f_{0}} \, \times \frac{\gamma(\alpha_h, \beta_h \times (T - T_h)}{\Gamma(\alpha_h)}$, where $f_{0}$ the percentage of mass that the halo has attained at present day with respect its final value, $\alpha_h$/$\beta_h$ parameters described at the 4th and 5th bullet points of this conculsion section that are determined from N-body merger trees and other analytical models, $T$ is the comic time and $T_h$ is the time interval the halo embryo started growing. Adopting the parameters obtained from the observed CSFRD ($\alpha_h = 2.9$, $\beta_h = 0.5$, $f_{0} = 1$), we compare our model predictions of the average/{\it cosmic} MAHs with those derived by other authors who employed N-body simulations and different parameterizations \citep[][]{Zhao2009,McBride2009,Van2014,Correa2015b} to find good agreements with them (section \ref{sec_model}, Fig. \ref{figEpicG1}.).

\item We employ the above formalism to describe the MAHs with different $M_{0}$ derived from cosmological hydro-dynamical simulations (EAGLE), N-body simulations (Bolshoi) and analytical models \citep{Zhao2009,Correa2015b}. Our parameterization can nicely describe the MAHs of these haloes (section \ref{sec_data}, Figs. \ref{figEpicG2},  \ref{figEpicG4}, \ref{figEpicG5} and \ref{figEpicG7}) for the redshift range of interest ($z = 0-9$) and gives us the opportunity to describe the parameters of the CSFRD to halo properties.

\item $f_{0}$ is $M_{0}$ dependent and can be parameterized via a power-law, $f_{0} = 1- c \times M_{0}^{d}$. Models and simulations agree that on average low mass haloes attain their final masses faster than high mass haloes. Between $M_0 = 10^{1} -  10^{10} M_{\odot}$ on average they have attained all of their final mass ($f_{0} \approx 1$) already. Objects between $M_0 = 10^{10} -  10^{13} M_{\odot}$ have attained, on average, more than 95$\%$ of their final mass. The Average/cosmic MAH has $<f_{0}> = 0.95$ (subsection \ref{averages}) and this is the underlying reason why the star formation of the Universe, which is dominated mostly from haloes of $M_0 = 10^{11.5} - 10^{13} M_{\odot}$, has already entered a deceleration-stationary phase (subsection \ref{final} and top right panel of Fig. \ref{figEpicG454}).  

\item the parameter $\beta_{h}$ determines the rate at which halo growth decelerates and describes the mass dependent decrement of the available resources for halo growth ($e^{-\beta_{h} \, \times T}$) due to the depletion of the available resources that are accreted to the halo and the expansion of the Universe that isolates haloes decreasing their merging. At the same time its inverse, $\tau_h = 1/\beta_{h}$, gives us the timescale at which this depletion occurs (Fig. \ref{figEpicG454}). \citet{Katsianis2021b} demonstrated that the average/cosmic star formation timescale $<\beta_{\star}>$ is 0.44 to 0.5 Gyr$^{-1}$. The equivalent parameter for halo growth in our current work among different simulations and models is found to be similar, 0.3 Gyr$^{-1}$ to 0.52 Gyr$^{-1}$. We note that the timescale of  $<\tau_h> = 1/<\beta_{h}> = 2$ Gyr coincides with other {\it important} characteristic timescales of halo growth at $z = 0$, like the dynamical timescale. A difference found among models is if the $\beta_{h}$ parameter is $M_{0}$ dependent or not. For example, \citet{Zhao2009} implies to a radius/$M_{0}$ dependence of $\times M_{0}^{0.311}$ while the $\beta_{h}$ parameter derived from the \citet{Correa2015b} model is $M_{0}$ independent (subsection \ref{beta_h} and top right top panel of Fig. \ref{figEpicG454}).

\item The dependence of the $\alpha_h$ parameter with $M_0$ within the range of $M_0 = 10^{1}-10^{14} M_{\odot}$ is a power-law relation with the \citet{Zhao2009} model and the Bolshoi simulations having a turn-over at higer masses described by the $f_{0}$ parameter. The $\alpha_{\star}=2.4-2.9$ (yellow shaded area Fig. \ref{figEpicG454}) value obtained from the CSFRD given in \citet{Katsianis2021b} and subsection \ref{averages} is consistent with models especially with that of \citet{Zhao2009} and the Bolshoi simulations. The exponent $\alpha_h$ can be seen as the number of mergers/steps occurring up to the time that the expansion slows the growth of haloes and less mergers occur. This interpretation is consistent with other fields of science that a $\Gamma$ Growth happens including infectious deseases (like Covid-19) where the exponent parameter represents the number of generations that occurs up to the lock down-social distancing time \citep{Vazquez2020,ZiffandZiff} (subsection \ref{alpha_h} and bottom left panel of Fig. \ref{figEpicG454}).

\item Galaxies within the $\Lambda$CDM cosmology start growing after 150-300 million years after the Big Bang. The $T_{h}$ parameter in our formalism is the time interval between the Big Bang and the time the halo embryo was assembled and ready to host star formation. After this time halo and galaxy start growing together following a $\Gamma$ growth. We find that $T_{h}$ is $\sim 170 $ million years for most haloes in \citet{Zhao2009} model and the Bolshoi/EAGLE simulations, while there is a $M_{0}$ dependence when the model of \citet{Correa2015b} is employed which suggest values between 100-360 Million years (subsection \ref{TGr} and bottom right panel of Fig. \ref{figEpicG454}).
  
\end{itemize}

It is important to point out that currently we only tuned our model within a $\Lambda$CDM framework. It would be interesting to investigate our model within other dark matter frameworks, too. If there are any prominent differences, we can then devise some observables to probe the nature of dark matter or beyond. We were able to connect all the parameters required to describe the CSFRD and SFRs of galaxies from the \citet{Katsianis2021b} model to the properties of haloes (specifically $M_{0}$ and MAHs), providing a more complete picture than empirical fits of CSFRD. Interestingly we demonstrated that the $\Gamma$ growth pattern, which is one of the most typical growth patterns in nature (from infectious diseases to the growth of stars) occurs also for the growth of Dark matter haloes. This will enable us to connect easier both mathematically and intuitively the smallest and largest scales of our Cosmos.

\section*{Acknowledgments}
 
We thank the anonymous reviewer and the assistant editor for carefully reading our manuscript and providing us with useful suggestions and comments that improved significantly the quality of our work. A.K has been supported by the Shanghai Jiao Tong University, Yang-Yang award, the 100 talents program of Sun Yat-sen University and the Tsung-Dao Lee Institute Fellowship. X.Y. is supported by the national science foundation of China (grant Nos. 11833005,11890692,11621303) and Shanghai Natural Science Foundation,grant No.15ZR1446700 and 111 project No. B20019. We acknowledge the science research grants from the China Manned Space Project with NO.  CMS-CSST-2021-A02.  A.K wants to thank Joy Liu for discussions on Astrophysics and Particle physics inspired by the book `` WE HAVE NO IDEA'' written by Jorge Cham and Daniel Whiteson.

\section*{Data Availability Statement}

No new data were generated or analyzed in support of this research.


\bibliographystyle{mn2e}	
\bibliography{Katsianis_MNRAS9}

\appendix

\section{Growth from small to large scales and the motivation for a $\Gamma$ Growth model}
\label{GAMMAMotivation}

\begin{figure*}
\centering
\includegraphics[scale=0.45]{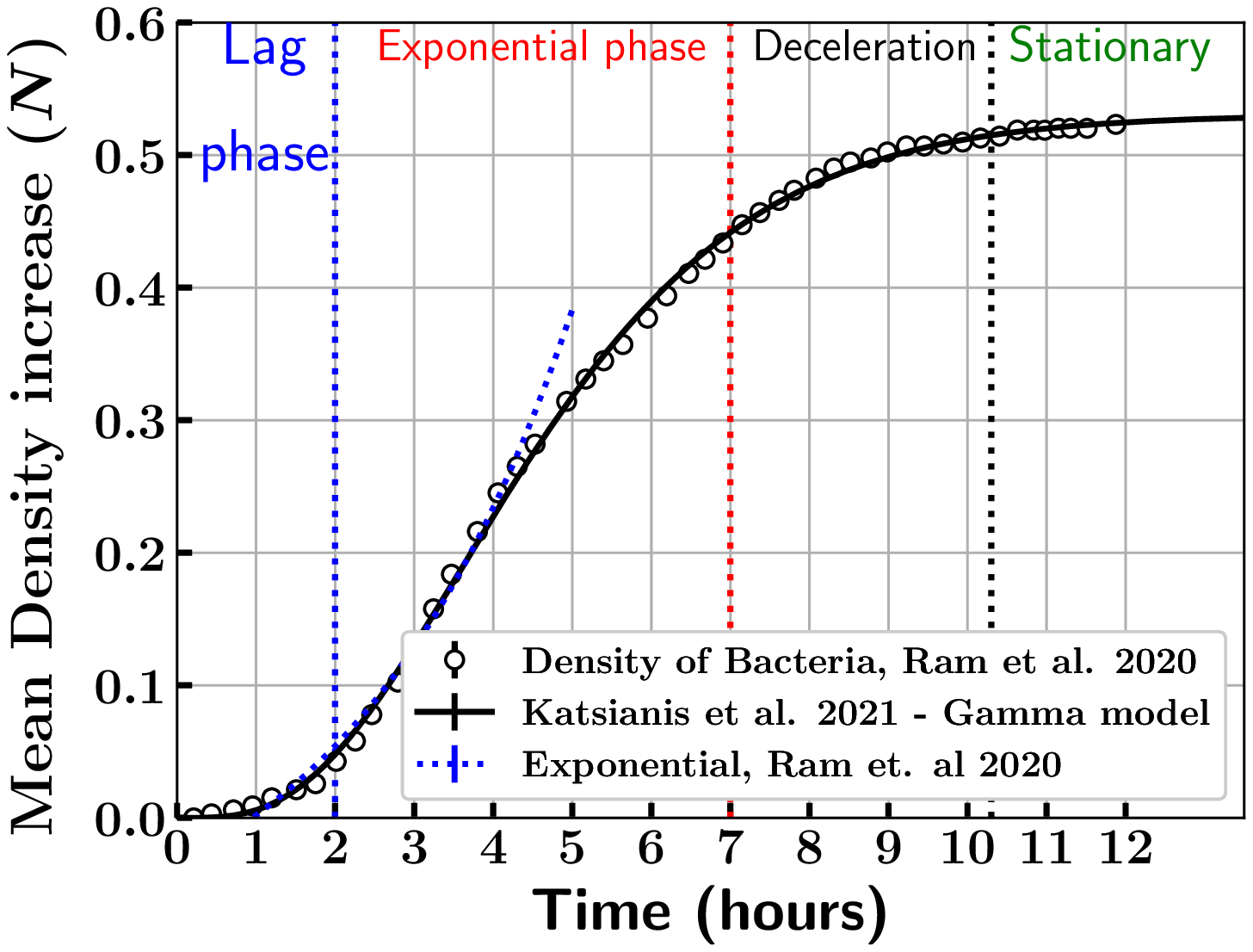}
\includegraphics[scale=0.45]{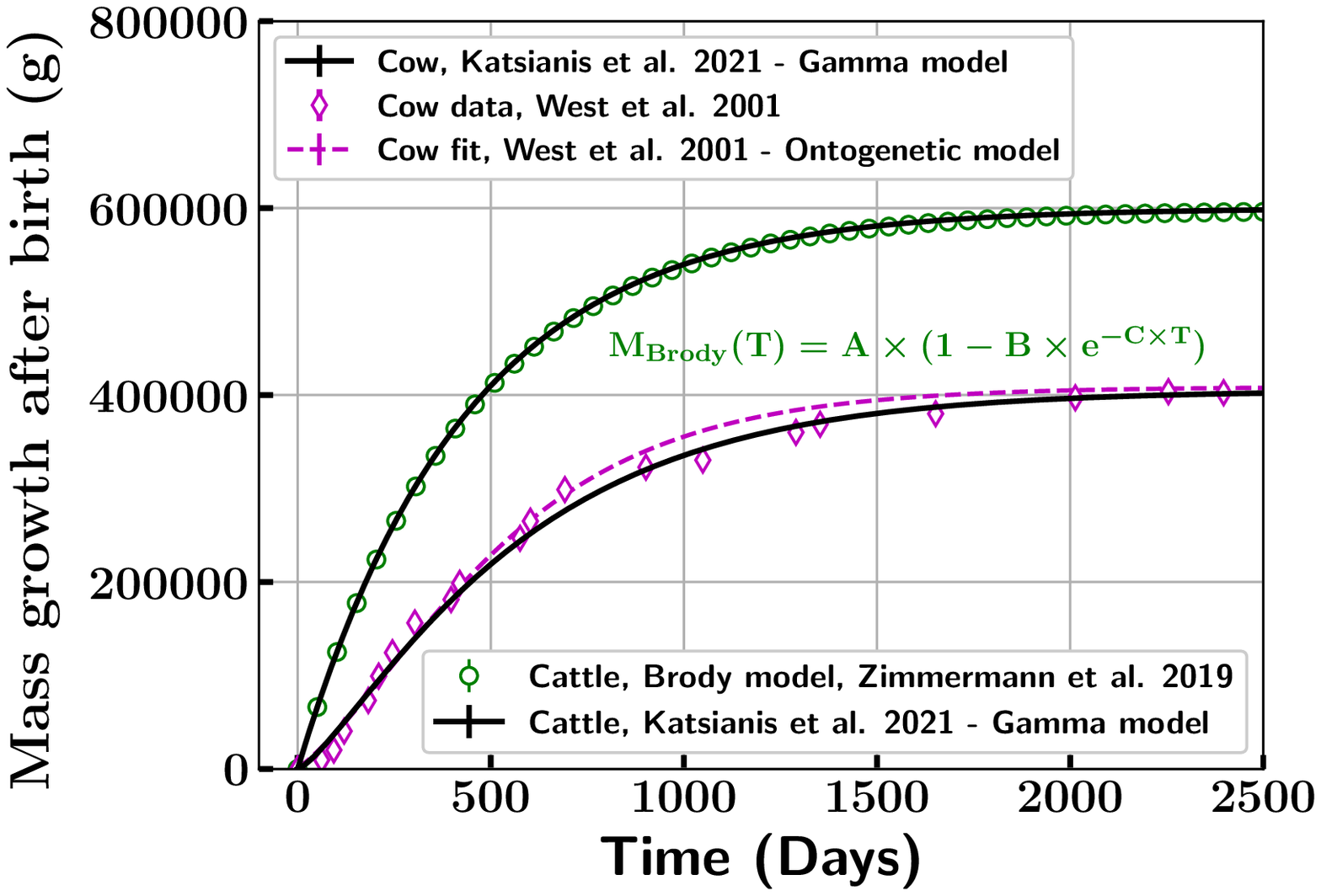}
\includegraphics[scale=0.45]{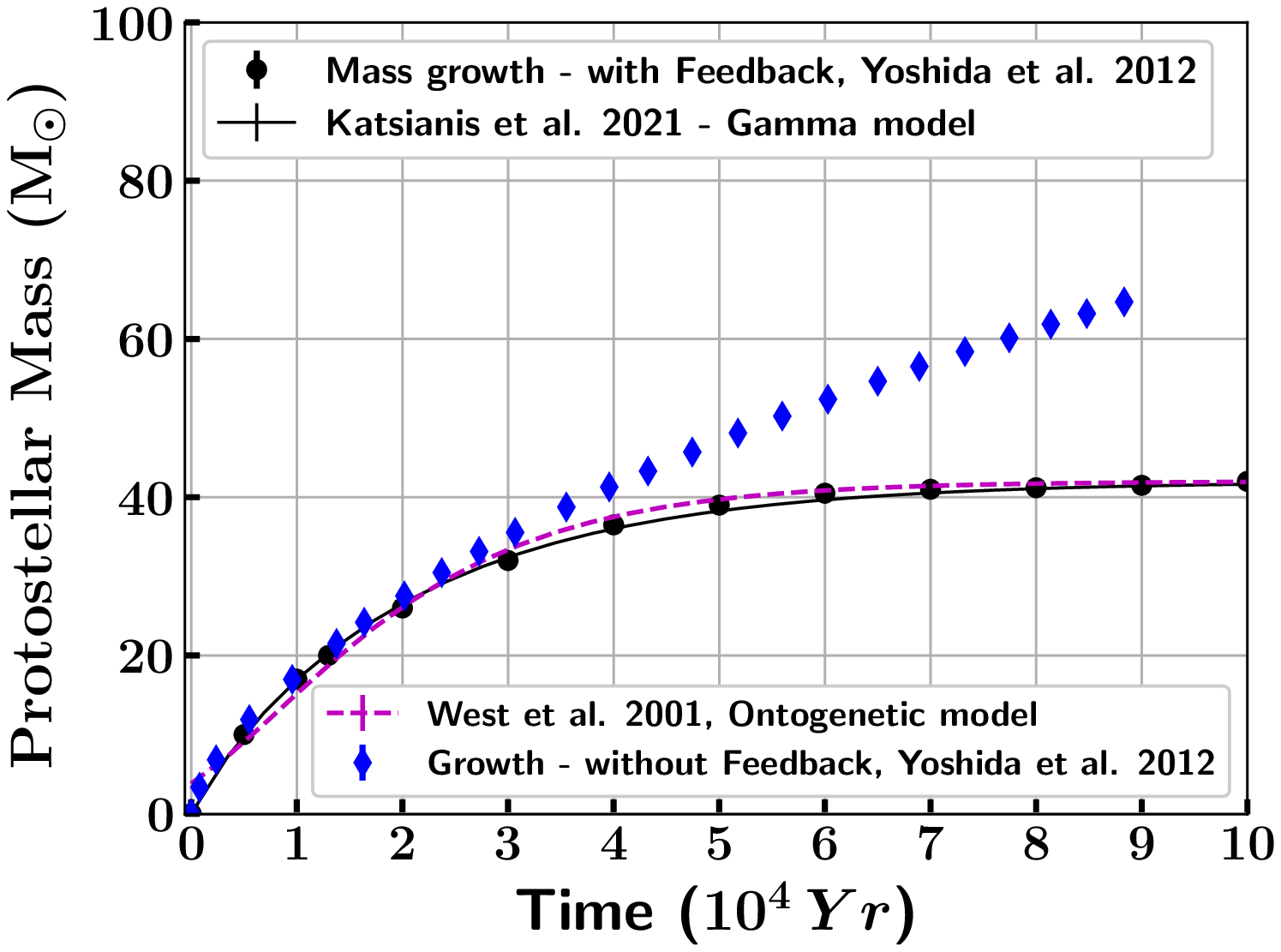}
\includegraphics[scale=0.45]{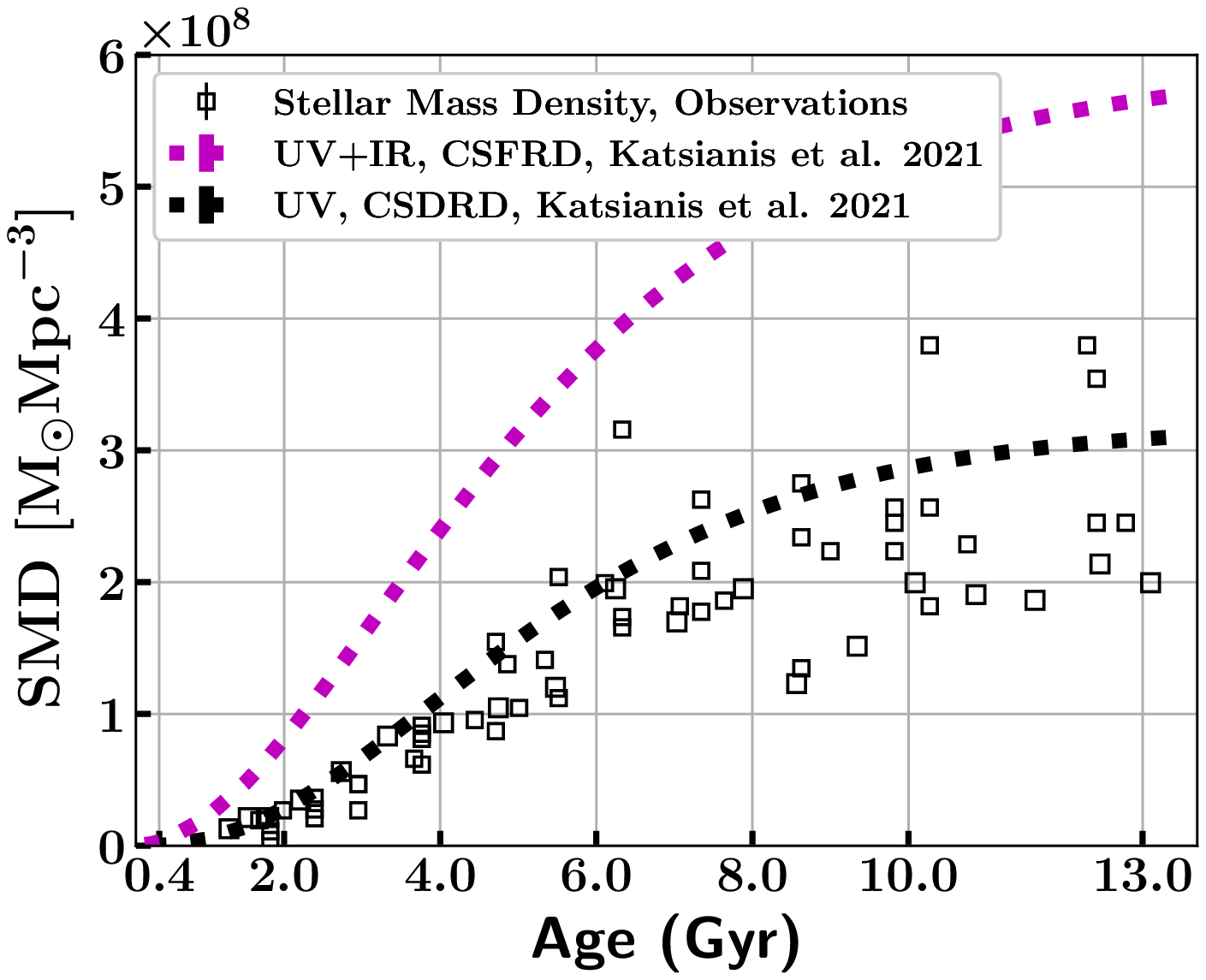}
\vspace{-1.0cm}
\caption{Top left Panel: Black open circles represent the data from the fluorescence experiments of E. Coli performed by \citet{Ram2019}, dotted line represents a fit for the Lag-Log Phase by \citet{Ram2019} and the black solid line a $\Gamma$ model (Eq. \ref{eq:GAMMAmodel}). Top right Panel: growth curve of Cows. The Magenta open diamonds represent the data from \citet{West2001} while the magenta dashed line their fit/growth model. The black solid line represents our fit to their data using a $\Gamma$ model (Eq. \ref{eq:GAMMAmodel}). The green open circles represent the average growth of 3044 Brahman Cattles by \citet{Zimmermann2019}. The black solid line represents our description via a $\Gamma$ model. Bottom left Panel: Protostellar mass growth given by \citet{Yoshoda2012}. The blue diamonds represent the growth without feedback, the black filled circles represent the results with feedback. The magenta dashed dotted line describe a fit using the model of \citet{West2001} while the black solid line our fit using a $\Gamma$ parameterization. Bottom right Panel: The different points represent the observations of the Stellar Mass Density from \citet{Madau2014} and \citet{Driver2018}. Black/magenta dotted line is the $\Gamma$ model of the UV/IR CSFRD given in \citet{Katsianis2021b}. We stress that these do not represent a fit to the Stellar mass density data but instead originate from analysis of the CSFRD. }
\label{figEpicG0Bacteria}
\end{figure*}

The Gamma function is an extremely useful tool that can be applied to a wide range of fields. Its usefulness stems from its ability to model the growth of a sample in a way that is simple and comprehensible. This simple model depends on two parameters $\alpha$/$\beta$ and the final value that is achieved after the growth is complete while it gives an opportunity to follow the evolution of a system analytically. In this appendix we first give an example of the Gamma function used in population and mass growth in other fields, then we describe how it can be applied to understand MAHs and the CSFRD better. 

Growth models in Biology are typically adopted in order to capture the evolution of the mass of an organism or the number density of a population. For example, different parameterizations can be used to model the number density of microbial growth of an isolated system. The above evolution comprises typically 4  phases: 
1) lag phase 2) exponential phase 3) deceleration phase 4) stationary phase. 
The parameterization has to grasp the fact that cell growth at the beginning has to adjust in the conditions (lag phase), then accelerates as there are ample resources which start to be consumed (exponential phase), then decelerates as resources become scarce (deceleration phase), and finally halts when resources are depleted (stationary phase). In the top left panel of Fig. \ref{figEpicG0Bacteria} the black open circles represent the data from the fluorescence experiments of E. Coli performed by \citet{Ram2019}.  According to the authors the growth from the lag phase to the early exponential phase can be described by an exponential fit. We demonstrate that we can describe broadly the growth from the lag phase to the stationary successfully using instead a $\Gamma$ Model. This parameterization is just the integration of a typical $\Gamma$ rate of the form $\frac{dP(T)}{dT} = P_{final} \times \frac{\beta^{\alpha}}{\Gamma(\alpha)} \, \times \, T^{\alpha-1} \times e^{-\beta \, T }$ and the number density of the population as a function of time can be written as: 
\begin{eqnarray}
\label{eq:GAMMAmodel}
{P(T)} =  P_{final} \, \times \frac{\gamma(\alpha, \beta \times T)}{\Gamma(\alpha)},
\end{eqnarray}
where $P_{final}$ is the number density of the population when it reaches its stationary phase, $\alpha$ is the parameter that determines the power-law and thus how the population increases, and $\beta$ is the parameter that describes the deceleration due to the consumption of the finite resources (which depends on how much the population has grown due to the consumption of these resources having a $e^{-\beta \, T }$ dependence) and has units of inverse time.

Growth curves are also typically used to describe the evolution of the mass of animals \citep{Karkach2006}. In the top right panel of Fig. \ref{figEpicG0Bacteria} we present the growth curve of different species of cows. The magenta open diamonds represent the data from \citet{West2001} while the magenta dashed line their fit/growth model. The magenta solid line represents our fit to their data using a $\Gamma$ model (Eq. \ref{eq:GAMMAmodel}). The green open circles represent the average growth of 3044 Braham  Cattles by \citet{Zimmermann2019}. A typical way to describe the growth of animals is the Brody model which is the inverse of the exponential growth curve that involves an asymptotic approach to the maximum size (top right panel of Fig. \ref{figEpicG0Bacteria}). The reason for this behavior (which is found among all species) is the developmental changes which are proportional to the attainment of more mass (i.e. on average the more massive the organism becomes the less fast it grows in mass something reflected in the exponential behavior). 
The black solid line represents our fit using Eq. \ref{eq:GAMMAmodel}.

Moving to Astronomy and primordial stars (protostars), at the moment of the birth of a tiny embryo ($\sim 0.01M_{\odot}$), the initial object is surrounded by a huge amount of the natal gas cloud. Initially it slowly accretes the surrounding gas due to its gravity but the protostar is not massive enough to exercise enough force to the surrounded matter so a slow growth occurs (resembles a lag phase). After this initial phase the protostar rapidly grows in mass by gathering the materials by its gravitational pull as the object is massive enough and there is an abundance of surrounding gas (exponential phase). However, these resources will start getting depleted the larger the star gets ({\it consumption of resource like in the case of the population of Bacteria outlined at the beginning of this section}) and the protostar growth starts being halted (deceleration phase). Once all the gas is depleted, the protostar mass will reach its maximum (stationary phase). The growth is not only halted by the depletion of resources but also by the strength of stellar feedback effect(s) against the accretion flow which are typically proportional to the attainment of  mass (for example a plausible feedback mechanism is gas heating by stellar radiation). Thus, there is a deceleration of the growth the more mass the protostar achieves ({\it like in the case of animal growth}). In the bottom left panel of Fig. \ref{figEpicG0Bacteria} we present the protostellar mass growth \citep{Yoshoda2012}. We demonstrate that we can describe this evolution successfully with a $\Gamma$ Model (and interestingly with a \citep{West2001} model which is originally constructed to just describe animal growth).

We have already demonstrated in a separate work that the evolution of the CSFRD and thus the cosmic stellar mass density can be described succesfully by Gamma forms \citep{Katsianis2021b}. We note that the bottom right panel of Fig. \ref{figEpicG0Bacteria} is the same as in \citet{Katsianis2021b} but here we present the results in a linear scale to emphasize the differences between IR CSFRD (Magenta dotted line) and UV CSFRD (black dotted line) and the uncertainties of the results due to the systematics of different SFR indicators. 

In conclusion, in our current work we decide to focus on the $\Gamma$ growth model (Eq. \ref{eq:GAMMAmodel}) as it is one of the simplest and most flexible parameterizations (useful from the scales of Bacteria Growth to the Cosmic star formation rate density growth) to describe the growth of numerous systems. It also interestingly implies a fractal growth pattern. The power-law (labelled sometimes in the literature as fractal) behavior is the result from a power-law distribution, a self-similar scale free structure or a process that occurs following an exponential distribution \citep{Askar2019,Mori2020,ZiffandZiff}. The exponential cutoff is related to the consumption and exhaustion of resources, which are proportional to the current growth stage of the system giving mathematically the exponential decline.

\label{lastpage}
\end{document}